\documentclass[nofootinbib,aps,superscriptaddress,preprint,floatfix]{revtex4}

\usepackage{graphicx,epsf}

\def\D0bar{\overline D{}^0}
\def\K0bar{\overline K{}^0}

\def\beq{\begin{equation}}
\def\eeq{\end{equation}}
\def\beqa{\begin{eqnarray}}
\def\eeqa{\end{eqnarray}}
\def\bea{\begin{eqnarray}}
\def\eea{\end{eqnarray}}
\def\beq{\begin{equation}}
\def\eeq{\end{equation}}

\def\Re{{\cal R \mskip-4mu \lower.1ex \hbox{\it e}\,}}
\def\Im{{\cal I \mskip-5mu \lower.1ex \hbox{\it m}\,}}
\def\be{\begin{equation}}
\def\ee{\end{equation}}

\def\Re{{\cal R \mskip-4mu \lower.1ex \hbox{\it e}\,}}
\def\Im{{\cal I \mskip-5mu \lower.1ex \hbox{\it m}\,}}
\def\ie{{\it i.e.}}
\def\eg{{\it e.g.}}

\def\etal{{\it et al.}}
\def\ibid{{\it ibid}.}
\def\sub#1{_{\lower.25ex\hbox{$\scriptstyle#1$}}}
\def\sul#1{_{\kern-.1em#1}}
\def\sll#1{_{\kern-.2em#1}}
\def\sbl#1{_{\kern-.1em\lower.25ex\hbox{$\scriptstyle#1$}}}
\def\ssb#1{_{\lower.25ex\hbox{$\scriptscriptstyle#1$}}}
\def\sbb#1{_{\lower.4ex\hbox{$\scriptstyle#1$}}}

\def\to{\rightarrow}
\def\dmix{\ifmmode D^0-\bar D^0 \else $D^0$-$\bar D^0$\fi}
\def\dm{\Delta M_D}

\def\dmd{\ifmmode \Delta M_D \else $\Delta M_D$\fi}
\def\mh{\ifmmode m\sbl H \else $m\sbl H$\fi}
\def\mch{\ifmmode M_{H^\pm} \else $M_{H^\pm}$\fi}
\def\mt{\ifmmode m_t\else $m_t$\fi}
\def\mc{\ifmmode m_c\else $m_c$\fi}
\def\mz{\ifmmode M_Z\else $M_Z$\fi}
\def\mw{\ifmmode M_W\else $M_W$\fi}
\def\mws{\ifmmode M_W^2 \else $M_W^2$\fi}
\def\mhs{\ifmmode M_H^2 \else $M_H^2$\fi}
\def\mzs{\ifmmode M_Z^2 \else $M_Z^2$\fi}
\def\mts{\ifmmode m_t^2 \else $m_t^2$\fi}
\def\mcs{\ifmmode m_c^2 \else $m_c^2$\fi}
\def\mchs{\ifmmode M_{H^\pm}^2 \else $M_{H^\pm}^2$\fi}
\def\ztwo{\ifmmode Z_2\else $Z_2$\fi}
\def\zone{\ifmmode Z_1\else $Z_1$\fi}
\def\mtwo{\ifmmode M_2\else $M_2$\fi}
\def\mone{\ifmmode M_1\else $M_1$\fi}
\def\tb{\ifmmode \tan\beta \else $\tan\beta$\fi}
\def\xw{\ifmmode x\sub w\else $x\sub w$\fi}
\def\ch{\ifmmode H^\pm \else $H^\pm$\fi}
\def\lum{\ifmmode {\cal L}\else ${\cal L}$\fi}
\def\inpb{\ifmmode {\rm pb}^{-1}\else ${\rm pb}^{-1}$\fi}
\def\infb{\ifmmode {\rm fb}^{-1}\else ${\rm fb}^{-1}$\fi}
\def\epem{\ifmmode e^+e^-\else $e^+e^-$\fi}
\def\ppb{\ifmmode \bar pp\else $\bar pp$\fi}

\newskip\zatskip \zatskip=0pt plus0pt minus0pt
\def\matth{\mathsurround=0pt}
\def\lsim{\mathrel{\mathpalette\atversim<}}
\def\gsim{\mathrel{\mathpalette\atversim>}}
\def\atversim#1#2{\lower0.7ex\vbox{\baselineskip\zatskip\lineskip\zatskip
  \lineskiplimit 0pt\ialign{$\matth#1\hfil##\hfil$\crcr#2\crcr\sim\crcr}}}

\def\Re{{\cal R \mskip-4mu \lower.1ex \hbox{\it e}\,}}
\def\Im{{\cal I \mskip-5mu \lower.1ex \hbox{\it m}\,}}
\def\ie{{\it i.e.}}
\def\eg{{\it e.g.}}

\def\etal{{\it et al.}}
\def\ibid{{\it ibid}.}
\def\sub#1{_{\lower.25ex\hbox{$\scriptstyle#1$}}}
\def\sul#1{_{\kern-.1em#1}}
\def\sll#1{_{\kern-.2em#1}}
\def\sbl#1{_{\kern-.1em\lower.25ex\hbox{$\scriptstyle#1$}}}
\def\ssb#1{_{\lower.25ex\hbox{$\scriptscriptstyle#1$}}}
\def\sbb#1{_{\lower.4ex\hbox{$\scriptstyle#1$}}}

\def\to{\rightarrow}

\def\rb{\ifmmode R_b\else $R_b$\fi}
\def\rc{\ifmmode R_c\else $R_c$\fi}
\def\ac{\ifmmode A_c\else $A_c$\fi}
\def\dmix{\ifmmode D^0-\bar D^0 \else $D^0$-$\bar D^0$\fi}
\def\dm{\ifmmode \Delta M_D \else $\Delta M_D$\fi}
\def\rb{\ifmmode R_b\else $R_b$\fi}
\def\mh{\ifmmode m\sbl H \else $m\sbl H$\fi}
\def\mch{\ifmmode M_{H^\pm} \else $M_{H^\pm}$\fi}
\def\mt{\ifmmode m_t\else $m_t$\fi}
\def\mc{\ifmmode m_c\else $m_c$\fi}
\def\mz{\ifmmode M_Z\else $M_Z$\fi}
\def\mw{\ifmmode M_W\else $M_W$\fi}
\def\mws{\ifmmode M_W^2 \else $M_W^2$\fi}

\def\mhs{\ifmmode m_H^2 \else $m_H^2$\fi}
\def\mzs{\ifmmode M_Z^2 \else $M_Z^2$\fi}
\def\mts{\ifmmode m_t^2 \else $m_t^2$\fi}
\def\mcs{\ifmmode m_c^2 \else $m_c^2$\fi}
\def\mchs{\ifmmode m_{H^\pm}^2 \else $m_{H^\pm}^2$\fi}
\def\ztwo{\ifmmode Z_2\else $Z_2$\fi}
\def\zone{\ifmmode Z_1\else $Z_1$\fi}
\def\mtwo{\ifmmode M_2\else $M_2$\fi}
\def\mone{\ifmmode M_1\else $M_1$\fi}
\def\bsg{\ifmmode b\to s\gamma\else $b\to s\gamma$\fi}
\def\tb{\ifmmode \tan\beta \else $\tan\beta$\fi}
\def\xw{\ifmmode x\sub w\else $x\sub w$\fi}
\def\ch{\ifmmode H^\pm \else $H^\pm$\fi}
\def\lum{\ifmmode {\cal L}\else ${\cal L}$\fi}
\def\inpb{\ifmmode {\rm pb}^{-1}\else ${\rm pb}^{-1}$\fi}
\def\infb{\ifmmode {\rm fb}^{-1}\else ${\rm fb}^{-1}$\fi}
\def\epem{\ifmmode e^+e^-\else $e^+e^-$\fi}
\def\ppb{\ifmmode \bar pp\else $\bar pp$\fi}

\def\be{\begin{equation}}
\def\ee{\end{equation}}

\begin{document}
\vspace{3.0cm}
\preprint{\vbox {\hbox{SLAC-PUB-12496}
\hbox{WSU--HEP--0701} \hbox{UH-511-1104-07}
%\hbox{hep-ph/yymmddd}
}}

\vspace*{2cm}

\title{\boldmath Implications of $D^0$-${\bar D}^0$ Mixing for New Physics}

\author{Eugene Golowich}
\affiliation{Department of Physics,
        University of Massachusetts\\[-6pt]
        Amherst, MA 01003}

\author{JoAnne Hewett}
\affiliation{Stanford Linear Accelerator Center, 
        Stanford University\\[-6pt]
        Stanford, CA 94309}

\author{Sandip Pakvasa}
\affiliation{Department of Physics and Astronomy\\[-6pt],
        University of Hawaii, 
        Honolulu, HI 96822}

\author{Alexey A.\ Petrov\vspace{8pt}}
\affiliation{Department of Physics and Astronomy\\[-6pt]
        Wayne State University, Detroit, MI 48201\\[-6pt] $\phantom{}$ }

\begin{abstract}
We provide a comprehensive, up-to-date analysis of possible New Physics 
contributions to the mass difference $\Delta M_D$ in $D^0$-${\overline D}^0$ 
mixing. We consider the most general low energy effective Hamiltonian 
and include leading order QCD running of effective operators.
We then explore an extensive list of possible New Physics models that
can generate these operators, which we organize as including 
Extra Fermions, Extra Gauge Bosons, Extra Scalars, Extra Space Dimensions 
and Extra Symmetries. For each model we place restrictions on the allowed 
parameter space using the recent evidence for observation of $D$ meson
mixing.  In many scenarios, we find strong constraints that
surpass those from other search techniques and provide an important
test of flavor changing neutral currents in the up-quark sector.
We also review the recent BaBar and Belle findings, and
describe the current status of the Standard Model predictions of 
$D^0$-${\overline D}^0$ mixing.
\vskip 1in
%\Large{Version: JLH 05/23/07}
%
\end{abstract}

\def\thepage{{}}
\maketitle
\def\thepage{\arabic{page}}

%%%%%%%%%%%%%%%%%%%%%%%%%%%%%%%%%%%%%%%%%%%%%%%%%%%%%%%%%%%%%%%%%%%%%
\section{Introduction}
Meson-antimeson mixing has traditionally been of importance 
because it is sensitive to heavy degrees of freedom that propagate in
the underlying mixing amplitudes. Estimates of the charm quark 
and top quark mass scales were inferred from the observation of mixing in 
the $K^0$ and $B_d$ systems, respectively, 
before these particles were discovered directly.

This success has motivated attempts to indirectly detect New Physics (NP) 
signals by comparing the observed meson mixing with predictions of 
the Standard Model (SM).  Mixing in the Kaon sector has historically
placed stringent constraints on the parameter space of theories
beyond the SM and provides an essential hurdle that must be passed in the
construction of models with NP.  However, anticipated breakthroughs 
from the B-factories and the Tevatron collider have not been borne out -- 
the large mixing signal in the $B_d$ and $B_s$ systems is
successfully described in terms of the SM alone (although 
the parameter spaces of various NP models have become increasingly 
constrained).  Short of awaiting LHCB and the construction of 
a super-B facility, there is one remaining example for possibly 
observing indirect signs of NP in meson mixing, the $D^0$ flavor 
oscillations.  In this case, the SM mixing rate is 
sufficiently small that the NP component might be able to 
compete~\cite{datta}.
There has been a flurry of recent experimental activity regarding 
the detection of $D^0$-${\bar D}^0$ 
mixing~\cite{babar,belle,staric,babar:2007aa},
which marks the first time Flavor Changing Neutral Currents (FCNC)
have been observed in the charged $+2/3$ quark sector.  With the potential
window to discern large NP effects in the charm 
sector~\cite{bghp,Burdman:2001tf}
and the anticipated improved accuracy for future mixing measurements,
the motivation for a comprehensive up-to-date theoretical analysis 
of New Physics contributions to $D$ meson mixing is compelling.
  
\subsection{Observation of Charm Mixing}
The heightened interest in $D^0$-${\bar D}^0$ mixing started with 
the almost simultaneous observations by the BaBar~\cite{babar} 
and Belle~\cite{belle} collaborations 
of nonzero mixing signals at about the per cent 
level,\footnote{Our definitions of the mixing parameters 
$x_{\rm D}$, $y_{\rm D}$, $y'_{\rm D}$ and $y_{\rm D}^{\rm (CP)}$ 
are standard and are given 
in Eqs.~(\ref{xy}),(\ref{y-defs}).}
\beqa
& & y_{\rm D}' = (0.97 \pm 0.44 \pm 0.31) \cdot 10^{-2} \qquad 
{\rm (BaBar)} \ , \\
& & y_{\rm D}^{\rm (CP)} = (1.31 \pm 0.32 \pm 0.25)\cdot 10^{-2} 
\qquad  {\rm (Belle)}\ \ .
\eeqa
This was soon followed by the announcement by the Belle collaboration 
of mixing measurements from the Dalitz plot analyses
of $D^0 \to K_S \pi^+ \pi^-$~\cite{staric}, 
\beq
x_{\rm D} = (0.80 \pm 0.29 \pm 0.17)\cdot 10^{-2}~, \ \ \qquad 
y_{\rm D} = (0.33 \pm 0.24 \pm 0.15)\cdot 10^{-2} \ \ .
\eeq
A preliminary fit to the current database\footnote{An updated 
fit\cite{asner} gives the
values $ x_{\rm D} = 8.4^{+3.2}_{-3.4} \cdot 10^{-3}$~,  
$y_{\rm D} = (6.9\pm 2.1) \cdot 10^{-3}$. These are essentially unchanged
from the HFAG preliminary results given above and used in our analysis; the
difference will not affect our numerical results.}
by the Heavy
Flavor Averaging Group (HFAG) gives~\cite{asner} 
\beqa\label{hfag}
& & x_{\rm D} = 8.7^{+3.0}_{-3.4} \cdot 10^{-3}~, \quad 
y_{\rm D} = (6.6\pm 2.1) \cdot 10^{-3} \ \ .
%\qquad (\cos\delta_{K\pi} = 1.09 \pm 0.66) \ \ , 
%& & x = 8.4^{+3.3}_{-3.0} \cdot 10^{-3}~, \quad 
%y = 6.4^{+1.6}_{-1.8} \cdot 10^{-3} \qquad (\cos\delta_{K\pi} = 
%1.0 \pm 0.1) \ \ , 
\eeqa
%where the associated value of the strong phase 
%$\cos\delta_{K\pi}$ corresponds 
%to the current CLEO-c determination~\cite{strongphasedet}.  
Since this paper addresses 
the issue of the mass splitting induced by mixing, our primary 
concern is with the signal for $x_{\rm D}$, seen here to be a 2.4 sigma 
effect.  This is below the generally accepted threshold for ``evidence'' 
and is more in the nature of a ``hint''.  However, we note that a 2.4 
sigma effect will automatically have a non-zero lower bound at 
$95\%$ confidence-level.  For the sake of 
reference, we cite the one-sigma window for the HFAG value of $x_{\rm D}$, 
\beq
5.4 \cdot 10^{-3} < x_{\rm D} < 11.7 \cdot 10^{-3} 
\qquad {\rm (one-sigma\ window)} \ \ ,
\label{xbnds}
\eeq
or equivalently for $\Delta M_{\rm D}$ itself, 
\beq
8.7 \cdot 10^{-15} \mbox{~GeV} < \Delta M_{\rm D} < 1.9 
\cdot 10^{-14}\mbox{~GeV}
\qquad {\rm (one-sigma\ window)} \ \ .
\label{mbnds}
\eeq

Let us briefly describe our strategy for dealing with the above 
HFAG values in light of both SM and NP contributions.  We shall 
argue in Sect.~III that the SM predictions, although indeed compatible 
with the observed range of values for the $D$ mixing parameters, 
contain significant 
hadronic uncertainties.  Moreover, we do not know the 
relative {\it phase} between the SM contribution and that from any 
NP model, so that $x_{\rm D}$ will lie between the extreme limiting 
cases of constructive and destructive interference.  In addition, since the
observation of $D$ mixing is new, the measurements will fluctuate with
future refinements in the analyses and as more data is collected.
To best deal with these realities, we will present our results 
by displaying a given NP prediction as a 
pure NP signal ({\it i.e.} as if there were no 
SM component) and for comparison, display curves of constant $x_{\rm D}$ 
for the five values 
\beq
x_{\rm D} = 15.0 \cdot 10^{-3}~,\qquad 
11.7 \cdot 10^{-3}~,\qquad 8.0 \cdot 10^{-3}~,\qquad 
5.0 \cdot 10^{-3}~,\qquad 3.0 \cdot 10^{-3} \ \ .
\label{xrange}
\eeq
This (approximately HFAG 2$\sigma$) range reveals 
the sensitivity of $x_{\rm D}$ to variations in the underlying 
NP parameter space.  We will then show the present constraints placed
on the NP model parameter space, by assuming that the NP contribution
cannot exceed the $1\sigma$ upper bound on $x_D$.  This procedure
mirrors that which is traditionally employed in obtaining bounds on
NP from $K^0$-$\overline K^0$ mixing.

\subsection{New Physics Possibilities} 
$D^0$-${\bar D}^0$ mixing at the observed level is much 
larger than the quark-level (`short-distance') SM 
prediction~\cite{Golowich:2005pt} but is in qualitative accord with 
hadron-level (`long-distance') SM expectations.  However, because 
the latter are beset with hadronic uncertainties, 
it cannot be rigorously concluded that only SM physics is being 
detected.  In this paper, we will consider a broad menu of NP 
possibilities.  As the operation of the LHC looms near, the number of 
potentially viable NP models has never been greater.  Our 
organizational approach to analyzing these is to address 
NP models with:  
\begin{enumerate}
\item Extra Fermions (Sect.~IV)
\subitem A: Fourth Generation 
\subitem B: Heavy Vector-like Quarks 
\subsubitem (1): $Q = -1/3$ Singlet Quarks
\subsubitem (2): $Q = +2/3$ Singlet Quarks
\subitem C: Little Higgs Models 
\item Extra Gauge Bosons (Sect.~V)
\subitem A: Generic Z' Models 
\subitem B: Family Symmetries
\subitem C: Left-Right Symmetric Model 
\subitem D: Alternate Left-Right Models from E$_6$ Theories
%\subitem D: Twin Higgs Model
\subitem E: Vector Leptoquark Bosons
\item Extra Scalars (Sect.~VI)
%\subitem Two-Higgs Doublet Models, 
\subitem A: Flavor Conserving Two-Higgs-Doublet Models
\subitem B: Flavor Changing Neutral Higgs Models
\subitem C: Scalar Leptoquark Bosons 
\subitem D: Higgsless Models
\item Extra Space Dimensions (Sect.~VII) 
\subitem A: Universal Extra Dimensions
\subitem B: Split Fermion Models
\subitem C: Warped Geometries
\item Extra Symmetries (Sect.~VIII) 
\subitem A: Minimal Supersymmetric Standard Model
\subitem B: Quark-Squark Alignment Models
\subitem C: Supersymmetry with R-Parity Violation 
\subitem D: Split Supersymmetry 
\end{enumerate}
In the above, we have chosen to consider only supersymmetry in 
Sect.~VIII due to its extensive literature and to cover other 
extended symmetries elsewhere in the paper.

Any NP degree of freedom will generally be associated with 
a generic heavy mass scale $M$, at which the NP interaction will 
be most naturally described.  At the scale $m_c$ of the charm mass, 
this description will have been modified by the 
effects of QCD.  These should not be neglected, so we perform 
our NP analyses at one-loop level for the strong interactions.  
The theoretical background for this is presented in Sect.~II.

Finally, in order to place the NP discussion within its proper context, 
it makes sense to first review SM charm mixing.  This is done in
Sect.~III.  The remainder of 
the paper then amounts to considering charm mixing with lots of `extras'. 
The paper concludes in Sect.~IX with a summary of our findings.

%The paper concludes in Sect.~IX with a summary of our findings, 
%including a comprehensive numerical Table for the convenience of the
%reader.

\subsection{Basic Formalism}
Let us first review some formal aspects 
of charm mixing.  The mixing arises from $|\Delta C|=2$ interactions 
that generate 
off-diagonal terms in the mass matrix for $D^0$ and $\D0bar$ mesons.
The expansion of the off-diagonal terms in the neutral $D$ mass
matrix to second order in the weak interaction is 
\beq\label{M12}
\left (M - \frac{i}{2}\, \Gamma\right)_{21} =
  \frac{1}{2M_D}\, \langle \D0bar | {H}_w^{|\Delta C|=2} | D^0 \rangle +
  \frac{1}{2M_D}\, \sum_n {\langle \D0bar | {H}_w^{|\Delta C|=1} | n 
  \rangle\, \langle n | {H}_w^{|\Delta C|=1} | D^0 \rangle 
  \over M_D-E_n+i\epsilon} \ \ ,
\eeq
where ${H}_w^{|\Delta C|=2}$ and ${H}_w^{|\Delta C|=1}$ 
are the effective $|\Delta C|=2$ and $|\Delta C|=1$ hamiltonians.

 The off-diagonal mass-matrix terms induce mass eigenstates 
$D_1$ and $D_2$ that are superpositions of the 
flavor eigenstates $D^0$ and $\D0bar$,
\begin{equation} \label{definition1}
    D_{{\rm 1}\atop{\rm 2}} 
= p\,  D^0 ~ \pm ~ q\,  \D0bar \ \ ,
\end{equation}
where $|p|^2 + |q|^2=1$.  The key quantities in $D^0$ mixing are 
the mass and width differences, 
\begin{equation} 
\Delta M_{\rm D} \equiv M_{\rm 1} - M_{\rm 2} 
\qquad  {\rm and} \qquad \Delta \Gamma_{\rm D} \equiv \Gamma_{\rm 1} - 
\Gamma_{\rm 2}  \ \ , 
\label{diffs}
\end{equation}
or equivalently their dimensionless equivalents, 
\beq
x_{\rm D} \equiv {\Delta M_{\rm D} \over \Gamma_{\rm D}}, \qquad {\rm and} 
\qquad y_{\rm D} \equiv {\Delta \Gamma_{\rm D} \over 2\Gamma_{\rm D}}\ \ ,
\label{xy}
\eeq
where $\Gamma_{\rm D}$ is the average width of the two 
neutral $D$ meson mass eigenstates.  Two quantities, 
$y_{\rm D}^{\rm (CP)}$ and $y_{\rm D}'$, 
which are actually measured in most experimental 
determinations of $\Delta \Gamma_{\rm D}$, are 
defined as 
\beqa
y_{\rm D}^{\rm (CP)} &\equiv& (\Gamma_+ - \Gamma_-)/  
(\Gamma_+ + \Gamma_-) = y_{\rm D} \cos\phi - x_{\rm D}\sin\phi
\left(\frac{A_m}{2}-A_{prod}\right) \ \ , \nonumber \\
y_{\rm D}' &\equiv& y_{\rm D} \cos \delta_{K\pi} - x_{\rm D} 
\sin\delta_{K\pi} \ \ ,
\label{y-defs}
\eeqa
where the transition rates $\Gamma_\pm$ pertain to decay into 
final states of definite CP, 
$A_{prod} = \left(N_{D^0} - N_{{\overline D}^0}\right)/
\left(N_{D^0} + N_{{\overline D}^0}\right)$ is the so-called 
production asymmetry of $D^0$ and $\overline{D}^0$ (giving 
the relative weight of $D^0$ and ${\overline D}^0$ in the 
sample) and $\delta_{K\pi}$ is the strong phase difference between 
the Cabibbo favored and double Cabibbo suppressed 
amplitudes~\cite{Falk:1999ts}.  The quantities 
$A_m$ and $\phi$ account for the presence of CP violation in 
$D^0$-${\bar D}^0$ mixing, with $A_m$ being related to the $q,p$ 
parameters of Eq.~(\ref{definition1}) as $A_m \equiv |q/p|^2 - 1$ 
and $\phi$ a CP-violating phase of $M_{21}$ (if one neglects 
direct CP-violation)~\cite{Bergmann:2000id}.  In practice,
$y_{\rm CP}$ is measured by comparing decays of $D^0$ into a state of
definite CP, such as $K^+K^-$, to decays of $D^0$ into a final state 
which is not a CP-eigenstate (such as $K\pi$) whereas $y'$ is 
extracted from a time-dependent analysis of the 
$D\to K\pi$ transition~\cite{Bergmann:2000id}.

The states $D_{{\rm 1}\atop{\rm 2}}$ allow for effects of 
CP violation.  However, CP violation in $D^0$ 
mixing is negligible in the Standard Model and there is no 
evidence for it experimentally~\cite{babar,Nir:2007ac,PDG}.
Many New Physics scenarios contain new phases which can induce
sizable CP violation in the $D$ meson sector.  Nonetheless, a
thorough investigation of such effects is beyond the scope of the
present paper.  Therefore, we shall work in the 
limit of CP invariance (so that $p=q$) for the remainder of this 
paper.  
Throughout, our phase convention will be 
\begin{equation} \label{phase}
 {\cal C} {\cal P}~   D^0 =  + ~  {\overline{D}}^0 \ \ .
\end{equation}
Then $ D_{{\rm 1}\atop{\rm 2}} $ become the CP eigenstates 
$ D_{\pm}$ with  
${\cal C}{\cal P} ~ D_{\pm} = \pm ~ D_{\pm} $.

Keeping in mind the neglect of CP-violation and also 
the phase convention of Eq.~(\ref{phase}), we relate 
the mixing quantities $x_{\rm D}$ and $y_{\rm D}$ to the mixing matrix as 
\begin{eqnarray}
\label{ope}
& & x_{\rm D} = \frac{1}{2M_{\rm D}\Gamma_{\rm D}}\, {\rm Re}\, 
\left[ 2\langle \D0bar | {H}^{|\Delta C|=2}\, 
| D^0 \rangle +  \langle \D0bar |
    \,i\! \int\! {\rm d}^4 x\, T \Big\{
    {\cal H}^{|\Delta C|=1}_w (x)\, {\cal H}^{|\Delta C|=1}_w(0) \Big\}
    | D^0 \rangle \right] \ \ , \nonumber \\
%  \sum_n {\langle \D0bar | {H}_w^{|\Delta C|=1} | n 
%  \rangle\, \langle n | {H}_w^{|\Delta C|=1} | D^0 \rangle 
%  \over M_D-E_n+i\epsilon} \right] \ \ ,
& & y_{\rm D} = \frac{1}{2 M_{\rm D}\Gamma_{\rm D}}\, 
{\rm Im}\, \langle \D0bar |
    \,i\! \int\! {\rm d}^4 x\, T \Big\{
    {\cal H}^{|\Delta C|=1}_w (x)\, {\cal H}^{|\Delta C|=1}_w(0) \Big\}
    | D^0 \rangle \ \ , 
\end{eqnarray}
where ${\cal H}^{|\Delta C|=1}_w(x)$ is the weak hamiltonian density 
for $|\Delta C|=1$ transitions and $T$ denotes the time-ordered product.
There is no contribution to $y_{\rm D}$ from the local $|\Delta C|=2$ 
term, as it has no absorptive part.  New Physics contributions 
to $y_{\rm D}$ have already been addressed in Ref.~\cite{Golowich:2006gq}, 
so the primary thrust of this paper will be to focus on $x_{\rm D}$.  

The next step, in Sect.~II, is to expand the time-ordered product 
of Eq.~(\ref{ope}) in local operators of increasing dimension 
(higher dimension operators being suppressed by powers of 
$\Lambda_{\rm QCD}/m_c$).  

%%%%%%%%%%%%%%%%%%%%%%%%%%%%%%%%%%%%%%%%%%%%%%%%%%%%%%%%%%%%%%%%%%%%%
\section{Generic Operator Analysis of $D^0$-${\bar D}^0$ Mixing}

Though the particles present in models with New Physics may not 
be produced in charm quark decays, their effects can nonetheless 
be seen in the form of effective operators generated by the exchanges
of these new particles. Even without specifying
the form of these new interactions, we know that their effect 
is to introduce several $|\Delta C|=2$
effective operators built out of the SM degrees of freedom.

\subsection{Operator Product Expansion and Renormalization 
Group}\label{RGSect}

By integrating out new degrees
of freedom associated with new interactions at a scale $M$, we are
left with an effective hamiltonian written in the form of a
series of operators of
increasing dimension. Operator power counting then tells us
the most important
contributions are given by the operators of the lowest possible
dimension,
$d=6$ in this case. This means that they must contain only
quark degrees of freedom.
Realizing this, we can write the complete basis of
these effective operators,
which can be done most conveniently in terms of chiral quark fields,
\beq\label{SeriesOfOperators}
\langle f | {\cal H}_{NP} | i \rangle =
G \sum_{i=1} {\rm C}_i (\mu) ~
\langle f | Q_i  | i \rangle (\mu) \ \ ,
\eeq
where the prefactor $G$ has the dimension of inverse-squared mass, the 
%(associated with the NP scale), the
${\rm C}_i$ are dimensionless Wilson coefficients,\footnote{Throughout 
this paper, we shall denote Wilson coefficients for $|\Delta C| = 1$
operators as $\{ {\rm c}_i \}$ and those for 
$|\Delta C| = 2$ operators as $\{ {\rm C}_i \}$.} 
and the $Q_i$ are the effective operators:
\beqa
\begin{array}{l}
Q_1 = (\overline{u}_L \gamma_\mu c_L) \ (\overline{u}_L \gamma^\mu
c_L)\ , \\
Q_2 = (\overline{u}_L \gamma_\mu c_L) \ (\overline{u}_R \gamma^\mu
c_R)\ , \\
Q_3 = (\overline{u}_L c_R) \ (\overline{u}_R c_L) \ , \\
Q_4 = (\overline{u}_R c_L) \ (\overline{u}_R c_L) \ ,
\end{array}
\qquad
\begin{array}{l}
Q_5 = (\overline{u}_R \sigma_{\mu\nu} c_L) \ ( \overline{u}_R
\sigma^{\mu\nu} c_L)\ , \\
Q_6 = (\overline{u}_R \gamma_\mu c_R) \ (\overline{u}_R \gamma^\mu
c_R)\ , \\
Q_7 = (\overline{u}_L c_R) \ (\overline{u}_L c_R) \ , \\
Q_8 = (\overline{u}_L \sigma_{\mu\nu} c_R) \ (\overline{u}_L
\sigma^{\mu\nu} c_R)\ \ .
\end{array}
\label{SetOfOperators}
\eeqa
In total, there are eight possible operator structures that
exhaust the
list of possible independent contributions to $|\Delta C|=2$ transitions.
Since these operators are generated at the scale $M$
where the New Physics is 
integrated out, a non-trivial operator mixing can occur
when we take into account
renormalization group running of these operators between the scales $M$
and $\mu$, with $\mu$ being the scale where the 
hadronic matrix elements are computed.
We shall work at the renormalization scale $\mu= m_c \simeq 1.3$~GeV.
This evolution is determined by solving the RG equations obeyed by
the Wilson coefficients,
\beq\label{AnomEq}
\frac{d}{d \log \mu} \vec C (\mu) = \hat \gamma^T \vec C (\mu)\ \ ,
\eeq
where $\hat \gamma$ represents the matrix of anomalous dimensions of
the operators in Eq.~(\ref{SetOfOperators}) (note the transposition).
Eq.~(\ref{AnomEq}) can be solved by transforming to the basis
where the transpose of the anomalous dimension matrix is diagonal,
integrating, and then transferring back
to the original basis $\vec C_i$. At leading order, we have
\beq
\vec C(\mu) = \hat U (\mu,M) \vec C(M)\ \ ,
\eeq
where $U (\mu,M)$ is the evolution matrix, obtained from
Eq.~(\ref{AnomEq})
by
\beq
\hat U (\mu_1,\mu_2) = \hat V
\left[r(\mu_1,\mu_2)^{{\vec\gamma}^{(0)}/2\beta_0}\right]_D
\hat V^{-1}\ \ .
\eeq
In the above, $\vec\gamma^{(0)}$ is the vector containing the diagonal
elements
of the diagonalized transposed matrix of the anomalous dimensions $\hat
\gamma^T$, ${\hat V}$ is the matrix that diagonalizes $\hat \gamma^T$ and
\beq
r(\mu_1,\mu_2) \equiv \frac{\alpha_s(\mu_1)}{\alpha_s(\mu_2)} \ \ .
\eeq
For completeness, we display the matrix of anomalous dimensions 
at leading-order (LO) in QCD~\cite{Ciuchini:1997bw}, 
\beqa
\hat\gamma=
\left(
     \begin{array}{c c c c c c c c}
6-\frac{6}{N_c}& 0& 0& 0& 0& 0& 0& 0 \\
0& \frac{6}{N_c}& 12& 0& 0& 0& 0& 0 \\
0& 0& -6 N_c+\frac{6}{N_c}& 0& 0& 0& 0& 0 \\
0& 0& 0& 6-6 N_c+\frac{6}{N_c}& \frac{1}{2}-\frac{1}{N_c}& 0& 0& 0 \\
0& 0& 0& -24-\frac{48}{N_c}& 6+2 N_c - \frac{2}{N_c}& 0& 0& 0 \\
0& 0& 0& 0& 0& 6-\frac{6}{N_c}& 0& 0 \\
0& 0& 0& 0& 0& 0& 6-6 N_c+\frac{6}{N_c}& \frac{1}{2}-\frac{1}{N_c} \\
0& 0& 0& 0& 0& 0& -24-\frac{48}{N_c}& 6+2 N_c - \frac{2}{N_c} \\
     \end{array}
  \right)
\nonumber
\eeqa
We note that Ref.~\cite{Ciuchini:1997bw} also includes the next-to-leading
order (NLO)
expressions for the elements in the anomalous dimensions matrix.  However,
we perform our calculations at LO here since the NLO corrections to 
the matching conditions in the various models of New Physics have generally
not been computed.

Due to the relatively simple structure of $\hat\gamma$,
one can easily write the evolution of each Wilson coefficient in
Eq.~(\ref{SeriesOfOperators}) from the New Physics scale
${\rm M}$ down to the hadronic scale $\mu$, taking into
account quark thresholds.  Corresponding to each of the eight
operators $\{ Q_i \}$ ($i=1,\dots,8$) is an RG factor
$r_i (\mu , M)$.  The first of these, $r_1(\mu,M)$, is given
explicitly by
\beqa
r_1(\mu,M)&=& \left(\frac{\alpha_s(M)}{\alpha_s(m_t)}\right)^{2/7}
\left(\frac{\alpha_s(m_t)}{\alpha_s(m_b)}\right)^{6/23}
\left(\frac{\alpha_s(m_b)}{\alpha_s(\mu)}\right)^{6/25} \ \ .
\label{wilson}
\eeqa
and the rest can be expressed in terms of $r_1(\mu,M)$ as
\beqa
\begin{array}{l}
r_2(\mu,M) = [r_1(\mu,M)]^{1/2} \ , \\
r_3(\mu,M) = [r_1(\mu,M)]^{-4} \ , \\
r_4(\mu,M) = [r_1(\mu,M)]^{(1 + \sqrt{241})/6} \ ,\\
r_5(\mu,M) = [r_1(\mu,M)]^{(1 - \sqrt{241})/6} \ ,
\end{array}
\qquad
\begin{array}{l}
\phantom{xxxx} \\
r_6(\mu,M) = r_1(\mu,M) \ , \\
r_7(\mu,M) = r_4(\mu,M) \ , \\
r_8(\mu,M) = r_5(\mu,M) \ \ . 
\end{array}
\eeqa
The RG factors are generally only weakly dependent on the NP scale
$M$ since it is taken to be larger than the top quark mass, $m_t$,
and the evolution of $\alpha_s$ is slow at these high mass scales.
In Table~\ref{tab:LO}, we display numerical values for the
$r_i (\mu , M)$ with $M = 1,2$~TeV and $\mu = m_c \simeq 1.3$~GeV.
Here, we compute $\alpha_s$ using the one-loop evolution and matching 
expressions for perturbative consistency with the RG evolution of the
effective hamiltonian.
\begin{table}[t]
\begin{tabular}{c||c|c|c|c|c}
\colrule\hline 
$M{\rm (TeV)}$ & $r_1(m_c, M)$ & $r_2(m_c, M)$ & $r_3(m_c, M)$ 
& $r_4(m_c, M)$ & $r_5(m_c, M)$ \\
\colrule \colrule
$1$ & $0.72$ & $0.85$ & $3.7$ & $0.41$ & $2.2$ \\
\colrule
$2$ & $0.71$ & $0.84$ & $4.0$ & $0.39$ & $2.3$ \\
\colrule\hline
\end{tabular}
\vskip .05in\noindent
\caption{Dependence of the RG factors on the heavy mass scale $M$.}
\label{tab:LO}
\end{table}

\subsection{Operator Matrix Elements}
We will need to evaluate the $D^0$-to-${\bar D}^0$ matrix elements of
the eight dimension-six basis operators. In general, this implies
eight non-perturbative parameters that would have to be evaluated
by means of QCD sum rules or on the lattice. We choose those
parameters (denoted by $\{B_i\}$) as follows,
\begin{eqnarray}\label{ME}
& & \begin{array}{l}
\langle Q_1 \rangle = {2 \over 3} f_{\rm D}^2 M_{\rm D}^2 B_1\ , \\
\langle Q_2 \rangle = - {5 \over 6} f_{\rm D}^2 M_{\rm D}^2 B_2 \ , \\
\langle Q_3 \rangle = {7 \over 12} f_{\rm D}^2 M_{\rm D}^2 B_3 \ ,\\
\langle Q_4 \rangle = - {5 \over 12} f_{\rm D}^2 M_{\rm D}^2 B_4 \ ,
\end{array}
\quad \qquad
\begin{array}{l}
\langle Q_5 \rangle = f_{\rm D}^2 M_{\rm D}^2 B_5 \ ,\\
\langle Q_6 \rangle = {2 \over 3} f_{\rm D}^2 M_{\rm D}^2 B_6 \ ,\\
\langle Q_7 \rangle = - {5 \over 12} f_{\rm D}^2 M_{\rm D}^2 B_7 \ ,\\
\langle Q_8 \rangle = f_{\rm D}^2 M_{\rm D}^2 B_8 \ \ ,
\end{array}
%\nonumber
\end{eqnarray}
where $\langle Q_i \rangle \equiv \langle {\bar D}^0
| Q_i | D^0 \rangle$, and $f_D$ represents the $D$ meson decay constant.
By and large, the compensatory $B$-factors $\{ B_i \}$ are unknown, 
except in vacuum saturation and in the heavy quark limit; there, one 
has $B_i \to 1$.

Since most of the matrix elements in Eq.~(\ref{ME}) are not known, 
we will need something more manageable in order to obtain numerical
results.  The usual approach to computing matrix elements is to
employ the vacuum
saturation approximation. However, because some of the 
$B$-parameters are known, we would like to introduce a 
`modified vacuum saturation' (MVS), where
all matrix elements in Eq.~(\ref{ME}) are written in terms of (known)
matrix elements of $(V-A)\times (V-A)$ and $(S-P)\times (S+P)$ matrix
elements $B_{\rm D}$ and $B_{\rm D}^{\rm (S)}$, 
%\equiv B_{\rm S}$,
%
\begin{eqnarray}\label{ME_MVS}
& & \begin{array}{l}
\langle Q_1 \rangle = {2 \over 3} f_{\rm D}^2 M_{\rm D}^2 B_D \ ,\\
\langle Q_2 \rangle = -{1 \over 2} f_{\rm D}^2 M_{\rm D}^2 B_D
   - \displaystyle{1 \over N_c} f_{\rm D}^2 M_{\rm D}^2 
{\bar B}_{\rm D}^{\rm (S)} \ ,\\
\langle Q_3 \rangle = \displaystyle{1 \over 4 N_c} 
f_{\rm D}^2 M_{\rm D}^2 B_D
   + {1 \over 2} f_{\rm D}^2 M_{\rm D}^2 {\bar B}_{\rm D}^{\rm (S)} \ ,\\
\langle Q_4 \rangle = - \displaystyle{2 N_c - 1 \over 4 N_c} 
f_{\rm D}^2 M_{\rm D}^2 {\bar B}_{\rm D}^{\rm (S)} \ ,
\end{array}
\qquad \qquad
\begin{array}{l}
\langle Q_5 \rangle = \displaystyle{3 \over N_c} 
f_{\rm D}^2 M_{\rm D}^2 {\bar B}_{\rm D}^{\rm (S)} \ , \\
\langle Q_6 \rangle = \langle Q_1 \rangle \ , \\
\langle Q_7 \rangle = \langle Q_4 \rangle \ , \\
\langle Q_8 \rangle = \langle Q_5 \rangle \ \ ,
\end{array}
%\nonumber
\end{eqnarray}
where we denote $N_c=3$ as the number of colors and, 
as in Ref.~\cite{Golowich:2005pt}, define 
\beq\label{bbar}
{\bar B}_{\rm D}^{(S)} 
\equiv B_{\rm D}^{\rm (S)} \cdot {M_{\rm D}^2 \over (m_c+m_u)^2}  
\eeq
as well as 
\beq\label{eta}
\eta \equiv {{\bar B}_{\rm D}^{(S)} \over B_{\rm D}} \ \ .
\eeq

%We shall use both
%Eq.~(\ref{ME}) and Eq.~(\ref{ME_MVS}) in due course.  
In our numerical
work, we take $B_{\rm D}^{\rm (S)}=B_{\rm D}=0.82$, which is the
most recent result from the quenched lattice calculation~\cite{Gupta:1996yt},
and use the CLEO-c
determination $f_D = 222.6 \pm 16.7 ^{+2.3}_{-2.4}~{\rm MeV}$~\cite{cleoc}.
We urge the lattice community to perform an evaluation of the 
$\{ B_i\}$ parameters defined in Eq.~(\ref{ME}) for the full 
operator set relevant to $D$ meson mixing.

%%%%%%%%%%%%
%%%%%%%%%%%%%%%%%%%%%%%%%%%%%%%%%%%%%%%%%%%%%%%%%%%%%%%%%%%%%%%%%%%%%
\section{Standard Model Analysis}

Theoretical predictions of $x_{\rm D}$ and $y_{\rm D}$ within 
the Standard Model span several orders of magnitude.
Roughly, there are two approaches, neither of which give very reliable
results because $m_c$ is in some sense intermediate between the 
heavy-quark and light-quark limits.  Consider, for example, 
$\Delta\Gamma_{\rm D}$ as given in Eq.~(\ref{ope}) 
\begin{eqnarray}
\Delta\Gamma_{\rm D} 
= \frac{1}{M_{\rm D}}\, {\rm Im}\, \langle \D0bar |
    \,i\! \int\! {\rm d}^4 x\, T \Big\{
    {\cal H}^{|\Delta C|=1}_w (x)\, {\cal H}^{|\Delta C|=1}_w(0) \Big\}
    | D^0 \rangle \ \ .
\nonumber 
\end{eqnarray}
To utilize this relation, one inserts intermediate states between the 
$|\Delta C| = 1$ weak hamiltonian densities ${\cal H}^{|\Delta C|=1}_w$.  
This can be done using either quark or hadron degrees of freedom.  
Let us consider each of these possibilities in turn.  

\subsection{Quark-level Analysis}
The ``inclusive'' (or quark-level) approach is based on the
operator product expansion (OPE). In the $m_c \gg \Lambda$ limit, where
$\Lambda$ is a scale characteristic of the strong interactions, $\Delta
M$ and $\Delta\Gamma$ can be expanded in terms of matrix elements of local
operators~\cite{datta,inclusive} of increasing dimensions suppressed by
powers of inverse charm quark mass. 
\begin{figure}[t]
\includegraphics[width=14pc]{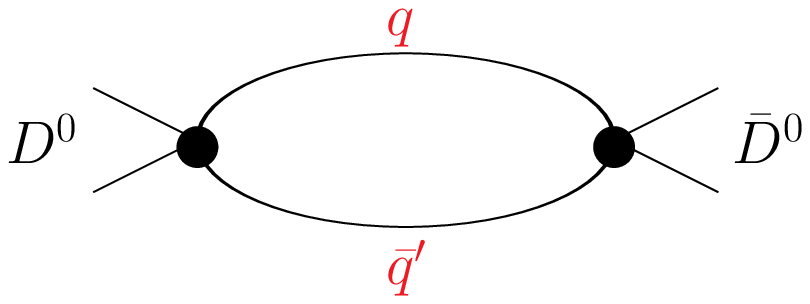}
\caption{Loop diagram for $D^0 \to {\overline D}^0$.}
\label{fig:loop}
\end{figure}
An instructive example concerns a recent analysis of the leading 
dimension $D=6$ case~\cite{Golowich:2005pt} in which the width 
difference $y_{\rm D}$ is calculated in terms of 
quarks ({\it cf} Fig.~\ref{fig:loop}) and the mass 
difference $x_{\rm D}$ is then found 
from dispersion relations.  The calculation is carried out as 
an expansion in QCD, including leading-order ${\cal O}(\alpha_s^0)$ 
and next-to-leading order ${\cal O}(\alpha_s^1)$ 
contributions with
\begin{equation}
x_{\rm D} = x_{\rm D}^{({\rm LO})}+x_{\rm D}^{({\rm NLO})}\quad\quad
{\rm and}\quad\quad
y_{\rm D} = y_{\rm D}^{({\rm LO})}+y_{\rm D}^{({\rm NLO})}\ \ .
\end{equation}
Here, LO and NLO denote only
the corrections at that order and not the full quantity computed to that
order.  Because $m_b > M_{\rm D}$, $\Delta \Gamma_{\rm D}$ 
experiences no $b$-quark contribution.\footnote{We ignore 
here the $b$-quark contribution to $\Delta M_{\rm D}$; its 
numerical contribution 
is subleading ($|V_{ud}V_{cd}| \simeq |V_{us}V_{cs}| \simeq 0.22$ whereas 
$|V_{ub}V_{cb}| \simeq 1.8\cdot 10^{-4}$).} This leaves only 
$s{\overline s}$, $d{\overline d}$ and $s{\overline d} + d{\overline s}$ 
intermediate states contributing to 
the mixing diagram of Fig.~\ref{fig:loop}.  Taking 
$m_d = 0$, the mixing loop-functions will depend on $z \equiv m_s^2/m_c^2 
\simeq 0.006$.  Table~\ref{tab:qkmix} examines in detail the 
loop-functions for $\Delta\Gamma_{\rm D}$ and shows the results of
carrying out an expansion in powers of $z$.  We see that
the contributions of the individual intermediate states in the mixing diagram 
are {\it not} intrinsically 
small -- in fact, they begin to contribute at ${\cal O}(z^0)$.  
However, flavor 
cancellations remove all contributions through ${\cal O}(z^2)$ for
$\Delta\Gamma_{\rm D}$, 
so the net result is ${\cal O}(z^3)$.  
Charm mixing clearly experiences a remarkable GIM suppression! 
The corresponding result for $\Delta M_{\rm D}$ turns out to be 
${\cal O}(z^2)$.   Summarizing, the leading dependences 
in $z$ for the dimension six contributions are 
\beq
y_{\rm D}^{\rm (LO)} \sim z^3\qquad x_{\rm D}^{\rm (LO)} 
\sim z^2\qquad y_{\rm D}^{\rm (NLO)} \sim
z^2\qquad x_{\rm D}^{\rm (NLO)} \sim z^2 \ \ .
\label{order}
\eeq
\begin{table}[t]
%\newcommand{\m}{\hphantom{$-$}}
%\newcommand{\cc}[1]{\multicolumn{1}{c}{#1}}
%\renewcommand{\tabcolsep}{2pc} % enlarge column spacing
%\renewcommand{\arraystretch}{1.2} % enlarge line spacing
%\begin{tabular}{@{}llll}
\begin{tabular}{cccc}
\hline\hline
Int. State & ${\cal O}(z^0)$ & ${\cal O}(z^1)$ & ${\cal O}(z^2)$\\
\hline
$s{\overline s}$ & $1/2$ & $-3z$ & $3z^2$ \\
$d{\overline d}$ & $1/2$ & $0$ & $0$ \\
$s{\overline d}+d{\overline s}$ & $-1$ & $3z$ & $-3z^2$ \\
\hline
Total & $0$ & $0$ & $0$ \\ \hline\hline
\end{tabular}\\[2pt]
\vskip .05in\noindent
\caption{Flavor cancellations in $\Delta\Gamma_{\rm D}$.}
\label{tab:qkmix}
\end{table}
The source of this $z$-dependence is understood 
as follows.  The mixing amplitude is 
known to vanish in the $m_d = m_s = 0$ 
limit, so the breaking of chiral symmetry and of SU(3) flavor symmetry  
play crucial roles.  Thus, a factor of 
$m_s^2$ comes from an $SU(3)$ violating 
mass insertion on each internal quark line and 
another from an additional mass insertion on each line to 
compensate the chirality flip from the first insertion.  This 
mechanism of chiral suppression accounts for the $z^2$ dependence 
of $x_{\rm D}^{\rm (LO)}$, $y_{\rm D}^{\rm (NLO)}$ and 
$x_{\rm D}^{\rm (NLO)}$. The case of  
$y_{\rm D}^{\rm (LO)}$ requires yet another 
factor of $m_s^2$ to lift the helicity suppression 
for the decay of a pseudoscalar meson into a massless fermion pair.

%\vspace{-1pc}

Let us next display the LO expressions for $y_{\rm D}$ and $x_{\rm D}$ 
(to leading order also in $z$)~\cite{Golowich:2005pt}, 
\begin{eqnarray}
y_{\rm D}^{{\rm (LO)}[z^3]} &=& {G_F^2 m_c^2 f_D^2 M_D \over 
3 \pi \Gamma_D} ~\xi_s^2~
z^3~ \left( c_2^2 - 2 c_1 c_2 - 3 c_1^2 \right) \left[ 
B_{\rm D} - {5 \over 2} {\bar B}_{\rm D}^{(S)}
\right] \ \ , \nonumber \\
x_{\rm D}^{{\rm (LO)}[z^2]} &=& 
{G_F^2 m_c^2 f_D^2 M_D \over 3 \pi^2 \Gamma_D} ~\xi_s^2~
z^2~ \bigg[  c_2^2 B_{\rm D} - {5 \over 4} (c_2^2 - 
2 c_1 c_2 - 3 c_1^2) {\bar B}_{\rm D}^{(S)}\bigg] \ \ ,
\label{yLO}
\end{eqnarray}
where $\xi_s \equiv V_{us}V^*_{cs}$ and $c_{1,2}$ are the relevant Wilson
coefficients.  
For our numerical computations, we adopt the values used in 
Ref.~\cite{Golowich:2005pt}, 
\beq
m_c = 1.3~{\rm GeV}~, \quad 
%B_{\rm D} = 0.82~, \quad 
c_1 = -0.411~, \quad c_2 = 1.208~, \quad \alpha_s = 0.406 \ \ .
\label{values}
\eeq
%In addition, we allow for a range of the ratio 
%$XXX<{\bar B}_{\rm D}^{(S)}/B_{\rm D}<XXX$. 
Numerical results for the LO and NLO contributions, where a discussion
of the NLO effects can be found in Ref.~\cite{Golowich:2005pt},
({\it cf} Table~\ref{tab:SD}) reveal that $y_{\rm D}$ is given 
by $y_{\rm NLO}$ to a reasonable approximation (due to the
$z$ dependence discussed above) whereas $x_{\rm D}$ is greatly 
affected by destructive interference between $x_{\rm LO}$ and 
$x_{\rm NLO}$.  The net effect is to render $y_{\rm D}$ and 
$x_{\rm D}$ of similar small 
magnitudes, at least through this order of analysis.
\begin{table}[t]
\begin{tabular}{l||c||c|c} 
\colrule\hline 
 &LO & NLO & LO + NLO (Central Values) \\
\colrule \hline
$y_{\rm D}$ & $-(5.7 \to 9.5) \cdot 10^{-8}$
& $(3.9 \to 9.1)\cdot 10^{-7}$
& $\simeq 6 \cdot 10^{-7}$  \\
$x_{\rm D}$ & $-(1.4 \to 2.4) \cdot 10^{-6}$ 
& $(1.7 \to 3.0)\cdot 10^{-6}$ & 
$\simeq 6 \cdot 10^{-7}$ \\
\colrule \hline
\end{tabular}
\vskip .05in\noindent
\caption{Results at dimension-six in the OPE.}
\label{tab:SD}
\end{table}

The quark-level prediction of $x_{\rm D}$ and $y_{\rm D}$ just described 
is a result of expanding 
in terms of {\it three} `small' quantities, 
$z$, $\Lambda/m_c$, and $\alpha_s$.  As a consequence, 
the use of an OPE to describe charm
mixing is not entirely straightforward because terms suppressed by 
higher powers of $m_c$ could nevertheless be important if 
they contained relatively fewer powers of $m_s$.  
However, at the next orders in the OPE one 
encounters ${\cal O}(z^{3/2})$ corrections multiplied by about a dozen matrix
elements of dimension-nine operators and ${\cal O}(z)$ corrections 
with more than twenty matrix elements of dimension-twelve operators. 
This introduces a multitude of unknown parameters for 
matrix elements that cannot be computed at this time. Simple dimensional 
analysis~\cite{Bigi:2000wn} suggests the magnitudes 
$x_{\rm D} \sim y_{\rm D} \sim 10^{-3}$, although 
order-of-magnitude cancellations or enhancements are 
possible.\footnote{Any effect of higher orders in $1/m_c$ or $\alpha_s(m_c)$ 
which could produce a $z^n$ contribution in the lowest possible
power $n=1$ could yield 
a dominant contribution to the prediction of $x_{\rm D}$ 
and $y_{\rm D}$~\cite{Falk:2001hx,ggp}.
Although the BaBar and Belle observations of $y \sim 10^{-2}$ could be 
ascribed to a breakdown of the OPE or of duality, it is clear that 
such a large value of $y_{\rm D}$ is by no means a generic 
prediction of OPE analyses.}

\subsection{Hadron-level Analysis}

The $D$ meson mass is not very large, so one might question whether the 
OPE approach discussed in the previous subsection can successfully 
describe $D^0$-${\overline D}^0$ 
mixing. This is especially so since the leading 
contribution in the $SU(3)$-breaking parameter $m_s$ enters only as a 
$\Lambda^4/m_c^4$ suppressed contribution in the $1/m_c$ expansion, which 
implies that one has to deal with a large number of unknown
operator matrix elements.

As an alternative, one might consider saturating the correlation 
functions of Eq.~(\ref{ope}) with exclusive hadronic states, switching 
to a purely hadronic description. This approach should be valid as the
mass of the $D$ meson lies in the
middle of a region populated by excited light-quark states. 
In principle, this ``exclusive'' (or hadronic) approach should sum 
over all possible intermediate hadronic multiplets.  
Since one has to deal with off-shell hadronic states in the
calculation of $x_{\rm D}$, some modeling is necessarily involved. By 
contrast, a calculation of $y_{\rm D}$ in this approach is less 
model-dependent. The usual approach to computing $x_{\rm D}$ is to 
first calculate $y_{\rm D}$ and then use a dispersion relation to 
obtain $x_{\rm D}$. This is appropriate, as the contribution due to 
$b$-flavored intermediate states (which appears in $x_{\rm D}$ but 
not $y_{\rm D}$) is negligibly small.

One possible approach would be to select a set of, say, two-body 
intermediate states~\footnote{The simplest intermediate state is a 
single-particle resonance contribution. Preliminary estimates of 
resonance contributions to $D^0$-${\bar D}^0$ mixing appear to be 
small~\cite{Golowich:1998pz}, although much remains to be learned
about the resonance spectrum in the vicinity of the $D^0$ mass.}, 
and write their contribution to mixing in terms of charged
pseudoscalar ($P^+P^-$) branching fractions~\cite{Donoghue:1985hh,lw},  
\begin{eqnarray}\label{LD}
& & y_{\rm D}^{(P^+P^-)} = {\cal B}_{[D^0 \to K^+K^-]} 
+ {\cal B}_{[D^0 \to \pi^+\pi^-]} 
- 2 \cos\delta_{K\pi} \left[ {\cal B}_{[D^0 \to K^-\pi^+]}\cdot 
{\cal B}_{[D^0 \to K^+\pi^-]} 
\right]^{1/2} \ \ ,
\end{eqnarray}
where $\delta_{K\pi}$ is as in Eq.~(\ref{y-defs}). One can use 
available experimental data on two-body branching ratios to estimate 
their contribution to $y_{\rm D}$. A dispersion relation 
then relates $y_{\rm D}$ to $x_{\rm D}$. However, 
the example above explicitly shows the 
cancellations between states that are present within a given $SU(3)$ 
multiplet. Such cancellations make this procedure 
very sensitive to experimental uncertainties. One would need 
to know the contribution of each decay mode with extremely high precision, 
and that is simply not feasible at this time.  Another possibility is 
to model $|\Delta C| = 1$ decays theoretically~\cite{Buccella:1994nf}. 
In this reference, $\Delta\Gamma_{\rm D}$ was determined 
in this manner and the result $y_{\rm D} \simeq 10^{-3}$ was found.  
This result is, however, smaller than the recent BaBar and Belle observations. 

Clearly, $D^0$ is not sufficiently light for its decays to be 
dominated by just two-body final states.  Multiparticle 
intermediate states must also be taken into account in 
$D^0$-${\overline D}^0$ mixing calculations.  In doing so, 
it is convenient to calculate 
the contribution of each $SU(3)$ multiplet separately, as $SU(3)$ 
symmetry produces substantial cancellations among members of the same 
multiplet as we saw above.  
This can be thought of as a long-distance version of the 
GIM mechanism. The surviving contribution is expected to 
be of second order in the $SU(3)$-breaking parameter $m_s$~\cite{Falk:2001hx}.
Denoting by $y_{F_R}$ a value that $y$ would take if elements of 
the final state $F$ belonging to $SU(3)$ representation $R$, or $F_R$, 
were the only channels open for D-decay, one can write $y_{\rm D}$ as 
a sum over all possible ${F_R}$'s weighted by the $D$-decay rate to each 
representation,
\begin{equation}\label{yLD}
y_D=\frac{1}{\Gamma_D} \sum_{F_R} y_{F_R} \left[
\sum_{n \in F_R} \Gamma(D \to n)\right] \ \ .
\end{equation}
It is possible to show that $y_{F_R}$ can be computed as~\cite{Falk:2001hx}
\begin{equation}\label{yFR}
y_{F_R}=\frac{\sum_{n \in F_R} 
\langle \overline{D}^0 | {\cal H}_w | n \rangle
\langle n | {\cal H}_w | D^0 \rangle}
{\sum_{n \in F_R}
\langle D^0 | {\cal H}_w | n \rangle
\langle n | {\cal H}_w | D^0 \rangle} \ \ .
\end{equation}
It should be noted that in the limit of $CP$-conservation and 
retaining phase space differences as the only source of $SU(3)$ 
breaking ({\it i.e.} neglecting $SU(3)$ breaking
in the matrix elements), $y_{F_R}$ can be computed without any 
hadronic parameters.  This is an appropriate approximation, as the 
main contribution comes from the multiparticle (four-particle) 
intermediate state multiplets. For those states, there are multi-kaon 
modes which are kinematically forbidden. In such cases, phase 
space effects alone can provide enough $SU(3)$ violation to induce 
$y_D\sim 10^{-2}$~\cite{Falk:2001hx}. In other words, such large 
effects in $y_{\rm D}$ appear for decays near the $D$ threshold, 
where an analytic expansion in $SU(3)$ violation is no longer
possible.  It is interesting that such effects from 
multiparticle states are not reproduced in the OPE calculation, 
as the resulting contribution does not come from short-distances.

The use of a dispersion relation for $x_{\rm D}$ then suggests 
it would receive contributions of a similar order of 
magnitude as those for $y_{\rm D}$~\cite{Falk:2004wg}. An important difference 
between the resulting values of $x_{\rm D}$ and $y_{\rm D}$ is that even
retaining phase space differences as the sole contributor to $SU(3)$ breaking
does not insure cancellation of the 
hadronic matrix elements. However, with some
reasonable model-dependent assumptions, one arrives at the 
conclusion that
$x_{\rm D} \sim y_{\rm D} \sim 1\%$~\cite{Falk:2004wg}. It is thus 
reasonable to believe that the observed $D^0$-${\bar D}^0$ mixing is 
reflecting Standard Model contributions.

\subsection{Comments} 

The above discussions show that, contrary to the $B$ system,
Standard Model estimates of $x_{\rm D}$ and $y_{\rm D}$ for the charm
system contain
significant intrinsic uncertainties.  On the other hand, SM values near 
those found by 
BaBar and Belle cannot be ruled out.  Therefore, it will be difficult 
to attribute a clear indication of New Physics to 
$D^0$-${\bar D}^0$ mixing measurements alone. This means that the only 
robust signal of New Physics in the charm system would be the
observation of large CP violation, which we will not consider here.  
Nonetheless, a thorough analysis 
of indirect New Physics contributions is of value, and we find
that large regions of parameter space can be excluded in many models,
placing additional restrictions on model building.
This will be useful in conjunction with corresponding direct searches
for New Physics at the LHC.

In what follows, we will take the approach that the New
Physics contributions cannot exceed the 1$\sigma$ experimental upper 
bound for $x_{\rm D}$. Keeping in mind that this upper limit is
likely to change as data samples increase and analyses mature,
we also display the effects of $x_{\rm D}< (15.0\,, 8.0\,, 5.0\, 
\ {\rm and}\ 3.0) \cdot
10^{-3}$ on the parameter space of New Physics scenarios.
These values are to be used as a guide for how our resulting constraints
may change in the future.
In addition, we will neglect the errors on the determinations
of the $D$ meson decay constant and B-factors; this will have a small
effect on our results given the present large uncertainty in the experimental
determination of $D^0$-$\overline D^0$ mixing.  
In all cases, we will neglect the possibility of
interference between the SM and New Physics contributions.  We now
turn to the examination of various scenarios for physics beyond the SM.

%%%%%%%%%%%%%%%
%%%%%%%%%%%%%%%%%%%%%%%%%%%%%%%%%%%%%%%%%%%%%%%%%%%%%%%%%%%%%%%%%%%%%
\section{Extra Fermions} 

The quark sector of the Standard Model can be modified in several
ways, and new fermions are predicted to exist in many extensions of the
SM.  They can be classified according to their electroweak quantum
number assignments; here we consider the possibilities of 
a sequential fourth generation quark doublet 
(Sect.~\ref{QDoubletSect}), heavy quark iso-singlets 
(Sect.~\ref{QSingletSect}) and non-SM quarks associated with 
Little Higgs models (Sect.~\ref{LHiggsSect}).
The contributions of such heavy quarks can remove the efficient GIM 
cancellation inherent in the short distance SM
computation and can give rise 
to $D^0$-${\bar D}^0$ mixing at the level of the current experimental limit.

%%%%%%%%%%%%%%%%%%%
\subsection{Fourth Generation Quark Doublet}\label{QDoubletSect}

A simple extension to the Standard Model is the addition of a fourth
family of fermions.  Precision electroweak data severely constrains this
possibility. The Particle Data Group~\cite{PDG} quotes a restriction
on the number of families to be 
$N_F=2.81\pm 0.24$ from the oblique $S$ parameter~\cite{Peskin:1991sw} 
alone. We note, however, that the LEP Electroweak Working 
Group~\cite{LEPEWWG} allows for a more generous range of the $S$ parameter
from their electroweak fit.
In either case, this restriction can be relaxed by allowing the $T$ parameter
to vary as well, or by adding other sources of new physics which would
participate in the electroweak fit such as
an extended Higgs sector~\cite{He:2001tp}.  The 
requirement of anomaly cancellation implies the 
existence of a fourth lepton family as well (almost degenerate to satisfy 
the $\Delta \rho$ constraint with $m_{\nu_4}>M_Z/2$) 
or an extra right-handed quark doublet. 
Direct collider searches
by the CDF and D0 Collaborations at the Tevatron currently place a
bound~\cite{PDG} on
the mass of a charged $-1/3$ fourth generation quark $b'$ of
$m_{b'}>128\,,190\,,199$~GeV if the $b'$-quark 
decays respectively via charged current interactions into leptons + jets, 
via FCNC with $b'\to bZ$, or is quasi-stable. We recall
that perturbative unitarity considerations~\cite{Chanowitz:1978uj} 
in $FF\to FF$ scattering
restricts the mass of sequential heavy flavors to be $m_F\lsim 500$ GeV.

Here, we review the contribution of a fourth generation of quarks to
$D$ mixing, keeping in mind that some other New Physics may also be
present in order to evade the precision electroweak constraints and that
it also may or may not contribute to the mixing.  
The primary motivation
for this discussion is to set up the formalism that will be used
in the following Sections.

The $Q = -1/3$ fourth generation quark contributes to $D$ mixing via a 
box diagram which also contains the SM $W^\pm$ bosons.  Note that since
the $b'$-quark is not kinematically accessible in charm-quark decay,
it will not contribute to the dispersive amplitude for $x_{\rm D}$ in 
Eq.~(\ref{ope}).  The $|\Delta C|=2$ hamiltonian at 
the $b'$ mass scale in the fourth generation model is~\cite{q4}
\begin{equation}
{\cal H}_{4^{th}} = {G_F^2M_W^2\over 4\pi^2}\sum_{i,j}\lambda_i\lambda_j
S(x_i,x_j)Q_1 \ \ ,
\label{smham}
\end{equation}
where $S(x_i,x_j)$ are the well-known Inami-Lim 
functions~\cite{Inami:1980fz} (given in the Appendix), $x_i=m_{q_i}/M_W)^2$, 
$\lambda_i\equiv V^*_{ci}V_{ui}$, and the sum runs over the internal 
quark flavors.  As discussed in the previous Section, there is a
strong GIM cancellation in $D$ meson mixing, which leaves a sizable 
contribution only from the heavy $b'$ quark and sets $i=j=b'$ in the 
above sum.  Performing the RG evolution, we obtain at the scale $m_c$
\begin{equation}
{\cal H}_{4^{th}}={G_F^2M_W^2\over 4\pi^2}\lambda_{b'}^2 
S(x_{b'},x_{b'}) r_1(m_c,M_W)Q_1 \ \ ,
\end{equation}
which in turn gives 
\begin{equation}
x_{\rm D}^{(4^{th})}={G_F^2 M_W^2 f_D^2 M_D \over 6\pi^2\Gamma_D}B_D
\lambda_{b'}^2  r_1(m_c,M_W) S(x_{b'},x_{b'}) \ \ .
\label{xdfour}
\end{equation}
It should be noted that for $r_1(m_c,M_W)$, only contributions below
$M_W$ are required.

\begin{figure}[htbp]
\centerline{
\includegraphics[width=6cm,angle=90]{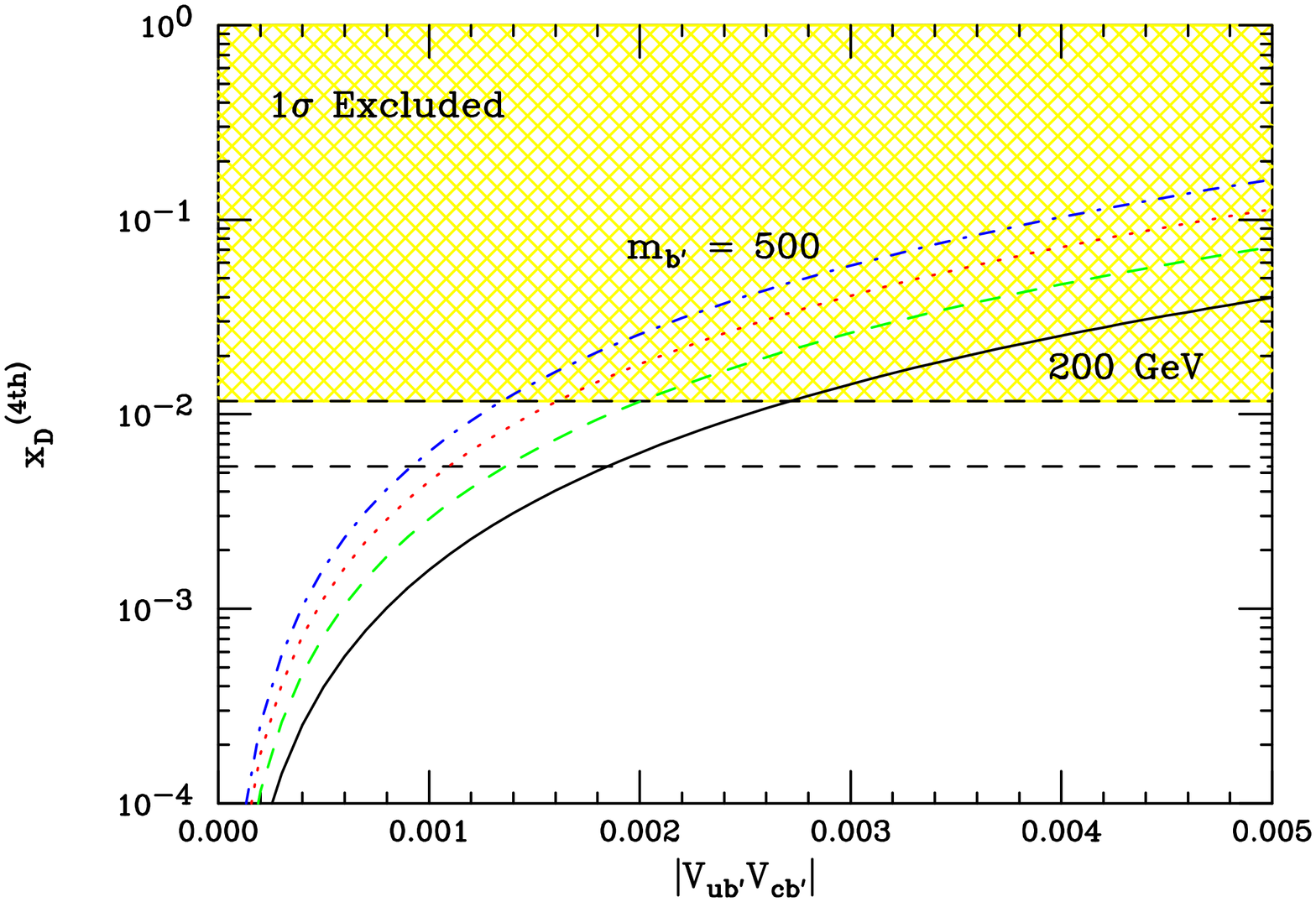}
%\hspace*{5mm}
\includegraphics[width=6cm,angle=90]{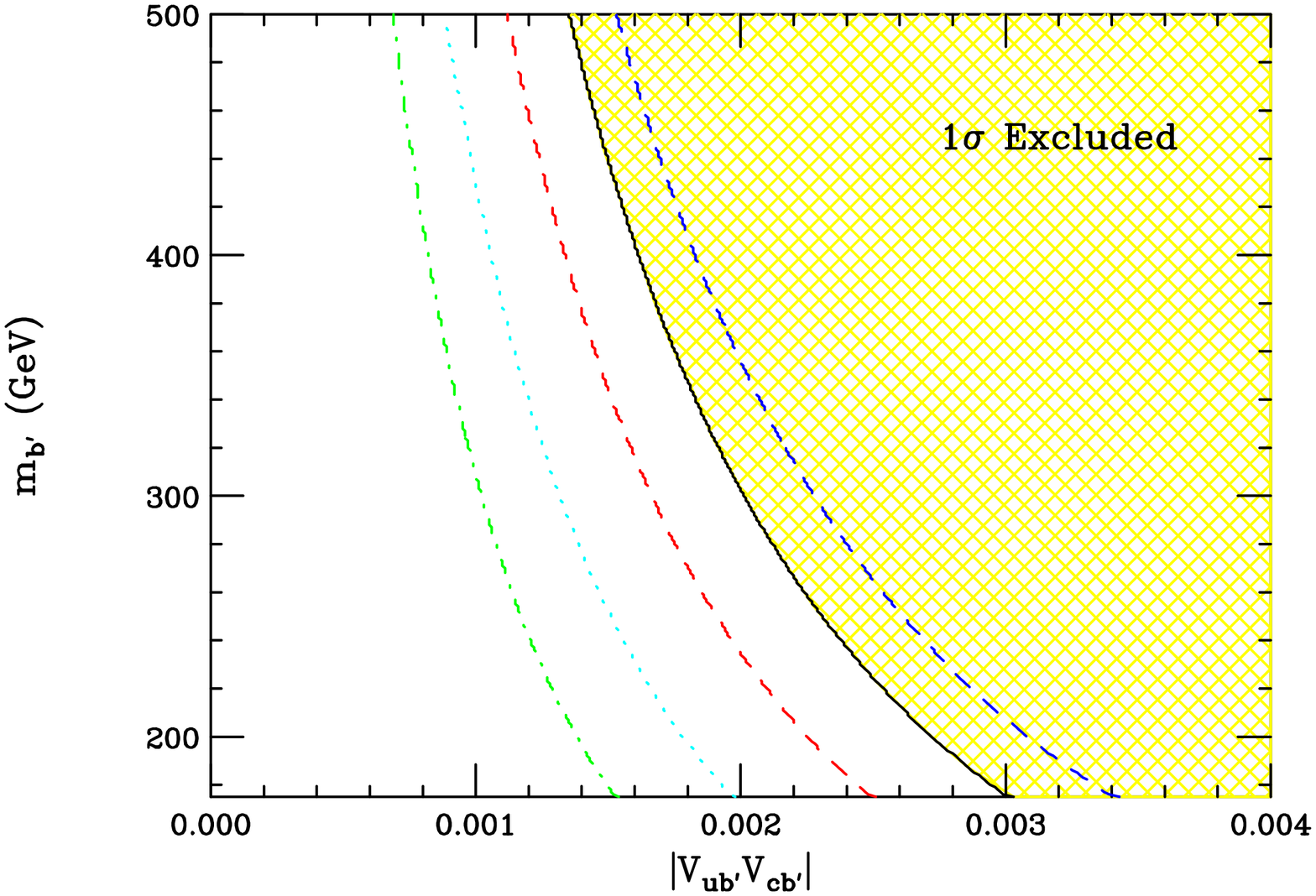}}
\vspace*{0.1cm}
\caption{Left:  $x_{\rm D}$ in the four generation model as a function of the
CKM mixing factor $|V^*_{cb'}V_{ub'}|$ for $b'$-quark masses
of 200, 300, 400, and 500 GeV from bottom to top.  The $1\sigma$ 
experimental bounds are as indicated, with the yellow shaded area 
depicting the region that is excluded.  \\
\noindent Right:  The present $1\sigma$ excluded region in the mass-mixing 
parameter plane, as well as possible future contours taking
$x_{\rm D}< (15.0\,, 8.0\,, 5.0\,, 3.0)\times 10^{-3}$,
corresponding to the blue dashed, red dashed, cyan dotted, and 
green dot-dashed curves, respectively.}
\label{4gen}
\end{figure}

The value of $x_{\rm D}^{(4^{th})}$ as a function of the CKM mixing
elements is displayed in
Fig.~\ref{4gen} for various values of the $b'$-quark mass. We see
that the $1\sigma$ experimental limit of $x_{\rm D}< 11.7\times 10^{-3}$ 
places sizable constraints
in the $b'$-quark mixing-mass parameter space.  We also show the
exclusion contours for possible future experimental bounds of 
$x_{\rm D}< (15.0\,, 8.0\,, 5.0\,, 3.0)\times 10^{-3}$ (corresponding
to the blue dashed, red dashed, cyan dotted, and green dot-dashed curves,
respectively) as discussed in the Introduction.  We note that the present
constraints on the CKM mixing parameters $|V_{ub'}V^*_{cb'}|\lsim 0.002$ 
are an order of magnitude stronger than those obtained from unitarity
considerations~\cite{PDG} of the CKM matrix.
%

%%%%%%%%%%%%%%%%%%%%
\subsection{Heavy Vector-like Quarks}\label{QSingletSect}

The next possibility of interest is the presence of heavy 
quarks which are $SU(2)_L$ singlets (so-called {\it vector-like} 
quarks)~\cite{branco}.  We will consider 
both charge assignments $Q = +2/3$ and $Q = -1/3$ for the heavy
quarks.  Both choices are well motivated, as such fermions appear 
explicitly in several models of physics beyond the Standard Model.  
For example, weak isosinglets with $Q = -1/3$ appear in
$E_6$ GUTs~\cite{q2,jlhtgr}, with one for each of the three
generations ($D$, $S$, and $B$).  Weak isosinglets with $Q = +2/3$ 
occur in Little Higgs theories~\cite{q3a,q3b} in which the 
Standard Model Higgs boson is a pseudo-Goldstone boson, and the heavy
iso-singlet $T$ quark cancels the quadratic divergences generated
by the top quark in the mass of the Higgs boson.

%%%%%%%%%%%%%%%%%%%%%%%%%%%%%%%%%%%%%%%%%%%%%%%%%%%%%%%%%%%%%%%%%%%%%
\subsubsection{$Q = -1/3$ Singlet Quarks}

\begin{figure}[tbp]
\centerline{
\includegraphics[width=4.5cm,angle=0]{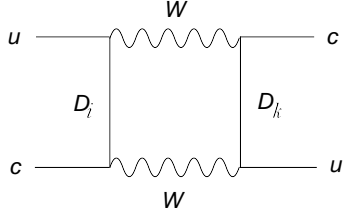}}
\vspace*{0.1cm}
\caption{Box contribution from heavy weak-isosinglet quarks.}
\label{qfig1}
\end{figure}

We first consider the class of models with $Q = -1/3$ down-type
singlet quarks.  For this case, the down quark mass matrix is 
a $4\times 4$ array if there is just one heavy singlet 
(or $6\times 6$ for three heavy singlets as 
in $E_6$ models).  As a consequence, the standard $3\times 3$ CKM 
matrix is no longer unitary.  Moreover, the weak charged current 
will now contain terms that couple up-quarks to the
heavy singlet quarks.  For three heavy singlets, we have 
\beq
{\cal L}_{\rm int}^{(ch)} = {g \over \sqrt{2}}~V_{i\alpha} 
W^\mu {\bar u}_{i,L} \gamma_\mu D_\alpha \ \ , 
\eeq
where $u_{i,L} \equiv (u,c,t)_L$ and $D_\alpha \equiv (D,S,B)$ 
refer to the standard up quark and heavy isosinglet down quark 
sectors.  The $\{ V_{i\alpha} \}$ are elements of a 
$3\times 6$ matrix, which is the product of the $3\times 3$ and
$6\times 6$ unitary matrices that diagonalize the $Q=+2/3$ and 
$Q=-1/3$ quark sectors, respectively.  The resulting
box diagram contribution to 
$\Delta M_{\rm D}$ from these new quarks is displayed in Fig.~\ref{qfig1}. 
Assuming that the contribution of one of the heavy quarks (say the 
$S$ quark) dominates, one can write an expression 
(similar to that in Eq.~(\ref{xdfour})) for $x_{\rm D}$~\cite{q4}, 
\beq
x_{\rm D}^{(-1/3)} \simeq 
{G_F^2 M_W^2 f_D^2 M_D \over 6 \pi^2 \Gamma_D}  B_D 
\left( V_{cS}^* V_{uS} \right)^2 r_1(m_c,M_W) f(x_S)  \ \ , 
\eeq 
where $x_S \equiv (m_S/M_W)^2$ and 
$f(x_S) \to x_S \left(1 + 6 \ln(x_S) \right)$ for large $x_S$.
The light-heavy mixing angles $\left|V^*_{cS} V_{uS}\right|^2$
should go as $1/m_S$ for large $m_S$ to keep the
contribution under control. The current bound on 
$\left|V^*_{cS} V_{uS}\right|^2$ from unitarity of the CKM matrix
is not very stringent, 
$\left|V^*_{cS} V_{uS}\right|^2 < 4\times 10^{-4}$~\cite{PDG}.
An $S$-quark mass in the range 0.2 to 1 TeV gives rise to a 
mixing contribution that can exceed the current experimental 
limit in Eq.~(\ref{hfag}). Hence a singlet heavy quark of charge $-1/3$
can give rise to $x_{\rm D}$ near the current experimental limit.

\begin{figure}[tbp]
\centerline{
\includegraphics[width=8cm,angle=0]{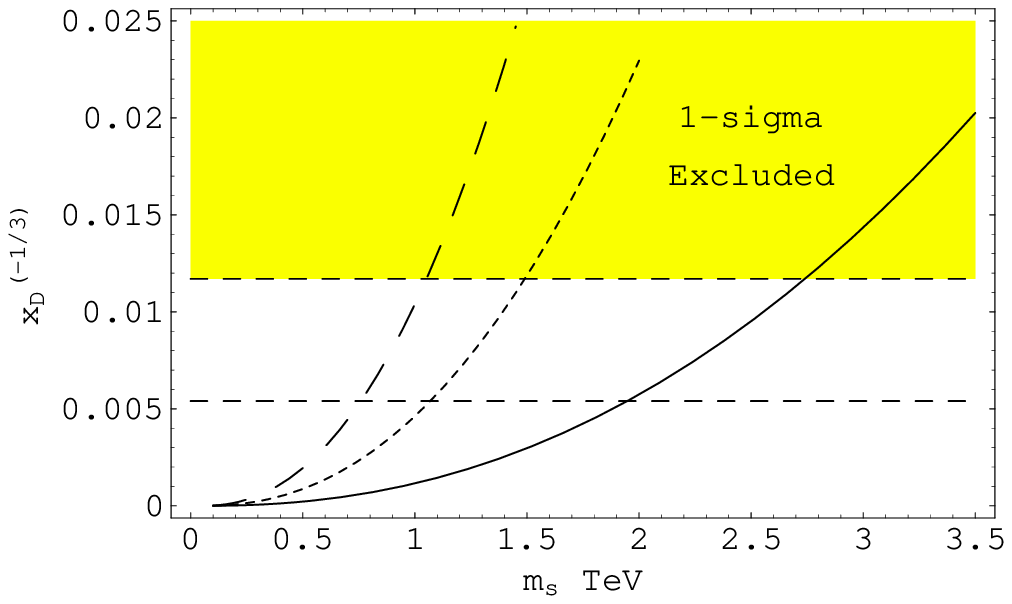}
\includegraphics[width=8cm,angle=0]{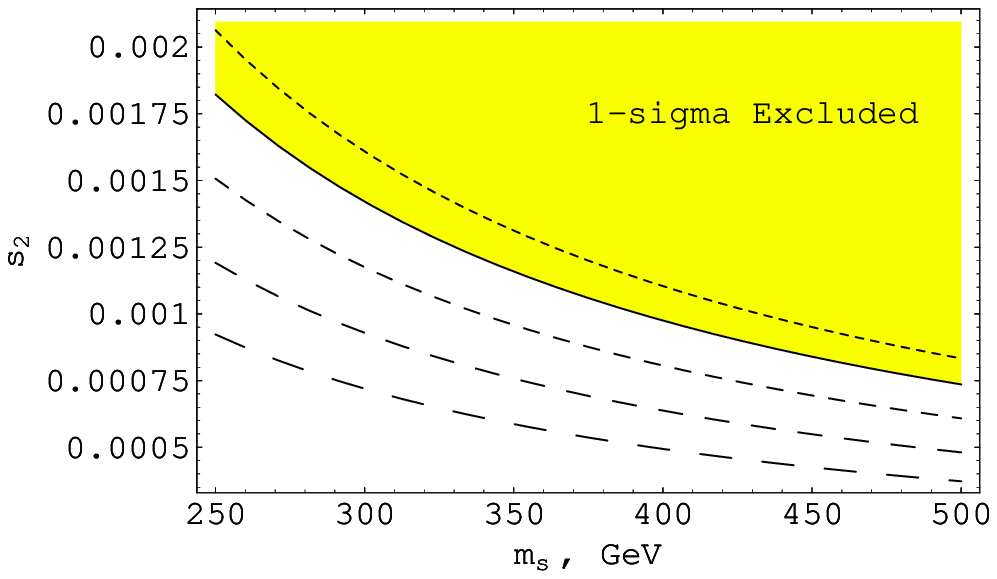}}
\vspace*{0.1cm}
\caption{Left: $x_{\rm D}$ in the singlet $Q=-1/3$ quark model as 
a function of the singlet quark mass for various values of the
mixing angle $s_2=0.0001\,, 0.0002\,, 0.0003$ corresponding to the
solid, dotted, and dashed curves, respectively.  The $1\sigma$
experimental bounds are as indicated, with the yellow shaded area
depicting the region that is excluded.\\
Right: The present $1\sigma$ excluded region (short-dashed curve) 
in the mass $m_S$ and mixing angle 
$s_2$ parameter plane for the singlet $Q=-1/3$ quark in the model of Bjorken 
\etal~\cite{q7} described in the text. Possible future contours are
also shown, taking
$x_{\rm D}< (15.0\,, 8.0\,, 5.0\,, 3.0)\times 10^{-3}$ from top
to bottom, 
corresponding to the solid, medium dashed, long dashed, and 
longer dashed curves, respectively.}
\label{SingletQuark13Fig}
\end{figure}

In the $E_6$-based model proposed by Bjorken {\it et al}~\cite{q7}, the 
$6\times 6$ mass matrix has an especially simple form. The resulting 
$6\times 6$ mass matrix has a pseudo-orthogonality property which
implies that the $3\times 3$ CKM matrix, although not unitary,
satisfies
\beq
\sum_{i=1}^3 \ \left(V_{\rm CKM}\right)_{bi}^* 
\left(V_{\rm CKM}\right)_{is} = 0\ \ .
\eeq
The analog of this condition in the up quark sector does not hold, and as a
result, there are no new FCNC effects in the down quark sector.  For 
$D^0$-${\bar D}^0$ mixing, the prediction is now that 
(recall capital lettering is used to denote the heavy quark) 
\beq
\left| V_{cS}^*V_{uS} \right|^2 = s_2^2 
\left| V_{cs}^*V_{us} \right|^2  \simeq s_2^2 \lambda^2 \ \ ,
\eeq
where $|V_{cs}^*V_{us}| \simeq \lambda \simeq 0.22$ and 
$s_2$ is the (small) mixing parameter describing the mixing between 
the light $s$ quark and the heavy $S$ quark. Using the 
experimental values in Eq.~(\ref{hfag}),
we can place bounds on $s_2$ for a given mass 
$m_S$, {\it e.g.} $s_2 \simeq 0.0009$ for 
$m_S \simeq 0.5$~ TeV. We present the constraints
on $s_2$ vs. $m_S$ in Fig.~\ref{SingletQuark13Fig}.
%

%%%%%%%%%%%%%%%%%%%%%%%%%%%%%%%%%%%%%%%%%%%%%%%%%%%%%%%%%%%%%%%%%%%%%
\subsubsection{$Q = +2/3$ Singlet Quarks}

Next, consider the possibility of weak isosinglet quarks having 
charge $Q=+2/3$.  These are present in some theories beyond the SM,
including for example, Little Higgs models which will be discussed
in more detail in the following section.  Here, we present the general
formalism for this scenario.

\begin{figure}[tbp]
\centerline{
\includegraphics[width=10cm,angle=0]{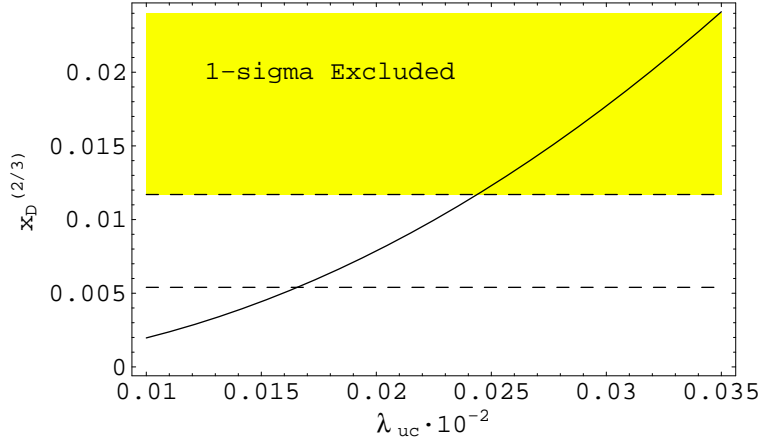}}
\vspace*{0.1cm}
\caption{Value of $x_{\rm D}$ as a function of the mixing parameter 
$\lambda_{uc}$ in units of
$10^{-2}$ in the $Q=+2/3$ quark singlet model.  The $1\sigma$ experimental
bounds are as indicated, with the yellow shaded area depicting the
region that is excluded.}
\label{SingletQuark23Fig}
\end{figure}

The presence of such quarks violate the Glashow-Weinberg-Paschos naturalness
conditions for neutral currents~\cite{gw77}.  Since their 
electroweak quantum number assignments are different than those for the
SM fermions, flavor changing neutral current interactions are generated
in the left-handed up-quark sector.  Thus, in addition to the charged current 
interaction 
\beq
{\cal L}_{\rm int}^{(ch)} = {g \over \sqrt{2}}~V_{\alpha i}
{\bar u}_{\alpha,L} \gamma_\mu d_{i,L} W^\mu\ \ , 
\eeq
there are also FCNC couplings with the $Z^0$ boson~\cite{branco}, 
\beq
{\cal L}_{\rm int}^{(ntl)} = {g \over 2\sqrt{2}\cos\theta_w}~\lambda_{ij} 
{\bar u}_{i,L} \gamma_\mu u_{j,L} Z^{0\mu}\ \ . 
\eeq
Here, $V_{\alpha i}$ is a $4\times 3$ mixing matrix with $\alpha$ running
over $1\to 4$, $i=1\to 3$, and with the CKM matrix comprising the first 
$3\times 3$ block.
\begin{figure} [tb]
\centerline{
\includegraphics[width=4cm,angle=0]{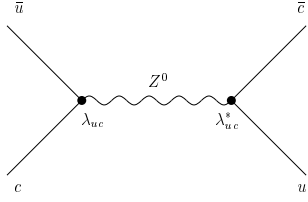}}
\caption{Tree-level contribution from $Z^0$-exchange.
\label{qfig2}}
\end{figure}
In this case, a tree-level contribution to $\Delta M_D$ is generated from 
$Z^0$-exchange as shown in Fig.~\ref{qfig2}.  This is represented 
by an effective hamiltonian of the form 
\beq
{\cal H}_{2/3} = {g^2 \over 8 \cos^2 \theta_w M_Z^2} 
\left( \lambda_{uc} \right)^2 {\bar u}_L \gamma_\mu 
c_L {\bar u}_L \gamma^\mu c_L  \ \ ,
\eeq
where unitarity demands
\beq
\lambda_{uc} \equiv - \left( V_{ud}^*V_{cd} + V_{us}^*V_{cs} + 
V_{ub}^*V_{cb} \right) \ \ .
\eeq
Taking the 1$\sigma$ ranges for the experimentally determined values
of the CKM elements~\cite{PDG} yields the constraint $\lambda_{uc}<0.02$.
This hamiltonian is just a particular case of a more general relation 
$\left( {\rm Eq.~(\ref{Zprime})}\right)$ appearing in the following 
section.\footnote{The specific correspondence is 
$C_R=0$, $C_L^2 \equiv C_{2/3}^2=g^2 \lambda_{uc}^2/(4 \cos^2\theta_w)$ and
$M_{Z^\prime}=M_Z$.} 
The QCD running from $\mu=M_Z$ to $\mu=m_c$ for this 
effective hamiltonian is trivial, leading to 
\beq
{\cal H}_{2/3} = {g^2 \over 8 \cos^2 \theta_w M_Z^2} 
\left( \lambda_{uc} \right)^2  r_1(m_c,M_Z) Q_1 \ \ ,
\eeq
where it should be noted that for $r_1(m_c,M_Z)$, only contributions 
below $M_Z$ are required.  This hamiltonian leads to 
\beq
x^{\rm (2/3)}_{\rm D} \ = \ {2G_F f_{\rm D}^2  M_{\rm D} 
\over 3\sqrt 2\Gamma_D}  B_D  
\left( \lambda_{uc} \right)^2 r_1(m_c,M_Z)  \ \ .
\eeq
We present this contribution to $x_{\rm D}$ from models with a 
singlet $Q=2/3$ quark 
in Fig.~\ref{SingletQuark23Fig}. Note that the bound on the mixing
$\lambda_{uc}$ from $D^0$-$\overline D^0$ mixing is roughly two 
orders of magnitude
better than that from unitarity constraints of the CKM matrix.

%

%%%%%%%%%%%%%%%
\subsection{Little Higgs Models}\label{LHiggsSect}

Little Higgs models~\cite{q3a,q3b,morelh} feature the Higgs as a pseudo
Nambu-Goldstone boson of an approximate global symmetry that is
broken by a vacuum expectation value (vev hereafter) 
at a scale of a few TeV.  This approximate symmetry protects
the Higgs vev through one-loop order relative to the ultra-violet 
(UV) cut-off of the theory which appears at a higher scale.  
The breaking of this symmetry is realized in such a way that the 
Higgs mass only receives quantum corrections at two-loops.
In these models the one-loop quadratic divergent contributions to the
Higgs mass in the SM are canceled by a new particle of the same spin.
These models thus predict the existence of new charged $Q = +2/3$ 
vector-like quarks, gauge bosons, and scalars at the TeV scale.

The most economical model of this type, in that it introduces the minimal
number of new fields, is known as the Littlest Higgs~\cite{q3a,q3b}.  
It is based on a nonlinear sigma model with $SU(5)$ global
symmetry that is broken to the subgroup $SO(5)$ by a vev $f$.  $f$ is
generated by strongly coupled physics at the UV scale $\lambda \sim 4\pi
f \sim 10 $~TeV.  The fourteen Goldstone bosons remaining after this symmetry
breaking yield a physical doublet and complex triplet under $SU(2)$,
which remain massless at this stage.
The $SU(5)$ contains a gauged subgroup $[SU(2)\times U(1)]^2$ which is
also broken by $f$ to the SM electroweak gauge group.  The remaining 
four Goldstone bosons are then eaten by a Higgs-like mechanism and give
mass, of order $f\sim 1$ TeV, to the gauge fields of the broken subgroups.
Masses for the complex triplet are generated at the TeV scale by one-loop
gauge interactions.  The neutral component of the complex doublet
plays the role of the SM Higgs, which receives its mass at two-loops from a 
Coleman-Weinberg potential, giving $\mu^2\sim f^2/16\pi^2$.  Thus the
natural scale for $f$ is around a TeV; if $f$ is much higher, the Higgs
mass must again be finely tuned and this model no longer addresses
the hierarchy problem.

The minimum physical spectrum of this model below a TeV is thus that of the 
SM with a single light Higgs.  At the TeV scale, there are four new
gauge bosons (an electroweak triplet and singlet), the scalar triplet,
and a single $Q=+2/3$ vector-like quark $T$.  Other variants of Little
Higgs models may expand this particle content at the TeV scale.

In general, the new vector-like $T$ quark can contribute to
$D^0$-$\overline D^0$ mixing.  Since it has different 
electroweak quantum numbers
from the $Q=+2/3$ quarks in the SM, FCNC interactions will be
induced in the left-handed up quark sector.  This generates a tree-level
$Z$ boson exchange contribution to $D$ mixing as depicted in 
Fig.~\ref{qfig2}.  This was considered in Ref.~\cite{Lee:2004me},
where a specific ansatz for the $4\times 4$ up quark mass matrix was
employed, leading to a quite small contribution to $\Delta M_D$.  In
general, however, one expects the quark mixing to be of order $v/f$ and
the contribution of the flavor-changing $Z$ interaction
induced by the existence of the $T$ quark can be sizable, as 
discussed in the previous section.  Current data can be used to constrain
the mass and mixing of the $T$ quark, and the results in 
Fig.~\ref{SingletQuark23Fig} are applicable in this case.  
Tree-level contributions to $D$ mixing with the exchange of the new heavy 
neutral gauge bosons will likewise be generated, as the fermionic 
coupling for these fields is also proportional to the  
fermion's third component of weak isospin.  The operator structure is
exactly the same as that given in the previous section,
but the magnitude of these contributions
will be suppressed relative to the SM $Z$ boson exchange 
by the heavy mass of the new neutral gauge bosons.  In addition,
there could also be new contributions to rare $D$ meson decays, as
discussed recently in the work of Chen \etal~\cite{Chen:2007yn}.

It has been shown~\cite{lhewp} that a global fit to the precision 
electroweak data set places a significant limit, which is roughly 
parameter independent, on the vev $f$ in the Littlest Higgs model 
of $f\gsim 4$~TeV.  Variants of this model, employing different 
global symmetries, can reduce this constraint somewhat~\cite{lhewp2}. 
In addition, a discrete symmetry, called T-parity, can be 
introduced~\cite{tparity} to alleviate the electroweak bounds. 
This symmetry is analogous to R-parity in supersymmetry and
has the consequence that
T-parity is conserved in interactions and that the lightest T-odd particle 
is stable.  This provides a natural Dark Matter candidate in these models.
Recent work by Hill and Hill~\cite{hillhill} has shown that
anomalies in topological interactions can break this discrete symmetry
and thus T parity is no longer an exact symmetry; the resulting phenomenology
has yet to be worked out.  However, the Ultra-Violet (UV) completion of
the theory may or may not allow for the terms which break T-parity, and
thus a general statement on the presence of this discrete symmetry in the
low-energy theory cannot be made.

In addition to the tree-level contribution discussed above, 
Little Higgs models with T parity can give rise to 
a loop contribution to $D^0$-$\overline D^0$ mixing involving 
the exchange of the heavy gauge bosons and new mirror fermions, which
are present in this form of the model.  In fact,  
three vector-like doublets of mirror fermions are introduced in Little
Higgs models with T parity in order to evade compositeness 
constraints~\cite{tparity}. The effective hamiltonian relevant to 
$D^0$-${\overline D}^0$ mixing for this contribution is~\cite{Hubisz:2005bd}
\beq
{\cal H}_{LH}=\frac{G_F^2 M_W^2}{16 \pi^2}
\frac{ v^2}{f^2} \sum_{i,j}
\xi^{(D)}_i \xi^{(D)}_j F_{LH}(z_i,z_j) Q_1   \ \ .
\eeq
Here, $\xi^{(D)}$ corresponds to the relevant elements of the weak mixing
matrix in the mirror fermion sector which parameterizes the flavor interactions
between the SM and mirror fermions.  The quantity $F_H$ 
(given in the Appendix) is the loop function
computed in Ref.~\cite{Hubisz:2005bd}; it depends on 
$z_i\equiv m^2_{Mi}/M^2_{W_H}$, where $m_{Mi}$ is the mass of the 
$i^{th}$ mirror quark doublet and $W_H$ represents the heavy charged
gauge boson mass.  The $Q_1$ operator appears since the heavy gauge
bosons $W_H$ have purely left-handed interactions. 
The RG running of this hamiltonian is trivial and leads to a 
factor of $r_1(m_c,M)$.
The resulting contribution to the mass difference is
\beq
x^{\rm (LH)}_{\rm D} = \frac{G_F^2 M_W^2 f_D^2 M_D}{24 \pi^2\Gamma_D}  
B_D \frac{ v^2  }{f^2 } \sum_{i,j}
\xi^{(D)}_i \xi^{(D)}_j F_H(z_i,z_j)r_1(m_c,M) \ \ .
\eeq
This has recently been computed in Ref.~\cite{Blanke:2007ee} 
in light of the recent experimental measurement of $D^0$-${\bar D}^0$ 
mixing, where it is found that this contribution can saturate 
the experimental bounds.  

In Little Higgs models with T parity, there is an additional
tree-level contribution arising from the interaction vertex 
$Z_H~\bar q~T^{(-)}$ where $Z_H$ represents either of the heavy neutral
gauge fields and $T^{(-)}$ is the odd T-parity quark.  $T^{(-)}$
couples to the weak eigenstate of the T-parity even quark $T^{(+)}$,
which receives its mass from the same $Q=+2/3$ Yukawa term that is 
responsible for the up quark masses.  This induces mixing between
the quarks in the up quark sector, resulting in 
FCNC interactions of the SM quarks with the
exchange of the new heavy neutral gauge fields.  The generic 
formalism for this contribution will be discussed in the next section,
however this particular contribution is thought to be 
small~\cite{Hubisz:2005bd}.

%%%%%%%%%%%%
%%%%%%%%%%%%%%%%%%%%%%%%%%%%%%%%%%%%%%%%%%%%%%%%%%%%%%%%%%%%%%%%%%%%%
\section{Extra Gauge Bosons} 

Many theories with physics beyond the SM have extended electroweak
gauge symmetries, whose hallmark are the existence of new heavy
neutral and charged gauge bosons.  We note that scenarios with extended
gauge symmetries also generally 
contain new fermions which are required for anomaly
cancellation, as well as an extended Higgs sector to facilitate
the extended symmetry breaking.  The additional heavy gauge bosons
are produced directly at hadron colliders via the Drell-Yan mechanism,
and the search limit on the masses of new $W',Z'$ gauge bosons with SM
couplings is approaching 1 TeV from Run II data at the
Tevatron. The lower bound on the mass of a SM-coupled $Z'(W')$ is 
923(965) GeV from CDF(D0)~\cite{tevzp}.  The LHC will be able to
search for these particles with masses up to $5$ TeV~\cite{tgrtasi}.
There are many such models that yield large FCNC effects in the
up-quark sector.

%%%%%%%%%%%%%%%
\subsection{Generic ${\rm Z'}$ Models}\label{ZprimeSect}

It is possible that a new heavy $Z'$ boson has flavor changing couplings
in the up quark sector.  Here, we examine
a generic tree-level FCNC interaction that mediates $D$ mixing 
via $Z'$-exchange, analogous to the transition depicted in Fig.~\ref{qfig2}.
While the discussion presented here is quite general, many string-inspired 
models have extra $U(1)$ gauge 
symmetries that lead to extra $Z'$-bosons with
possible flavor-changing 
couplings~\cite{jlhtgr,Langacker:2000ju,Arhrib:2006sg}.

The effective four-fermion hamiltonian just below the $Z'$ scale is 
\begin{eqnarray}\label{Zprime}
{\cal H}_{Z^\prime} &=& {1 \over 2 M^2_{Z^\prime}} \left[ 
{\rm C_L}^2 ~ Q_1 + 2 {\rm C_L  C_R} ~ Q_2 + {\rm C_R}^2 ~ Q_6 \right] \ \ ,
\label{4fham}
\end{eqnarray} 
where the dimensionless flavor changing couplings ${\rm C_{L,R}}$ are 
the model dependent inputs to the calculation.  
This is the most general effective hamiltonian and assumes 
flavor-changing interactions occur in both the left- and right-handed sectors. 
We first perform a 
general analysis and will then consider some particular occurrences of
a $Z^\prime$ in the Sections below.

We introduce the Wilson coefficients ${\rm C}_{1,2,6}(M_{Z^\prime})$
by matching at the $Z'$ mass scale, 
%({\it cf} Fig.~\ref{fig:tree}),
%
\bea
{\rm C}_1(M_{Z^\prime})&=& {\rm C_L}^2~, \qquad 
{\rm C}_2(M_{Z^\prime}) = 2 {\rm C_L C_R}~, \qquad 
{\rm C}_6(M_{Z^\prime}) = {\rm C_R}^2 \ \ , 
\eea
with all other Wilson coefficients being zero at this scale. 
Assuming that $M_{Z^\prime} > m_t$ and 
performing the RG running of Eq.~(\ref{Zprime}), 
we obtain the effective hamiltonian at the scale $\mu=m_c$,
\begin{eqnarray}\label{ZprimeMu}
{\cal H}_{Z^\prime}=\frac{1}{2 M_{Z'}^2} 
\left[ \
{\rm C}_1(m_c) ~ Q_1  + {\rm C}_2(m_c) ~ Q_2 + {\rm C}_3(m_c) ~Q_3 + 
{\rm C}_6(m_c) ~Q_6 \right]\ ,
\end{eqnarray}
with
\bea
& & {\rm C}_1(m_c) = r_1(m_c,M_{Z^\prime}) {\rm C}_1(M_{Z^\prime})\ , 
\nonumber \\
& & {\rm C}_2(m_c)= r_2(m_c,M_{Z^\prime}) {\rm C}_2(M_{Z^\prime})  \ , 
\nonumber \\
& & {\rm C}_3(m_c)= \frac{2}{3} \left[
r_2(m_c,M_{Z^\prime}) - r_3(m_c,M_{Z^\prime})
\right] {\rm C}_2(M_{Z^\prime})\ , 
\nonumber \\
& & {\rm C}_6(m_c)= r_6(m_c,M_{Z^\prime}) {\rm C}_6(M_{Z^\prime}) \ \ .
\label{zwilsons}
\eea
The presence of $Q_3$ in Eq.~(\ref{ZprimeMu}) is due to 
operator mixing in the RG running.  Finally, as a check note 
that for the case of no evolution ($r_i \to 1$) 
we obtain the expected behavior 
${\rm C}_i(\mu) \to {\rm C}_i(M_{Z^\prime})$.

Upon evaluating the $D^0$-to-${\bar D}^0$ matrix elements, we obtain 
the $Z'$ tree contribution to $x_{\rm D}$, 
%
%\begin{eqnarray}\label{xZprime}
%x^{\rm (Z')}_{\rm D}=\frac{1}{6} \frac{f_D^2}{\Gamma_D} \frac{M_D}{M_{Z'}^2} 
%\left[2 \left(C_1 (m_c) B_1 + C_6 (m_c) B_6\right)
%- \frac{1}{2} \left( C_2 (m_c) B_2 + {5\over 2} 
%C_3 (m_c) B_3 \right) \right] \ .
%\end{eqnarray}
%
%Since most of the B-parameters in Eq.~(\ref{xZprime}) are not known,
%we employ the modified vacuum saturation approximation introduced above. 
%In this case,
%
\begin{eqnarray}\label{zpr}
x^{\rm (Z')}_{\rm D} &=& {f_D^2 M_D  \over 2 M_{Z'}^2 \Gamma_D} 
B_D \left[ 
{2 \over 3} \left( C_1 (m_c) + C_6 (m_c) \right) - 
C_2 (m_c) \left( {1\over 2} + {\eta \over 3} \right) \right. \nonumber \\
& & \left. \hspace{3.0cm} + 
C_3 (m_c) \left( \frac{1}{12} + \frac{\eta}{2} \right) \right] \ ,
\end{eqnarray}
where we have made use of Eqs.~(\ref{ME_MVS}),(\ref{eta}). 
%$\eta$ is defined in Eq.~(\ref{eta}). 
%
%\beq
%\eta \equiv {M_D^2 \over (m_c + m_u )^2} \cdot {}~.  
%\label{eta}
%\eeq
%
%In our numerical work, we take $B_D^{(S)} = B_D$.  

\begin{figure}[tbp]
\centerline{
\includegraphics[width=11cm,angle=0]{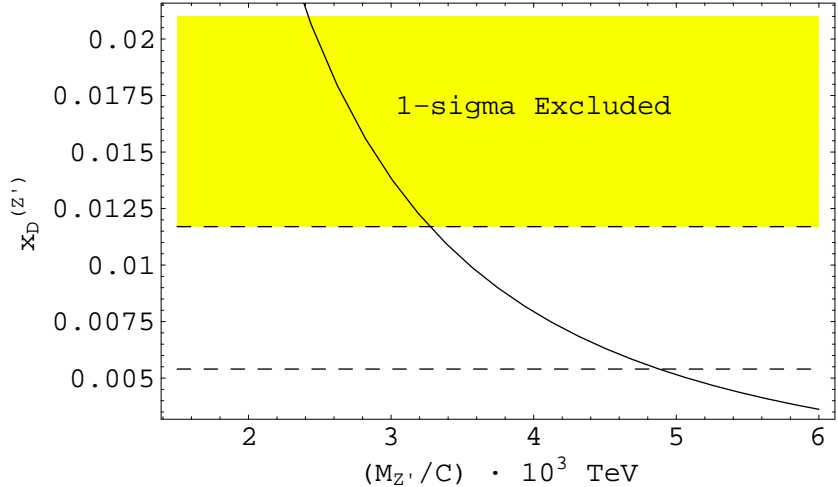}}
\vspace*{0.1cm}
\caption{$x_{\rm D}$ in a generic $Z'$ model as a function of the
$Z'$ boson mass (normalized by the flavor changing coupling constant)
for the case $C_L=C_R=C$.  
The $1\sigma$
experimental bounds are as indicated, with the yellow shaded area
depicting the region that is excluded.}
\label{ZprimeFig}
\end{figure}

Eq.~(\ref{zpr}) can be used to relate the input parameters of $Z'$ models 
($C_{\rm L}$ and $C_{\rm R}$) to some value of $x_{\rm D}$.  Taking 
$x_{\rm D} < 11.7\times 10^{-3}$,
particularly simple expressions are obtained for the limiting cases:
\begin{enumerate}
\item $C_{\rm R} = 0$ (the case with $C_{\rm L}$ replaced by 
$C_{\rm R}$ yields identical limits): 
\bea
{M_{Z^\prime} \over C_L} = \left(
{f_D^2 M_D B_D r_1(m_c,M_{Z^\prime}) \over 3 x_{\rm D} \Gamma_D}
\right)^{1/2} > 8.9 \times 10^5~\mbox{GeV}\ \
, 
\label{case1}
\eea
\item $C_{\rm L} = C_{\rm R} \equiv C$: 
\bea
{M_{Z^\prime} \over C} = 
\left(
{f_D^2 M_D B_D \kappa \over 2 x_{\rm D} \Gamma_D}
\right)^{1/2} > 3.4 \times 10^6~\mbox{GeV} \ \ ,
\label{case2}
\eea
where $\kappa \equiv |4 r_1(m_c,M_{Z'})/3 
- 8r_2(m_c,M_{Z'})/9 - 
(1 + 6 \eta)r_3(m_c,M_{Z'})/9|$. 
\end{enumerate}
In the above, we have fixed the slowly varying $r_1 (m_c,M_{Z'})$
and $\kappa (m_c,M_{Z'})$ at the middle of their ranges.  
The results in Eqs.~(\ref{zpr})-(\ref{case2}) which 
constrain combinations of the flavor-changing couplings 
and the $Z'$ mass (see Fig.~\ref{ZprimeFig}), 
can be applied to fit the needs of the NP model builder. 
For example, in Eq.~(\ref{case1}) 
the choice $C_{\rm L} = 1$, $C_{\rm R} = 0$ 
implies $M_{Z'} > 8.9 \times 10^2$~TeV or alternatively 
taking $M_{Z'} = 1$~TeV yields\footnote{After the work described 
in this section was completed, we received Ref.~\cite{hv} in which 
similar results were discussed.}  the bound 
$C_{\rm L}  < 1.1 \times 10^{-3}$.
%

%%%%%%%%%%%%%%%
\subsection{Family Symmetries}

One class of new physics possibilities to be explored in the TeV range is
family (horizontal) symmetries. The gauge sector of the Standard Model 
lagrangian,
\beq
{\cal L}_{SM} = {\cal L}_{QCD} + {\cal L}_{SU(2)\times U(1)}
+{\cal L}_H\ \ ,
\eeq
actually exhibits a very large global symmetry {\it viz.} 
$SU(3)_L\times SU(3)_R\times U(1)\times U(3)_L\times U(3)_R$. The hope is 
then that some subgroup $G$ of this large symmetry is shared by the Higgs
sector ${\cal L}_H$
and the gauge symmetry of the full lagrangian becomes 
$SU(3)_C\times SU(3)_L\times U(1)$.  The group $G$ 
acts on the families horizontally~\cite{Wilczek:1978xi}, and , of course, 
eventually $G$ has to be 
broken, preferably spontaneously. 

The symmetry $G$ can be implemented locally, so there will be flavor-changing
interactions mediated by new gauge bosons. The symmetry is broken 
spontaneously, making the gauge boson massive with new scalar fields 
being introduced in addition
to the standard Higgs field. 

\begin{figure}[tbp]
\centerline{
%\hspace{-1.cm}
\includegraphics[width=9cm,angle=0]{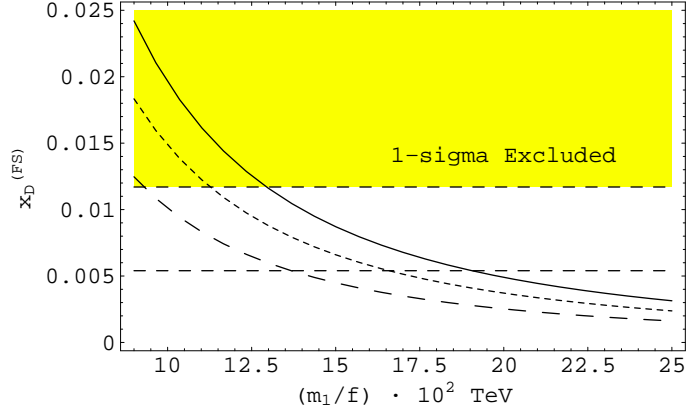}}
%\hspace{1.0cm}
%\includegraphics[width=6cm,angle=0]{family.eps}}
\vspace*{0.1cm}
\caption{$x_{\rm D}^{\rm (FS)}$ as a function of $m_1/f$ for
the gauge boson mass ratios $m_1/m_2 = 0.01\,, 0.5\,, 0.7$ corresponding
to the solid, dotted, and dashed curves, respectively.  The $1\sigma$
region for $x_{\rm D}$ is also shown, with the yellow shaded region
depicting the excluded region.}
%Right: $1\sigma$ excluded region (short-dashed curve) for 
%the masses of family symmetry-mediating gauge bosons discussed in 
%the text.  Possible future contours taking
%$x_{\rm D}< (15.0\,, 8.0\,, 5.0\,, 3.0)\times 10^{-3}$ from top
%to bottom, 
%corresponding to the solid, medium dashed, long dashed, and 
%longer dashed curves, respectively, are also shown.} 
\label{Family}
\end{figure}

As a prototype, let us consider a very 
simple model~\cite{Monich:1980rr}.  We consider the group $SU(2)_G$ 
acting only on the first two left-handed families (it may be regarded 
as a subgroup of an $SU(3)_G$, which is broken). Spontaneous 
breaking of $SU(2)_G$ makes the gauge
bosons $G_i$ massive. The LH doublets
\beq
\left(
\begin{array}{c}
u^0 \cr 
d^0
\end{array}
\right)_L \qquad {\rm and} \qquad 
\left(
\begin{array}{c}
c^0 \cr 
s^0
\end{array}
\right)_L \ \ ,
\eeq
transform as $I_G=1/2$ under $SU(2)_G$, as do the lepton doublets
\beq
\left(
\begin{array}{c}
\nu_e^0 \cr 
e^0
\end{array}
\right)_L  \qquad {\rm and} \qquad 
\left(
\begin{array}{c}
\nu_\mu^0 \cr 
\mu^0
\end{array}
\right)_L \ \ ,
\eeq
and the right-handed fermions are singlets under $SU(2)_G$.
The superscript refers to the fact that these are weak eigenstates and not mass
eigenstates. The couplings of fermions to the family gauge bosons
$G$ is given by
\beq
{\cal L} = f \left[
{\overline{\psi}_{d^0}}_L \gamma_\mu \vec{\tau}\cdot 
\vec{G}_\mu {\psi_{d^0}}_L +
{\overline{\psi}_{u^0}}_L \gamma_\mu \vec{\tau}\cdot 
\vec{G}_\mu {\psi_{u^0}}_L +
{\overline{\psi}_{\ell^0}}_L \gamma_\mu \vec{\tau}\cdot 
\vec{G}_\mu {\psi_{\ell^0}}_L
\right] \ \ ,
\eeq
where $f$ denotes the coupling strength and $\vec\tau$ are the generators
of $SU(2)_G$.  We define the mass basis by
\beq
\left(
\begin{array}{c}
d \cr 
s
\end{array}
\right)_L =
U_d \left(
\begin{array}{c}
d^0 \cr 
s^0
\end{array}
\right)_L, \qquad \
\left(
\begin{array}{c}
u \cr 
c
\end{array}
\right)_L =
U_u \left(
\begin{array}{c}
u^0 \cr 
c^0
\end{array}
\right)_L, \ \qquad 
\left(
\begin{array}{c}
e \cr 
\mu
\end{array}
\right)_L =
U_\ell \left(
\begin{array}{c}
e^0 \cr 
\mu^0
\end{array}
\right)_L \ \ .
\eeq
In the limit of CP-conservation each of the three $2\times 2$ matrices $U_d$, 
$U_u$, and $U_\ell$ is characterized by one angle:
$\theta_d$, $\theta_u$, and $\theta_\ell$ 
(where the Cabibbo angle is $\theta_c = \theta_u - \theta_d$). 
One then finds for the couplings in the fermion mass basis:
\bea
& & {\cal L} = f \Bigl[
G_{1\mu} 
\Bigl\{
\sin2\theta_d \left(
\overline{d}_L \gamma_\mu d_L - \overline{s}_L \gamma_\mu s_L \right)
+ \sin2\theta_u \left(
\overline{u}_L \gamma_\mu u_L - \overline{c}_L \gamma_\mu c_L \right)
\nonumber \\
&+& \sin2\theta_l \left(
\overline{e}_L \gamma_\mu e_L - \overline{\mu}_L \gamma_\mu \mu_L \right)
+ \cos2\theta_d \left(
\overline{d}_L \gamma_\mu s_L + \overline{s}_L \gamma_\mu d_L \right)
\nonumber \\
&+& \cos2\theta_u \left(
\overline{u}_L \gamma_\mu c_L + \overline{c}_L \gamma_\mu u_L \right)
+ \cos2\theta_l \left(
\overline{e}_L \gamma_\mu \mu_L + \overline{\mu}_L \gamma_\mu e_L \right)
\Bigr\}
\nonumber \\
&+& i G_{2\mu}
\left\{
\left(\overline{s}_L \gamma_\mu d_L - \overline{d}_L \gamma_\mu s_L \right)
+ \left(\overline{c}_L \gamma_\mu u_L - \overline{u}_L \gamma_\mu c_L \right)
+ \left(\overline{\mu}_L \gamma_\mu e_L - \overline{e}_L 
\gamma_\mu \mu_L \right)
\right\}
\nonumber \\
&+& G_{3\mu}
\Bigl\{
\cos2\theta_d \left(
\overline{d}_L \gamma_\mu d_L - \overline{s}_L \gamma_\mu s_L \right)
+ \cos2\theta_u \left(
\overline{u}_L \gamma_\mu u_L - \overline{c}_L \gamma_\mu c_L \right)
\nonumber \\
&+& \cos2\theta_l \left(
\overline{e}_L \gamma_\mu e_L - \overline{\mu}_L \gamma_\mu \mu_L \right)
- \sin2\theta_d \left(
\overline{d}_L \gamma_\mu s_L + \overline{s}_L \gamma_\mu d_L \right)
\nonumber \\
&-& \sin2\theta_u \left(
\overline{u}_L \gamma_\mu c_L + \overline{c}_L \gamma_\mu u_L \right)
- \sin2\theta_l \left(
\overline{e}_L \gamma_\mu \mu_L + \overline{\mu}_L \gamma_\mu e_L \right)
\Bigr\}
\Bigr]\ \ .
\eea
For simplicity we assume that after symmetry breaking the gauge 
boson mass matrix is diagonal to a good approximation in which 
case $G_{i\mu}$ are physical eigenstates and any mixing between
them is neglected.  This lagrangian clearly introduces tree-level 
FCNC interactions and gives a contribution to
$D^0$-${\bar D}^0$ mixing of
\beq
{\cal H}_{FS} (m_i) = f^2 \left(
\frac{\cos^2 2\theta_u}{m_1^2} + \frac{\sin^2 2\theta_u}{m_3^2} 
- {1\over m_2^2}\right)
\overline{u}_L \gamma_\mu c_L \overline{u}_L \gamma^\mu c_L \ \ .
\eeq
A simple symmetry breaking pattern (see, \eg,~\cite{Monich:1980rr}) 
leads to
$m_1=m_3\neq m_2$. Since the effective hamiltonian only involves the
operator $Q_1$, the
RG-running is simple and leads to the following structure
at the $m_c$ scale,   
\beq
{\cal H}_{FS} (m_c) = f^2 r_1(m_c,M) 
\frac{m_2^2-m_1^2}{m_1^2 m_2^2}
\overline{u}_L \gamma_\mu c_L \overline{u}_L \gamma^\mu c_L \ \ ,
\eeq
where $M$ is the smaller of the new gauge boson masses 
$m_1$ and $m_2$. This leads to
a value for $x^{\rm (FS)}_{\rm D}$ of
\beq\label{FlavorSymmetry}
x^{\rm (FS)}_{\rm D} = \frac{2f_D^2 M_DB_D}{3\Gamma_D}{f^2\over m_1^2} 
\left(1 - {m_1^2\over m_2^2} \right)
 r_1(m_c,M)  \ \ .
\eeq
Using the available experimental data on $D^0$-${\bar D}^0$ 
mixing parameters, yields
constraints on the masses of the family symmetry-mediating gauge 
bosons. They are presented in Fig.~\ref{Family} for $m_i/f$. 
%

%One can easily estimate an upper bound on a possible contribution of this
%sort to $D^0$-${\bar D}^0$ mixing. For simplicity, let us 
%disregard QCD running
%and compare the effects in the charm and strange systems,
%
%\beq
%\left(\Delta M_D\right)_{FS} = 
%\frac{f_D^2 M_D B_D}{f_K^2 M_K B_K}
%\cdot \left(\Delta M_K\right)_{FS} \ \ . 
%\eeq
%
%Since the contribution to $\left(\Delta M_K\right)$ from the $G$-exchange
%should be less than the experimental value,
%
%\beq
%\left(\Delta M_D\right)_{FS} \leq 10 ~ \left(\Delta M_K\right)_{exp}
%\simeq 1.8 \cdot 10^{-11}~\mbox{MeV} \ \ .
%\eeq
%
%This corresponds to a contribution to $x^{(FS)}_{\rm D}$ 
%up to 2\% and can easily be 
%larger than the SM contribution.

%%%%%%%%%%%%%%%%%%%%%%%%%%%%%%
\subsection{Left-Right Symmetric Model}

A puzzling feature of the SM is the left-handed nature of the 
electroweak interactions.  A long-standing possible remedy,
known as the Left-Right Symmetric Model (LRM)~\cite{mohap},
seeks to restore parity at high energies by enlarging the gauge
symmetry to $SU(2)_L\times SU(2)_R\times U(1)_{B-L}$. This model can be
embedded into an $SO(10)$ (or $E_6$) GUT structure which then provides a
natural mechanism (seesaw) for generating light neutrino masses.  A
supersymmetric version of a left-right symmetric $SO(10)$ GUT model
yields the correct prediction~\cite{tgrdesh} for
$x_w\equiv \sin^2\theta_w(M_Z)$ and $\alpha_s(M_Z)$, while allowing for
the masses of the
new gauge bosons ($Z_R$ and $W_R^\pm$) associated with the $SU(2)_R$
symmetry to be of order a few TeV or less.  Light masses for the new
gauge bosons can also be obtained in models with a horizontal
symmetry~\cite{Kiers:2005gh}.
Manifest left-right 
symmetry dictates that the right-handed gauge coupling take on the
same value as the left-handed SM coupling 
$g_L$ and that the elements of the right-handed CKM matrix
be equal to their left-handed counterparts.  In this case, the direct
search for new gauge bosons at the Tevatron places the bound~\cite{tevzp}
of $M_{R}> 788$~GeV on the mass of the charged right-handed gauge boson.

The $Z_R$ has flavor conserving couplings in this model, and thus
does not mediate $D^0$-$\overline D^0$ mixing.  The charged right-handed
gauge field, however, can participate in virtual exchange 
in a box diagram, in association with the SM $Q=-1/3$ quarks, and 
gives a contribution to meson mixing. In fact, the strongest bound on the
mass of the $W_R$, which is $M_{R}\gsim 1.6$~TeV in the limit of manifest
left-right symmetry, is derived from its
contribution to $K^0$-$\overline K^0$ mixing~\cite{irvine}.

However, there is no compelling theoretical reason to adopt manifest
left-right symmetry and in more general models 
the elements of the right-handed 
analog of the CKM matrix can take on any values, while still respecting
unitarity.  In addition the ratio of gauge couplings can vary~\cite{cmp}
between $x_w/(1-x_w)=0.55<g_R/g_L\equiv\kappa < 1-2$.  In this case where
manifest left-right symmetry is dropped, the bounds
from Kaon mixing are softened to $M_{W_R}\gsim 300 $~GeV~\cite{langsar} 
and the direct
collider searches are significantly weakened~\cite{tgrwr}.

The $|\Delta C|=2$ hamiltonian at the right-handed mass scale
is given by
\begin{equation}
{\cal H}_{LRM}={G_F^2M_W^4\over 4\pi^2M_R^2}[\kappa^2
V^R_{ub}V^{R*}_{cb}V^L_{ub}V^{L*}_{cb}J(x_b^W,\beta)Q_2
+\kappa^4(V^R_{ub}V^{R*}_{cb})^2S(x_b^R)Q_6]\ \ ,
\end{equation}
with $x_b^i=m_b^2/M_i^2$, $\beta=M_W^2/M_R^2$, $V^{L,R}$ denote the
left- and right-handed CKM matrix, and the quantities $S(x)$ 
(an Inami-Lim function) and $J(x,\beta)$ are given in the Appendix. 
The first term in this hamiltonian corresponds to the exchange of one
$W_R$ and one Standard Model $W$ boson in the box diagram, while
the second term represents the contribution where only the $W_R$ 
participates.
Here, we ignore mixing between the left- and right-handed gauge bosons.
The identification of the matching conditions at the high scale,
$C_{2,6}(M_R)$, are obvious.  The RG evolution to the charm scale yields
the effective hamiltonian at $\mu=m_c$
\begin{equation}
{\cal H}_{LRM}= {1\over 2M_R^2}[C_2(m_c)Q_2+C_3(m_c)Q_3+C_6(m_c)Q_6]\ \ ,
\end{equation}
where operator mixing has induced the dependence on $Q_3$ similar
to the generic $Z'$ case discussed above, with
\begin{eqnarray}
C_2(m_c) & = & r_2(m_c,M_R)C_2(M_R)\ \ , \nonumber \\
C_3(m_c) & = & {2\over 3}[r_2(m_c,M_R)-r_3(m_c,M_R)]C_2(M_R) \ \ , \\ 
C_6(m_c) & = & r_6(m_c,M_R)C_6(m_c,M_R)\ \ . \nonumber
\end{eqnarray}
Evaluating the hadronic matrix elements yields
\begin{equation}
x_{\rm D}^{\rm (LRM)} = {f_D^2M_DB_D\over 24M_R^2\Gamma_D}
\left[ -10C_2(m_c)+7C_3(m_c)+8C_6(m_c) \right]
\end{equation}
upon employing the vacuum saturation approximation and taking $\eta=1$ in 
Eq.~(\ref{eta}).

\begin{figure}[t]
\centerline{
\includegraphics[width=6cm,angle=90]{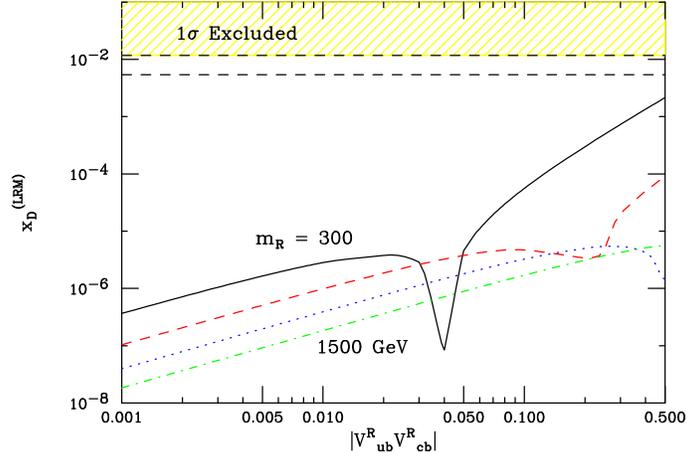}}
\vspace*{0.1cm}
\caption{$x_{\rm D}$ in the Left-Right Symmetric model with
non-manifest left-right symmetry as a function of the right-handed
CKM mixing factor $|V^R_{ub}V^{R*}_{cb}|$ for $M_R=300\,, 600\,,
1000\,, 1500$~GeV from top to bottom.  The $1\sigma$ 
experimental bounds are as indicated, with the yellow shaded area 
depicting the region that is excluded.}
\label{lrmfig}
\end{figure}

It is clear that in the case of manifest left-right symmetry, the
combination of the small $(V^L_{ub}V^{L*}_{cb})$ CKM elements and the 
$M_R^{-2}$ suppression will result in a very small value for 
$x_{\rm D}^{\rm(LRM)}$.  However, it is possible that for
non-manifest left-right symmetry, where the right-handed CKM
elements $(V^R_{ub}V^{R*}_{cb})$ may take on larger values, that
a significant effect may be generated.  We examine this scenario,
taking $(V^R_{ub}V^{R*}_{cb})$ to lie in the range $0.001 - 0.5$,
where $0.5$ is the maximum value that this quantity can attain while
respecting unitarity of the right-handed CKM matrix.  Our results
are shown in Fig.~\ref{lrmfig} for various values of $M_R$, 
taking $\kappa\equiv g_R/g_L=1$.  We see, that even for the most
extreme values of the parameters, the LRM contribution to 
$D^0$-$\overline D^0$ mixing never reaches the experimentally
determined value.  Since both of the loop functions $J(x,\beta)$ and
$S(x)$ go as $m_b^2$, the suppression from the small internal
quark masses dominates this result.  The dip in the curves results
from interference due to operator mixing.

This exercise shows that a generic scenario with a new heavy
charged gauge boson that participates in the box diagram for
$D$ mixing will not induce sizable contributions to the neutral
$D$ meson mass difference, unless it is accompanied by new heavy
$Q=-1/3$ quarks.  The recently proposed Twin-Higgs models~\cite{roni}, which
are based on a $SU(2)_R\times S(2)_L\times U(1)_{B-L}$ gauge
symmetry, contain an extended top quark sector with heavy
$Q=+2/3$ quarks and would give significant contributions to
$K\,, B_{d,s}$ mixing, but not to $\Delta M_{\rm D}$.

%%%%%%%%%%%%%%%
\subsection{Alternate Left-Right Model from $E_6$ Theories}

An alternative to the conventional Left-Right Symmetric Model discussed
above is possible in supersymmetric $E_6$ Grand Unified 
Theories~\cite{jlhtgr}.
This model is also based on the low-energy gauge group
$SU(2)_L\times SU(2)_R\times U(1)$, 
but makes use of ambiguous fermion assignments within the fundamental
representation of $E_6$~\cite{Babu:1987kp}.  
The additional right-handed charged and neutral gauge fields 
in this model have different properties than in the traditional 
scenario.  A single generation
in $E_6$ theories contains 27, 2-component fermions (in contrast
to the 16 fermions per generation in SO(10)), and quantum number
ambiguities arise that allow the $T_{3L(R)}$ assignments 
to differ from their customary values for the
$\nu_{L,R}\,,e_L$, and $d_R$ fields.  In the quark sector, the 
right-handed up-quarks then form $SU(2)_R$ doublets with the 
exotic $Q = -1/3$ vector singlet quark, $D_R$, that  
is present in the ${\bf 27}$ representation of $E_6$,
while the $SU(2)_L$ doublet $(u,d)$ remains unchanged.  $D_L$
and $d_R$ are then singlet fields under all gauge symmetries.  
This allows, for example, 
the right-handed $W$ boson to couple the right-handed up-quark
sector to the singlet quark $D_R$.  Examination of the superpotential for this 
model shows that the $D_R$ takes on the quantum number assignment 
of a leptoquark, while the 
$W_R$ carries negative R-parity and non-zero lepton
number, and thus cannot mix with the $W_L$ of the SM or couple to
the down-quark sector.  The usual
constraints on right-handed gauge bosons from the $K_L-K_S$ mass
difference and polarized $\mu$ decay are thus evaded in this scenario.

These exotic particles can induce significant contributions to
$D^0$-$\bar D^0$ mixing~\cite{Ma:1987ji} via $W_R$ and $D_R$
exchange in a standard box diagram.  Note that since the heavy
$D_R$ quarks are not kinematically accessible in charm-quark
decay, there is no dispersive amplitude in this case. 
The interactions of the right-handed $W$ boson take the form
\begin{equation}
{\cal L}={g_R\over\sqrt 2}V^R_{ij}\bar u_i\gamma_\mu(1+\gamma_5)
D_jW_R^\mu\ \ ,
\end{equation}
where $i,j$ are generational indices and $V^R_{ij}$ is the
right-handed analog of the CKM quark mixing matrix governing the right-handed
charged currents.  The effective $|\Delta C|=2$ hamiltonian at the scale of the
right-handed interactions is
\begin{equation}
{\cal H}_{ALRM}={g_R^4\over 128\pi^2M_R^2}\sum_{i,j}
(V^{R}_{ui}V^{R*}_{ci})(V^{R}_{uj}V^{R*}_{cj}) S(x_i,x_j) Q_6\ \ ,
\end{equation}
where the sum over $i,j$ extends over the three generations of $D_R$
quarks, $S(x_i,x_j)$ are the standard Inami-Lim functions~\cite{Inami:1980fz}
(given in the Appendix) and $x_i=m^2_{D_{R,i}}/M_R^2$ 
with $M_R$ being the mass
of the new right-handed gauge boson.  Note that this expression mirrors
that in Eq.~(\ref{smham}) except for the presence of the right-handed
operator $Q_6$.  Matching at the scale $M_R$ yields
\begin{equation}
C_6(M_R)= {g_R^4\over 64\pi^2}\sum_{i,j}\ 
(V^{R}_{ui}V^{R*}_{ci})(V^{R}_{uj}V^{R*}_{cj}) S(x_i,x_j)\ \ .
\end{equation}
Performing the RG evolution we obtain at the charm quark
scale
\begin{equation}
{\cal H}_{LR}= {1\over 2M_R^2} C_6(m_c)Q_6\ \ ,
\end{equation} 
with $C_6(m_c)=r_6(m_c,M_R)C_6(M_R)$.  This yields the contribution to
the $D$ meson mass difference
\begin{equation}
x^{\rm (ALRM)}_{\rm D}={g_R^4f_D^2B_D M_D\over 192\pi^2\Gamma_D M_R^2}
r_6(m_c,M_R)\sum_{i,j}
(V^R_{ui}V^{R*}_{ci})(V^{R}_{uj}V^{R*}_{cj}) S(x_i,x_j)\ \ .
\end{equation}

\begin{figure}[htbp]
\centerline{
\includegraphics[width=6cm,angle=90]{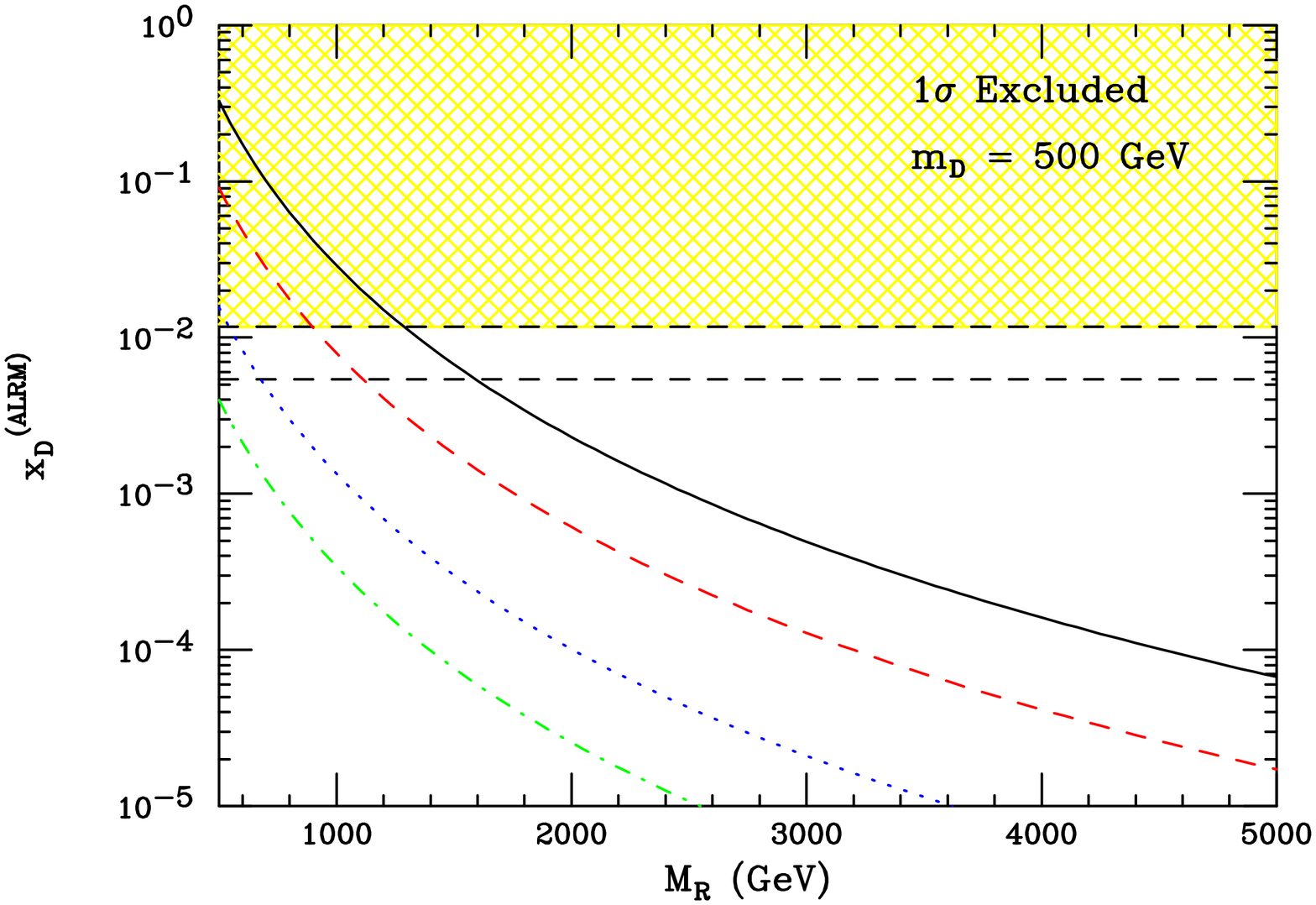}
%\hspace*{5mm}
\includegraphics[width=6cm,angle=90]{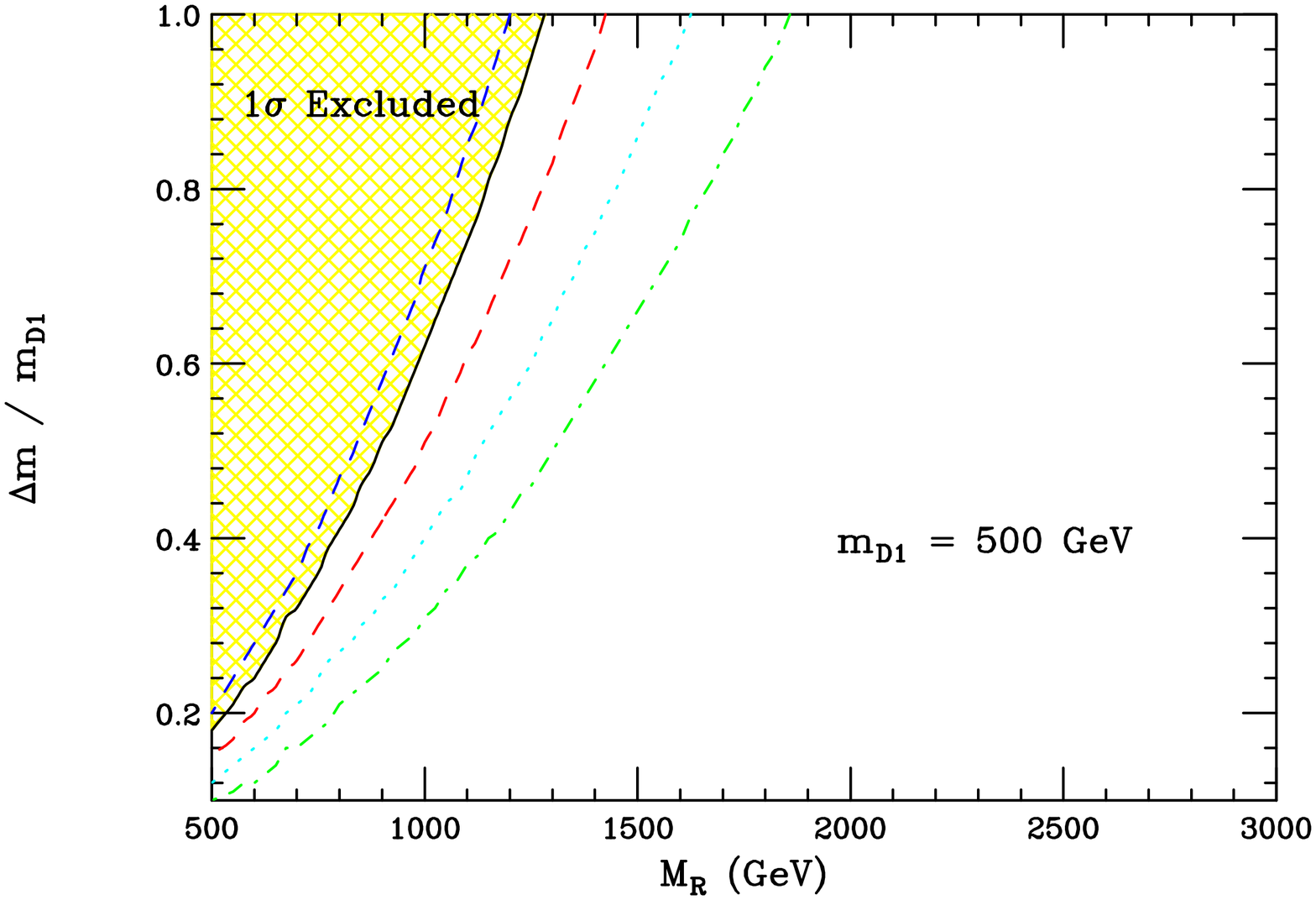}}
\centerline{
\includegraphics[width=6cm,angle=90]{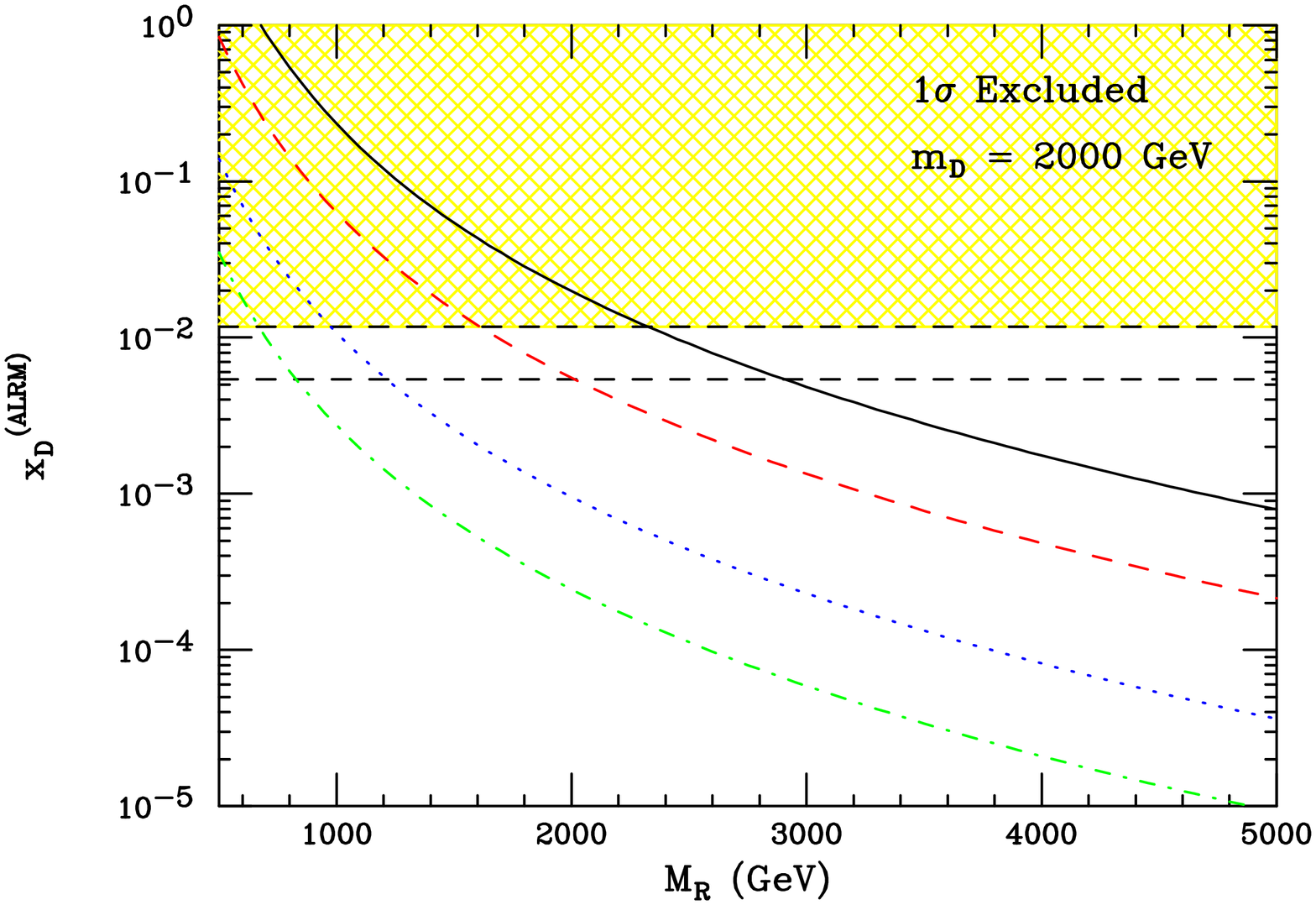}
%\hspace*{5mm}
\includegraphics[width=6cm,angle=90]{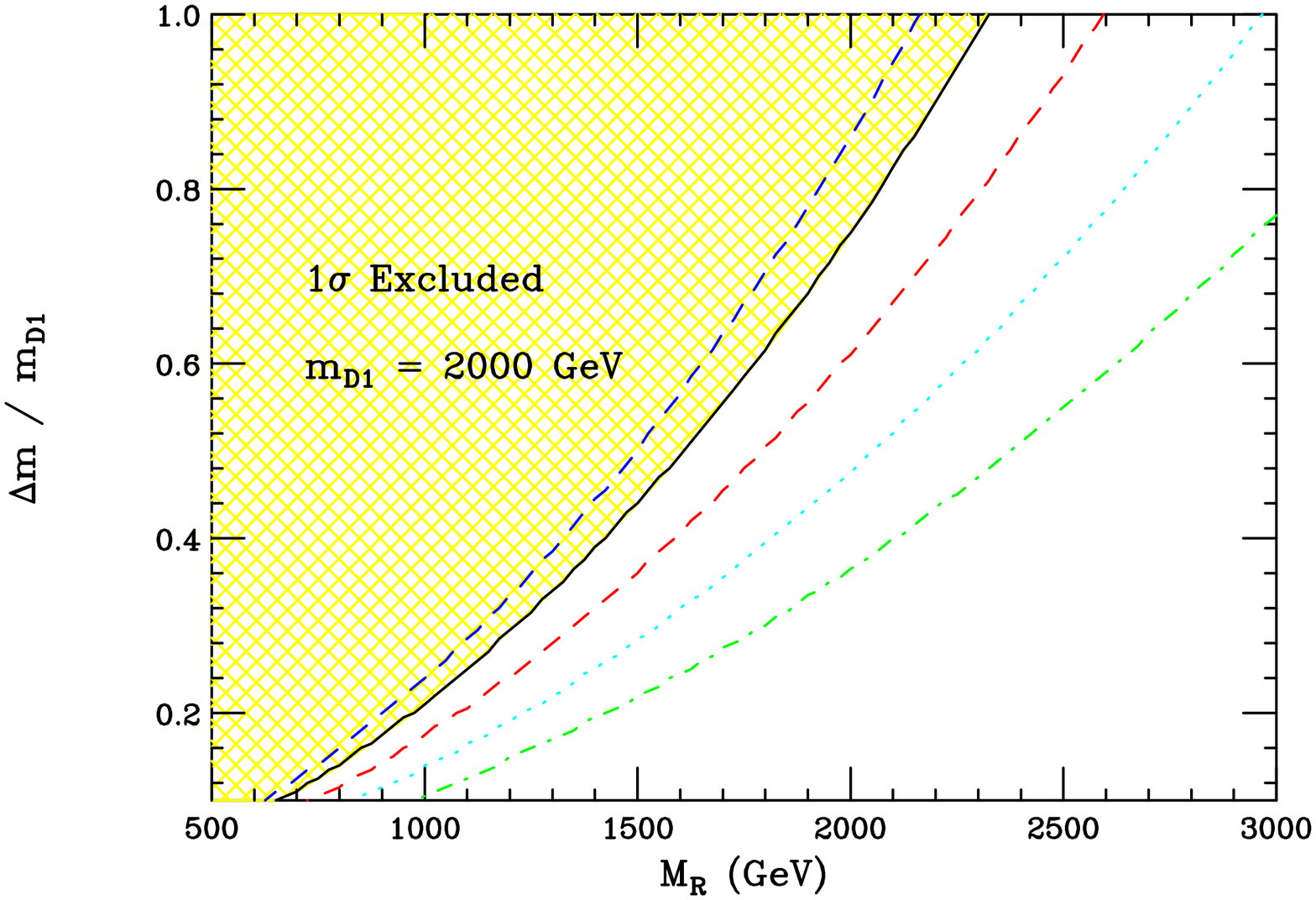}}
\vspace*{0.1cm}
\caption{
Left:  $x_{\rm D}$ in the alternate left-right model as a function of the
mass of the $W_R$ for various values of the singlet quark mass splittings,
$\Delta/m = 0.1\,, 0.2\,, 0.5\,, 1.0$ from bottom to top.  The $1\sigma$ 
experimental bounds are as indicated, with the yellow shaded area 
depicting the region that is excluded.  \\
\noindent Right: The present $1\sigma$ excluded region in the $M_R$ -\
mass splitting 
parameter plane, as well as possible future contours taking
$x_{\rm D}< (15.0\,, 8.0\,, 5.0\,, 3.0)\times 10^{-3}$,
corresponding to the blue dashed, red dashed, cyan dotted, and 
green dot-dashed curves, respectively.  The mass of the first generation
singlet quark $D_1$ is taken to be 500 and 2000 GeV as labeled.}
\label{alrmfig}
\end{figure}

The magnitude of these contributions is determined by the form of the
right-handed quark mixing matrix, the degeneracy of the 3
generations of $D_R$ quarks, as well as the right-handed mass scale. 
If the quarks are fully degenerate, then
a right-handed GIM mechanism is operative due to the unitarity of $V^R$
and this contribution to $D$ mixing vanishes.  If there are mass
splittings between the three generations of $D_{R,i}$, 
then the observed value of $D$ mixing can 
place bounds on the size of these splittings.  Here, we will examine 
the case where $V^R$ takes on the form of the left-handed
quark mixing matrix, {\it i.e.}, it displays the hierarchal structure
of the CKM matrix,
and derive constraints on this mass splitting as a function of the
right-handed mass scale.  Our results are presented in Fig.~\ref{alrmfig}.
The value of $x_{\rm D}^{\rm (ALRM)}$ as a function of $M_R$, the mass
of the right-handed charged gauge boson is displayed for various
mass splittings, $\Delta m/m_{D_1}$, where $\Delta m \equiv m_i-m_j$ is
taken to be constant between the first and second as well as second
and third generations.  We also display the constraints the present $1\sigma$
experimental bound of $x_{\rm D}<11.7\cdot 10^{-3}$, as well as
contours for future possible values of $x_{\rm D}$, places on the
$M_R$ - mass splitting parameter plane.  The results are shown for
two representative values of the first generation $D$-quark mass,
$m_{D_1}=500$ and 2000 TeV.  We see that the mass of the $W_R$ is
restricted to be $M_R\gsim 1-5=2.0$ for large values of the quark mass
splittings.

%%%%%%%%%%%%%
\subsection{Vector Leptoquark Bosons}

In most recent papers on the subject, scalar leptoquarks (particles with both 
quark and lepton quantum numbers) have usually been associated with 
$R$-parity-violating SUSY scalars (to be considered in 
Sect.~\ref{SUSYRsection}).  However, vector 
leptoquarks (VLQs) are also a possibility~\cite{brwlq}. For example, they 
naturally arise in Grand Unified Theories, where quarks 
and leptons belong to the same multiplet~\cite{Ross:1985ai}. Many 
New Physics models where leptoquarks 
are introduced as fundamental vector particles imply that they 
serve as gauge particles mediating quark and lepton-number-changing 
interactions resulting from GUT-model symmetry  
groups. Those symmetries are usually broken at a rather high scale, of order 
$10^{15}$~GeV, which implies that, barring fine-tuning, vector leptoquarks 
receive masses near the GUT symmetry-breaking scale. Yet, some 
models exist where leptoquarks receive masses at a lower scale. In addition,
more exotic constructions, such as preon (composite) models, 
could also contain 
vector leptoquarks. In those models, however,
leptoquarks are composite particles with masses that are of 
the order of compositeness scale. Thus, observations of effects of VLQs 
could potentially probe physics at a very high mass scale. It is for 
these reasons vector leptoquarks are searched for experimentally. 
Collider searches at the Tevatron for the direct production of vector
leptoquark pairs yield the constraint $m_{VLQ}>290$~GeV~\cite{d0lq} from
Run I data for second generation leptoquarks which decay into muons.

While there are a number of phenomenological studies of vector 
leptoquarks~\cite{Davidson:1993qk,Leurer:1993qx}, a general problem exists 
with placing constraints on VLQs from indirect measurements, and in 
particular from  $D^0$-$\overline{D}^0$ mixing. This is because 
their couplings are model-dependent. In particular, loop calculations 
with massive composites, {\it i.e.} non-gauge leptoquarks, receive 
contributions that are divergent and must be regulated by the 
compositeness scale. For the case of
gauge leptoquarks, and in the absence of a GIM-like mechanism in the 
leptoquark box diagram, one can choose a gauge (such as the Feynman gauge) 
to unambiguously compute the effects of leptoquark interactions.
In that gauge, however, one also must add contributions from unphysical 
states responsible for the generation of the VLQ masses.  In a specific
model, the interactions of the unphysical states are fixed and their 
contributions are readily computable.  However, this then becomes
a rather model-dependent procedure because VLQ masses can be generated by 
various means, including some version of the Higgs mechanism, 
or a Frogatt-Nielsen mechanism, etc.  Rather than rely on a specific model, 
in what follows, for generality,
we shall follow the approach of Ref.~\cite{Davidson:1993qk} 
and obtain bounds on the couplings of gauge VLQs by dropping the contributions 
from the unphysical states.  

\begin{figure}[tbp]
\centerline{
\includegraphics[width=8cm,angle=0]{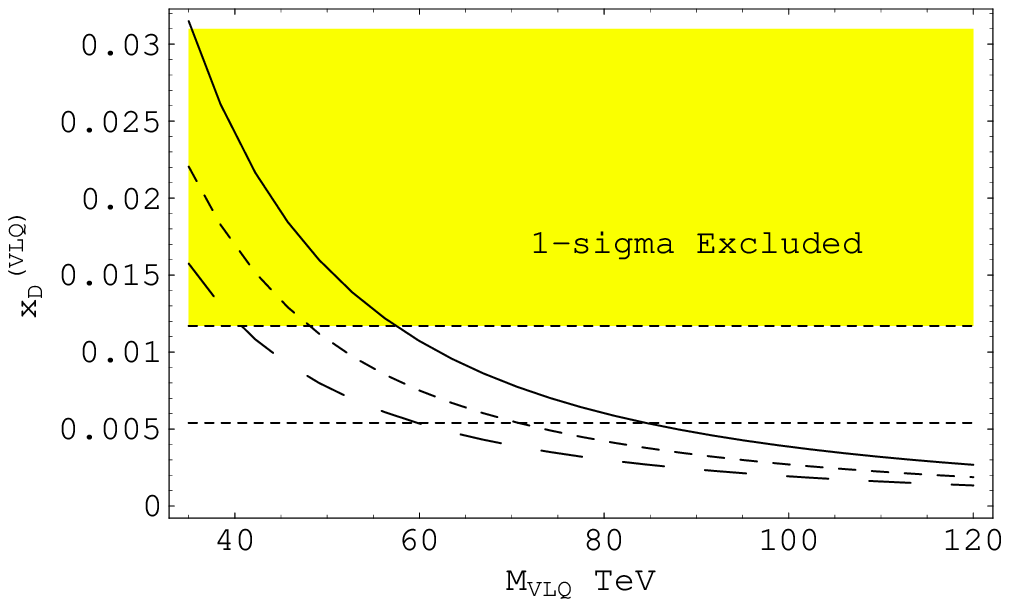}
\includegraphics[width=8cm,angle=0]{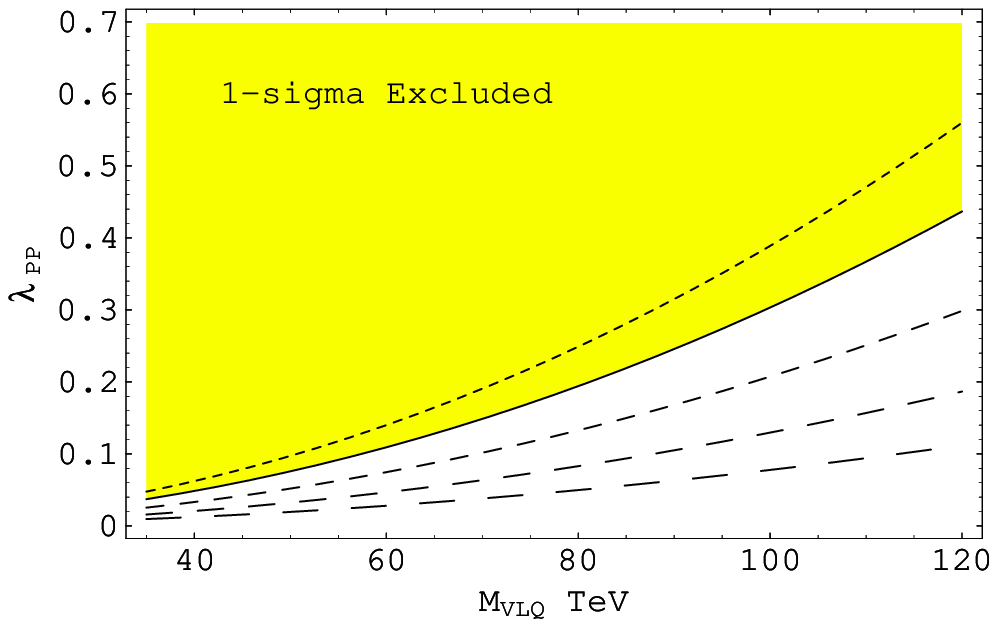}}
\vspace*{0.1cm}
\caption{Left: $x_{\rm D}$ in Vector Leptoquark models as a function of the
Vector Leptoquark mass $M_{VLQ}$, with 
$\lambda_{PP}=0.1$ (solid line), $\lambda_{PP}=0.07$ (short dash), 
and $\lambda_{PP}=0.05$ (long dash) for $P=L ~\mbox{or}~ R$.  The
$1\sigma$ experimental bounds are as indicated, with the yellow shaded
area depicting the region that is excluded.\\
Right: The present $1\sigma$ excluded region in the vector
leptoquark mass - coupling 
parameter plane, as well as possible future contours taking
$x_{\rm D}< (15.0\,, 8.0\,, 5.0\,, 3.0)\times 10^{-3}$,
corresponding to the blue dashed, red dashed, cyan dotted, and 
green dot-dashed curves, respectively.}
\label{VectorLeptoquarksFig}
\end{figure}

In general, a VLQ could couple to both left-handed and right-handed 
fermions, so we shall assume the general form of the coupling. 
We note, however, that there are stringent bounds~\cite{muchado} from 
low-energy data if leptoquarks couple to both left- and right-handed
states, and it is generally assumed that their couplings are chiral.
For a quark of flavor $q$ and a lepton $\ell$, 
we adopt the interaction vertex 
$i \gamma_\mu\left[\lambda^{\ell q}_L P_L + \lambda^{\ell q}_R 
P_R\right]$, which leads to the contribution to $D$ meson mixing, 
\begin{eqnarray}
x^{\rm (VLQ)}_{\rm D} &=& {1 \over 8 \pi^2 m_{LQ}^2 \Gamma_D M_D}
\bigg[\left(\lambda_{LL} \langle Q_1 \rangle + 
2 \lambda_{LR} \langle Q_2 \rangle 
+ \lambda_{RR} \langle Q_6 
\rangle\right) 
\nonumber \\
& & \hspace{1.5cm} +  \frac{10}{9}\frac{m_c^2}{m_{LQ}^2}
\left(\lambda_{LL} \langle Q_7 \rangle + 2 \lambda_{LR} \langle Q_3 \rangle +
\lambda_{RR} \langle Q_4 \rangle\right)
\bigg]
\nonumber \\
&=& {f_D^2 M_D B_D \over 12 \pi^2 m_{LQ}^2 \Gamma_D}  
\left[ \left(\lambda_{LL} + \lambda_{RR} \right) -
\frac{3}{2} \lambda_{LR} \left(1 + \frac{2}{3} \eta \right)
\right]\ \ ,
\end{eqnarray}
where $\lambda_{PP'}\equiv\sum_{ij} \left(\lambda^{\ell_i c}_{P} 
\lambda^{\ell_i u}_{P} \right)
\left(\lambda^{\ell_j c}_{P'} \lambda^{\ell_j u}_{P'} \right)$, and
we neglect ${\cal O}(m_c/m_{LQ})$ corrections in the last line. 
The resulting bounds on VLQ interactions are displayed in 
Fig.~\ref{VectorLeptoquarksFig}.
%

%%%%%%%%%%%%
%%%%%%%%%%%%%%%%%%%%%%%%%%%%%%%%%%%%%%%%%%%%%%%%%%%%%%%%%%%%%%%%%%%%%
\section{Extra Scalars} 

No known physical principle restricts the number of Higgs multiplets
that can participate in electroweak symmetry breaking.  In fact,
several theories beyond the SM, such as Supersymmetry and those with
extended gauge sectors, require an enlarged Higgs sector in order to
break the additional symmetries. Here we
examine the effect in $D^0$-$\overline{D}^0$ mixing of models with 
multiple Higgs doublets, with and
without flavor conservation, Higgsless models and models with
scalar leptoquarks.

%%%%%%%%%%%%%%%%%%%%%%%%%%%%%%%%%%%%%%%%%%%%%%%%%%%%%%%%%%%%
\subsection{Flavor Conserving Two-Higgs-Doublet Models}\label{2hdm}

\begin{figure} [tb]
\centerline{
\includegraphics[width=10cm,angle=0]{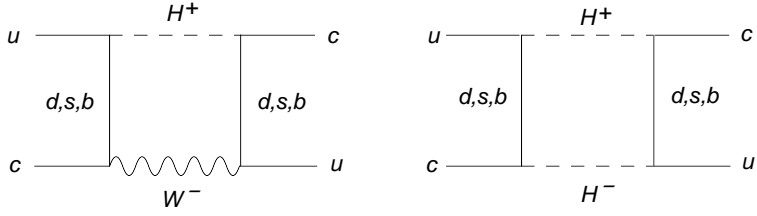}}
\caption{Box diagrams with charged Higgs contributions
to $D$ meson mixing.
\label{chhiggsdiag}}
\end{figure}

A simple extension of the SM is to enlarge the Higgs sector by one
additional $SU(2)$ doublet.  We first examine two-Higgs-doublet models 
that naturally avoid tree-level FCNC be requiring that all fermions of a
given charge receive their masses from only one Higgs doublet~\cite{gw77}. 
In one such model, known in the literature as Model II, one doublet ($\phi_2$)
gives mass to the up-type quarks, while the down-type
quarks and charged leptons receive their mass from the other 
doublet $\phi_1$.  This is the scenario that is present in supersymmetric
theories and is in fact required by supersymmetry in order to generate
masses for all the fermions.  Another model,
known as Model I, imposes a discrete symmetry such that
one doublet ($\phi_2$) generates masses for all 
fermions and the second ($\phi_1$) decouples from the fermion sector
In both cases, each doublet 
receives a vacuum expectation value $v_i$, subject to the constraint that
$v_1^2+v_2^2=v^2_{\rm SM}$. There are five physical scalars in these
models, $h^0\,, H^0\,, A^0$, and $H^\pm$.  The charged Higgs boson
can participate in the box diagram for \dm\ in exchange with the SM $Q=-1/3$
quarks as shown in Fig.~\ref{chhiggsdiag}.  The $H^\pm$ interactions with
the quark sector are governed by the lagrangian
\begin{equation}
{\cal L}={g\over 2\sqrt 2 M_W}H^\pm[V_{ij}m_{u_i}A_u\bar u_i(1-\gamma_5)d_j
+V_{ij}m_{d_j}A_d\bar u_i(1+\gamma_5)d_j]+ h.c. \ \ ,
\end{equation}
with $A_u=\cot\beta$ in both models and $A_d=-\cot\beta(\tan\beta)$ in Model 
I(II), where $\tan\beta\equiv v_2/v_1$.  The $H^\pm$ can have a large
contribution~\cite{bsg} to the rare decay $b\to s\gamma$, and in Model II
the branching fraction for this process sets the bound~\cite{bsg2}
$M_{H^\pm}\gsim 295$ GeV at 95\% C.L..  
This constraint is relaxed when other sources
of New Physics also contribute to $b\to s\gamma$, such as in supersymmetry.
In this case, the lower limit on the charged Higgs mass is 78.6 GeV from
LEP II data~\cite{PDG}.

It is clear that the contributions to \dm\ from the first term in this 
lagrangian (which are proportional to $m_{c,u}\cot\beta$) will only be
sizable for extremely small values of $\tan\beta$; this region is,
however, already excluded~\cite{bhp} from, \eg, $b\to s\gamma$ and 
$B_d^0$-$\overline B_d^0$ mixing.  In addition, since the contributions
in Model I go as $\cot\beta$ multiplied by small mass factors, 
the effects in this case are also restricted to be small.  However,
the term proportional to $m_{b,s}\tan\beta$ in Model II has the
potential to generate a significant contribution to $D$ mixing in the
large $\tan\beta$ limit.  We will thus work in this limit here.
Restrictions on the size of $\tan\beta$ can be obtained by requiring
that the $\bar tbH^+$ coupling remain perturbative.  If we demand that
this coupling not exceed the value of the strong coupling constant,
$g_s$, we find that $\tan\beta\lsim 100$.

The effective hamiltonian governing $D$ meson mixing in the large 
$\tan\beta$ limit takes the form
\begin{equation}
{\cal H}_{2HDM}= {G_F^2M_W^2\over 4\pi^2}\sum_{i,j} \lambda_i\lambda_j
\{\tan^4\beta A_{HH}(x_i,x_j,x_H) + \tan^2\beta A_{WH}(x_i,x_j,x_H)\}
Q_1\ \ ,
\end{equation}
where $\lambda_i=V_{ui}V^*_{ci}$ as usual, $x_i=m_i^2/M_W^2$, 
the sum extends over
$i,j=s,b$, the loop functions can be found in Ref.~\cite{bhp} and
are given in the Appendix.  The operator structure
is the same as in Eq.~(\ref{smham}); here, the $Q_1$ operator
appears due to the presence of the fermion propagator.
Note that this structure is quite different than for the
case of $B_{d,s}$ mixing~\cite{gino} in the large $\tan\beta$ limit.
This is simply due to the helicity structure of the couplings
when the charged $-1/3$ quarks are internal.
The QCD evolution to the charm quark scale is simple and results in
the factor of $r_1(m_c,M_{H^\pm})$.  The resulting contribution to
the mass difference is
\begin{eqnarray}
x_{\rm D}^{\rm (2HDM)} & = & {G_F^2M_W^2\over 6\pi^2\Gamma_D}
f_D^2M_DB_Dr_1(m_c,M_{H^\pm})
\\
& & \hspace{0.3cm} \times \sum_{i,j}\lambda_i\lambda_j
\left[ \tan^4\beta A_{HH}(x_i,x_j,x_H)
+\tan^2\beta A_{WH}(x_i,x_j,x_H)\right] \ \ . \nonumber
\end{eqnarray}

\begin{figure}[htbp]
\centerline{
\includegraphics[width=6cm,angle=90]{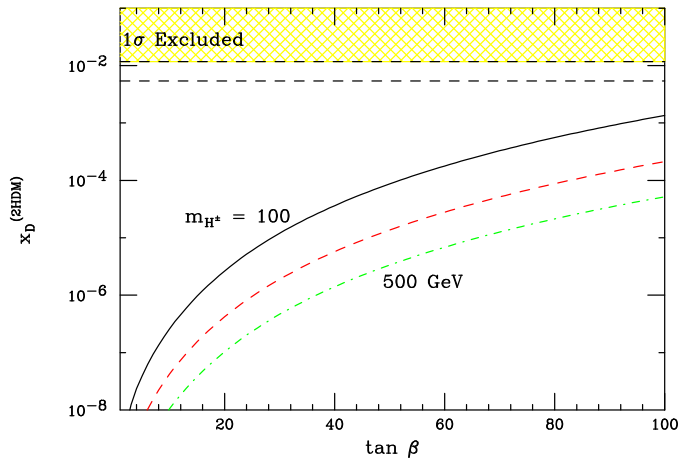}}
\vspace*{0.1cm}
\caption{$x_{\rm D}^{\rm (2HDM)}$ in the flavor conserving two-Higgs-doublet 
model as a function of $\tan\beta$ for charged Higgs boson masses of
$m_{H^\pm}=100\,, 250$, and 500 GeV, corresponding to the solid, dashed
red, and dashed-dot green curves, respectively.
The $1\sigma$ 
experimental bounds are as indicated, with the yellow shaded region 
depicting the region that is excluded.  
\label{2hdmfig}}
\end{figure}

Our results for $x_{\rm D}^{\rm (2HDM)}$ are displayed in Fig.~\ref{2hdmfig}
as a function of $\tan\beta$ for various values of the charged Higgs
mass.  We see that the effect is quite small and even at large values
of $\tan\beta$ the contributions from this model are at least an
order of magnitude below the experimental observation.

\begin{figure}[tbp]
\centerline{
\includegraphics[width=4.3cm,angle=0]{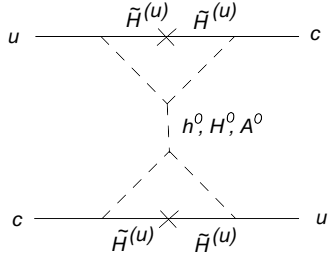}}
\vspace*{0.1cm}
\caption{The dipenguin diagram with neutral Higgs exchange.}
\label{doublepen}
\end{figure}

In the down quark sector, sizable effects in $B_{d,s}$ and $K$ meson mixing 
are obtained in the large $\tan\beta$ limit 
from a double penguin contribution with neutral Higgs exchange~\cite{gino}
as depicted in Fig.~\ref{doublepen}.  
In this limit, an effective
Yukawa interaction is induced for the down type quarks which includes
a contribution from the large Yukawa coupling of the top quark.  This
generates sizable one-loop FCNC couplings for the neutral Higgs 
in the down quark sector.  While the same effects occurs in the up quark
sector, the term that becomes significant at large $\tan\beta$ is now
proportional to the down quark Yukawa coupling and hence does not generate
a sizable effect in $D^0$-$\overline D^0$ mixing

%%%%%%%%%%%%%%%
\subsection{Flavor Changing Neutral Higgs Models}

It is well known that the existence of multiple Higgs doublets can lead
in general to tree level FCNC transitions~\cite{gw77}. In the down quark
sector, there are severe constraints on such couplings from Kaon
decays, but these do not necessarily lead to equally strong restrictions
in the up-quark sector.

The phenomenological 
requirement that FCNC effects in the down-strange sector must be very small
can be met in a variety of ways. For example, the imposition of 
global symmetries 
can make $\Delta S =1 $ FCNC vanish without affecting the $|\Delta C| = 1$ 
sector~\cite{fcnch}.  Another example is the Cheng-Sher 
ansatz~\cite{Cheng:1987rs}, 
where the flavor changing couplings of the neutral Higgs bosons are given by 
$\lambda_{h^0f_if_j}\simeq (\sqrt 2G_F)^{1/2} \sqrt{m_im_j}\Delta_{ij}$, 
with the $m_{i(j)}$ being the relevant fermion
masses and $\Delta_{ij}$ representing a combination of mixing angles.  

To keep our initial discussion general, we allow for $N$ Higgs scalars,
which have the interactions in the up quark sector
\begin{equation}
{\cal L} \sim \lambda^u_{ijn}\overline Q_{Li}u_{Rj}\phi_n\ \ ,
\end{equation}
where $Q_L$ represents the left-handed quark doublet and $u_R$ is the
singlet state.
If $M_{\rm H}$ is the mass of the lightest physical Higgs with 
flavor-changing couplings, 
the most general effective four-fermion hamiltonian just below the 
$M_{\rm H}$ scale is 
\begin{eqnarray}\label{NHiggs}
{\cal H}_{H}= - \frac{1}{2 M_{H^2}} \left[  
2 G_1 ~Q_3  + G_2 ~Q_7 + 
G_3 ~ Q_4  \right] \ ,  
\end{eqnarray}
where the couplings $G_{1,2,3}$ are model-dependent parameters.  As shown
in Ref.~\cite{hw}, 
\begin{eqnarray}
G_1 & = & \sum_{nmN} \lambda^{u*}_{12n}\lambda^u_{21m}A_{nN}A^*_{mN}\,,\\
G_2=G_3 & = & {1\over 2}\sum_{nmN}[\lambda^u_{21n}\lambda^u_{21m}
A_{nN}A_{mN}+\lambda^{u*}_{12n}\lambda^{u*}_{12m}A^*_{nN}A^*_{mN}]
\ ,\nonumber
\end{eqnarray}
at the $M_{\rm H}$ scale.  Here, $A_{nN}$ refers to the mixing matrix
which rotates the Higgs doublets $\Phi_{n}$ to their $N$ neutral physical
eigenstates.
Matching at the Higgs mass scale relates the Wilson coefficients 
to the three couplings $G_{1,2,3}$ via
\bea
C_3(M_H)&=& - 2 G_1 ~, \qquad C_7(M_H)= - G_2~, \qquad 
C_4(M_H) = - G_3\ \ ,
\label{scalar}
\eea
with all other Wilson coefficients being zero. 
Assuming that $M_{H_N} > m_t$ for all $N$ and computing the evolution of 
Eq.~(\ref{NHiggs}) to $\mu = m_c$ we obtain
\begin{eqnarray}\label{NHiggsMu}
{\cal H}_{H} &=& \frac{1}{2 M_{H^2}} \left[  
C_3(m_c) ~ Q_3 + C_4(m_c) ~ Q_4 
+ C_5(m_c) ~ Q_5 + C_7(m_c) ~ Q_7 
+ C_8(m_c) ~ Q_8 \right] \ ,  
\end{eqnarray}
with
\bea
C_3(m_c)&=& r_3(m_c,M_H) C_3(M_H)~,
\nonumber \\
C_4(m_c)&=& \left[
\left(\frac{1}{2}-\frac{8}{\sqrt{241}} \right) r_4(m_c,M_H) +  
\left(\frac{1}{2}+\frac{8}{\sqrt{241}} \right) r_5(m_c,M_H) 
\right]C_4(M_H)\ ,
\nonumber \\
C_5(m_c)&=& \frac{1}{8 \sqrt{241}} \left[
r_4(m_c,M_H) - r_5(m_c,M_H)\right] C_4(M_H)\ , 
\label{cs}
\\
C_7(m_c)&=& \left[
\left(\frac{1}{2}-\frac{8}{\sqrt{241}} \right) r_7(m_c,M_H) +  
\left(\frac{1}{2}+\frac{8}{\sqrt{241}} \right) r_8(m_c,M_H) 
\right] C_7(M_H) \ ,
\nonumber \\
C_8(m_c)&=& \frac{1}{8 \sqrt{241}} \left[
r_7(m_c,M_H) - r_8(m_c,M_H)\right] C_7(M_H)\ \ .
\nonumber 
\eea

\begin{figure}[tbp]
\centerline{
\includegraphics[width=10cm,angle=0]{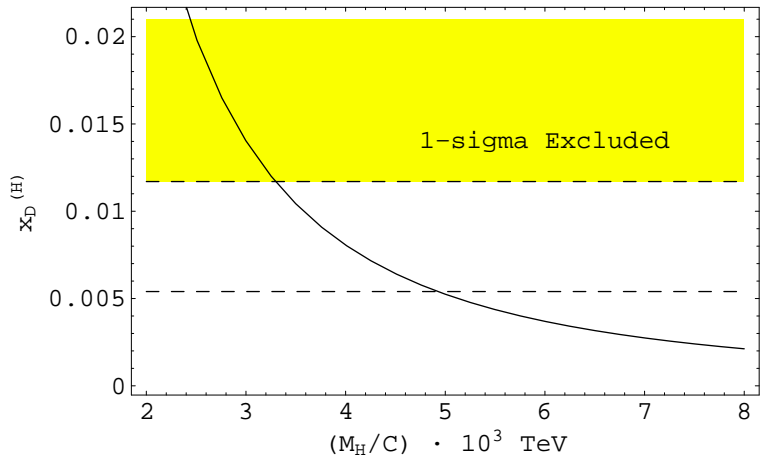}}
\vspace*{0.1cm}
\caption{$x_{\rm D}$ as a function of $M_H/C$ in models with no 
natural flavor conservation in the Higgs sector.  The $1\sigma$ 
experimental bounds are as indicated , with the yellow shaded area 
depicting the region that is excluded.}
\label{NeutralHiggsFig}
\end{figure}

The Higgs tree-level contribution to $x_{\rm D}$ is found by 
evaluating the $D^0$-to-${\bar D}^0$ matrix element, 
%
%\bea
%x_{\rm D}^{\rm (H)} = \frac{5}{24} \frac{f_D^2 M_D}{\Gamma_D M_H^2} 
%\left[
%- C_3(m_c) B_3 + C_4 (m_c) B_4  + C_7 (m_c) B_7
%- \frac{12}{5} \left(C_5(m_c) B_5 + C_8(m_c) B_8 \right)
%\right],
%\eea
%
which gives 
\bea
x_{\rm D}^{\rm (H)} = {5 f_D^2 M_D B_D \over 24 \Gamma_D M_{H}^2} \left[ 
\frac{1 + 6 \eta}{5} ~ C_3 (m_c) -
\eta \left(C_4 (m_c) + C_7 (m_c)\right) 
+ {12 \eta \over 5} \left( C_5(m_c) + C_8 (m_c)\right) \right].
\label{hggans}
\eea
where, again, we have used Eqs.~(\ref{ME_MVS}),(\ref{eta}). 
Together with Eqs.~(\ref{scalar}),(\ref{cs}), the above can be used 
to constrain the lightest Higgs mass and associated couplings.
As an example, let us assume that $|G_1| = |G_2| = C^2$ at the Higgs mass
scale in 
Eq.~(\ref{NHiggs}) and $M_H$ is the effective mass of the N Higgs scalars.
In that case, the restriction on possible values 
of the effective Higgs mass are presented in Fig~\ref{NeutralHiggsFig}.

\begin{figure}[htbp]
\centerline{
\includegraphics[width=6cm,angle=90]{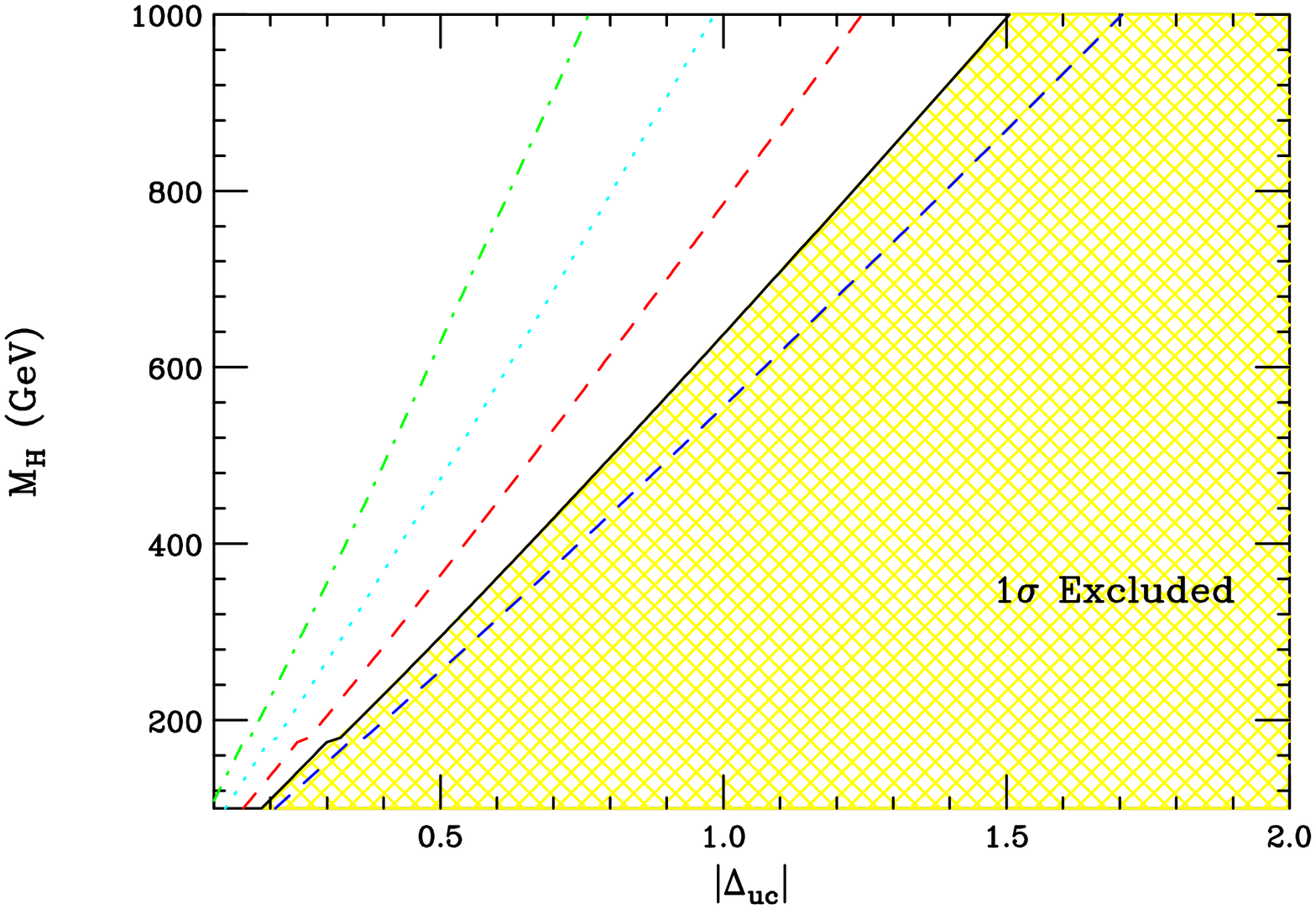}
%\hspace*{5mm}
\includegraphics[width=6cm,angle=90]{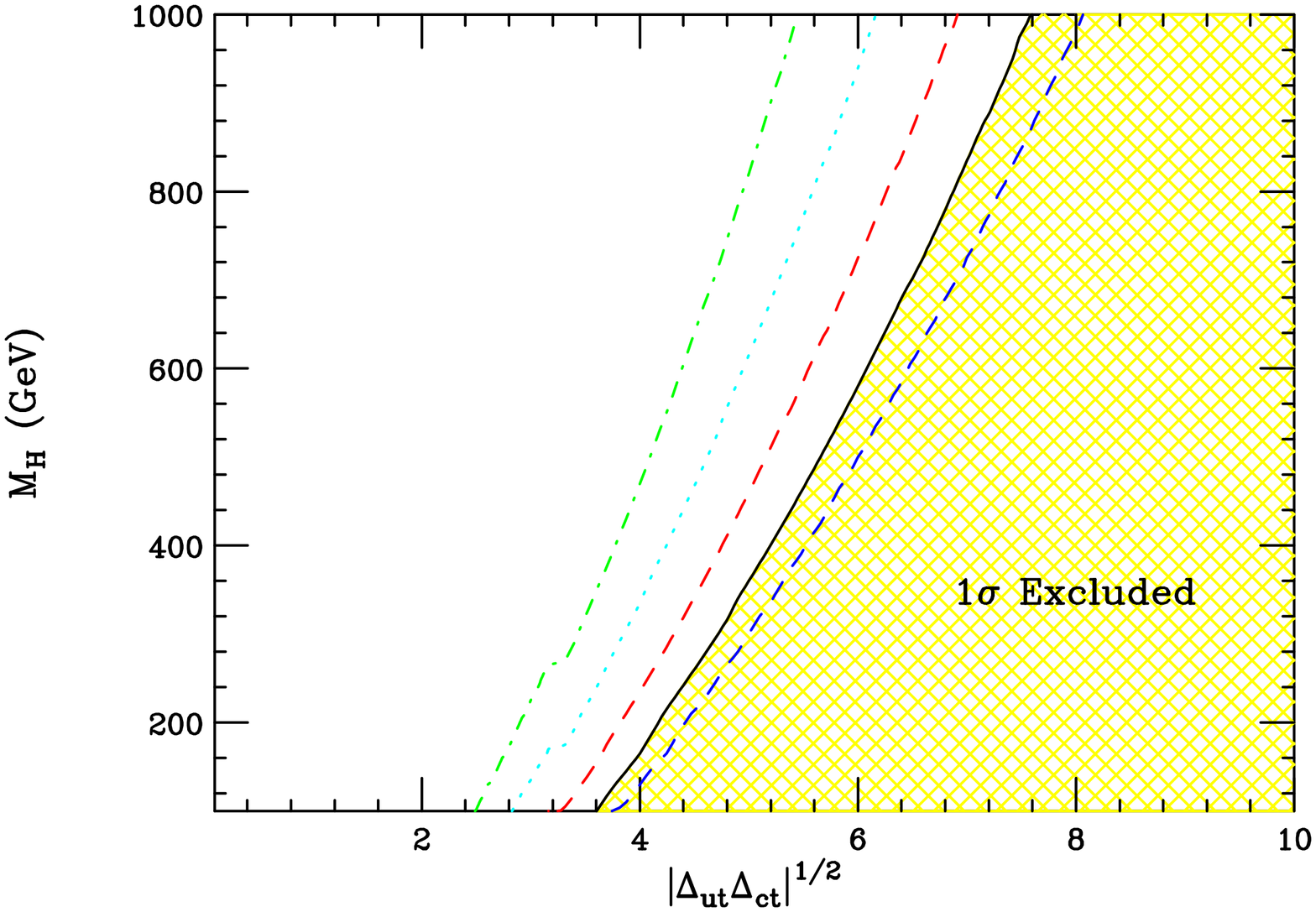}}
\vspace*{0.1cm}
\caption{Left: $1\sigma$ excluded region in the effective neutral
Higgs mass - coupling plane for the tree-level contribution to
$D^0$-$\overline D^0$ mixing in the Cheng-Sher ansatz.
\noindent Right: $1\sigma$ excluded region in the effective neutral
Higgs mass - coupling plane for the box diagram contribution to
$D^0$-$\overline D^0$ mixing in the Cheng-Sher ansatz.
In both figures, possible future contours taking
$x_{\rm D}< (15.0\,, 8.0\,, 5.0\,, 3.0)\times 10^{-3}$,
corresponding to the blue dashed, red dashed, cyan dotted, and 
green dot-dashed curves, respectively, are also displayed.}
\label{chengsher}
\end{figure}

We now return to the specific case of the Cheng-Sher ansatz.  Here, the
neutral Higgs bosons
can contribute to \dm\ through tree-level exchange 
as well as mediating $D$ meson mixing by $H^0$ and t-quark virtual
exchange in a box diagram.  The restrictions placed on the parameter
space of this model from the tree-level contribution 
are computed as described above,
and are presented for an effective Higgs mass $M_H$ as a function of
the coupling parameter $\Delta_{uc}$ in Fig.~\ref{chengsher}.  We see
that the form of the couplings, being proportional to the light quark masses,
result in reduced limits compared to those in Fig.~\ref{NeutralHiggsFig}
for the general case.  The box contribution with $H^0$, t-quark exchange
is described by the effective hamiltonian just below the $M_H$ scale of
\begin{equation}
{\cal H}_{CS}={G_F^2m_um_cm_t^2\Delta^2_{ut}\Delta^2_{ct}\over 8\pi^2
M_H^2}F_{tH}(x) [Q_1+Q_6]\ \ ,
\end{equation}
where $x=m_t^2/M_H^2$, $F_{tH}(x)$ is given in the Appendix, 
and the vector operators $Q_{1.6}$ are generated from the
fermion propagators.  The RG evolution and evaluation of the matrix
elements yields
\begin{equation}
x_{\rm D}^{\rm (CS)}={G_F^2m_um_cm_t^2\Delta^2_{ut}\Delta^2_{ct}\over
6\pi^2M_H^2\Gamma_D}f_D^2M_DB_D F_{tH}(x) r_1(m_c,M_H)\ \ .
\end{equation}
The resulting constraints from this contribution are displayed in 
Fig.~\ref{chengsher} in the effective Higgs mass - coupling parameter
plane.  We see that this box contribution only competes with those 
from the tree-level process for large values of $\Delta_{ij}$.  

%

%%%%%%%%%%%%%%%

%%%%%%%%%%%%%%%
\subsection{Scalar Leptoquark Bosons}\label{ScalarLQSect}

Leptoquarks are color triplet particles which couple to a lepton-quark pair
and are naturally present in many theories beyond the SM which relate
leptons and quarks at a more fundamental level.  Their
{\it a priori} unknown couplings can be parameterized 
as $\lambda^2_{\ell q}/4\pi=F_{\ell q}\alpha$.  Searches for the pair 
production of scalar leptoquarks at the Tevatron Run II yield the 
bounds~\cite{cdflq} $m_{LQ}\gsim 225$ GeV,
which are independent of the coupling strength $F_{\ell q}$

Scalar leptoquarks participate 
in $D$ meson mixing via virtual exchange inside a box 
diagram~\cite{Davidson:1993qk}, together with a charged lepton or neutrino.
Their interactions are analogous to those of R-parity violating
supersymmetric models with the terms in the superpotential proportional to 
$\lambda'$.  We thus refer to Section~\ref{SUSYRsection} 
for the details of the analysis 
for these contributions.  The resulting constraints on scalar leptoquark
are governed by the translation
\begin{equation}
F_{\ell u}F_{\ell c} = {\lambda'_{R_p(i2k)}\lambda'_{R_p(i2k)}
\over 4\pi\alpha} \qquad {\rm with} \qquad m_{\tilde d} =
m_{LQ}\ \ .
\end{equation} 

  %%%%%%%%%%%%%%%
\subsection{Higgsless Models}

A class of composite Higgs models which has been developed recently to
generate a naturally light Higgs, employs chiral symmetries of
'theory space'~\cite{Chivukula:2002ww} (see also 
Refs.~\cite{morelh,Arkani-Hamed:2002pa}). 
Such models involve the `deconstruction' of higher-dimensional 
field theories such that the low-energy effective field theory
resembles the Standard Model but has nice features such as the absence 
of quadratic divergences of the Higgs mass.  Here, the Higgs 
can be interpreted as a Goldstone boson of some interaction occuring
at higher energies.  This approach allows for the construction of 
realistic theories of 
electroweak symmetry breaking in four spacetime dimensions without 
any higher dimensional interpretation. 

This picture emerges from the AdS/CFT correspondence of the 
5-dimensional Higgsless model of Csaki \etal~\cite{csaki}.  In the
5-d framework, there is no physical Higgs boson and electroweak
symmetry breaking is generated via the boundary conditions for the
$5^{\rm th}$ dimension.  The gauge symmetry in the higher dimensional
space is $SU(2)_L\times SU(2)_R\times U(1)_{B-L}$ and the right-handed
gauge fields receive Planck scale masses.  The Kaluza-Klein towers
of the $\gamma$ and $Z$ bosons unitarize the $WW$ high energy
scattering amplitude~\cite{hitoshi}, although there is some
tension with precision electroweak data as to the precise energy
scale that the Kaluza-Klein states populate~\cite{ewhiggsless}.
The Standard Model fermion fields are localized within the
$5^{\rm th}$ dimension and also receive their masses from the
boundary conditions, with the exact value being dependent on their
position in the extra dimension~\cite{csakiferm}.  
The effects on $D$ mixing from 
this 5-dimensional picture are presented below in Section~\ref{warped}.
Here, we present our results for the AdS/CFT related framework
with a composite Higgs.

The key idea in the composite Higgs picture is that
the flavor physics responsible for generation of the Yukawa 
couplings can induce flavor-changing neutral currents~\cite{Chivukula:2002ww}. 
Applied to charm physics they generically lead to the following effective 
hamiltonian,
\begin{eqnarray}\label{Higgsless1}
{\cal H}_{\not{H}} \ &=& \ \sum_{{\cal C}=1,T^a} \bigl[
\left(c^c_L s^c_L \right)^2
\frac{g^2}{M^2}
(\overline{u}_L \gamma_\mu ~{\cal C}~ c_L) 
\ (\overline{u}_L \gamma^\mu ~{\cal C}~ c_L) 
\nonumber \\
& & \hspace{0.7cm}+ 2 \left(c^c_L s^c_L \right) \left(c^c_R s^c_R \right)
\frac{g^2}{M^2} \
(\overline{u}_L \gamma_\mu ~{\cal C} ~c_L) 
\ (\overline{u}_R \gamma^\mu ~{\cal C}~ c_R)
\\
& & \hspace{0.7cm} + \left(c^c_R s^c_R \right)^2
\frac{g^2}{M^2} \
(\overline{u}_R \gamma_\mu ~{\cal C}~ c_R) 
\ (\overline{u}_R \gamma^\mu ~{\cal C}~ c_R)
\bigr] \ \ ,
\nonumber 
\end{eqnarray}
where $g, M$ are respectively the gauge coupling and gauge boson 
mass of new flavor gauge interactions, and the mixing angles generate
different strengths for the gauge coupling.  In the sum over the 
color label ${\cal C}$, the case ${\cal C}= 1$ corresponds to 
color-singlet interactions, whereas 
${\cal C} = T^a$ refers to color-octet interactions for which 
$T^a \equiv \lambda^a/2$ are the generators of $SU(3)_C$.  
In addition, the angles $\theta_{L,R}^c$ that relate the gauge 
and mass eigenstates~\cite{Chivukula:2002ww} appear 
in factors of $c^c_{L,R}\equiv\cos\theta_{L,R}^c$,
$s^c_{L,R}\equiv\sin\theta_{L,R}^c$, where we take 
$\theta_{L,R}^c \sim \theta_C$, with $\theta_C$ being the Cabibbo angle.

\begin{figure}[tbp]
\centerline{
\includegraphics[width=10cm,angle=0]{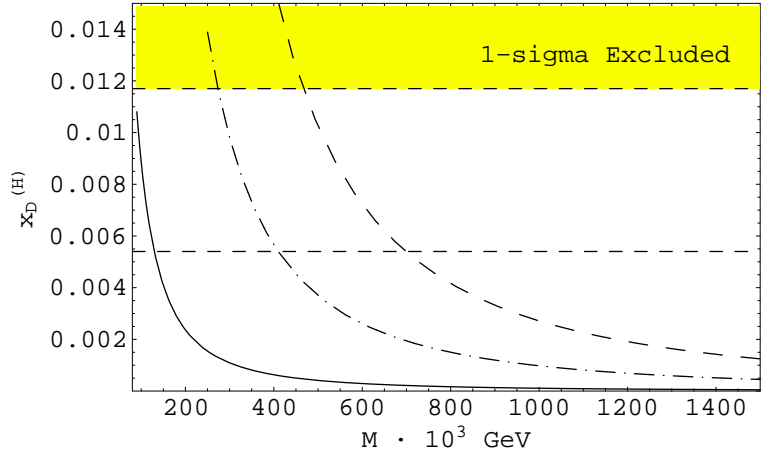}}
\vspace*{0.1cm}
\caption{$x_{\rm D}$ as a function of the new gauge boson 
mass $M$ in Higgless models for
$g=0.1$ (solid line), $g=0.3$ (dash-dot line), and $g=0.5$ (dashed line).
The $1\sigma$ experimental bounds are as indicated, with the yellow shaded
area depicting the region that is excluded.}
\label{HiggslessFig}
\end{figure}

The hamiltonian of Eq.~(\ref{Higgsless1}) can be easily transformed to contain 
the operators from the general basis of Eq.~(\ref{SetOfOperators}),
\begin{eqnarray}\label{Higgsless2}
{\cal H}_{\not{H}} \ &=& \ \left(c^c_L s^c_L\right)^2 \frac{g^2}{M^2}\cdot
\left(C_1^{\not H} Q_1 + C_2^{\not H} Q_2 +  
C_3^{\not H} Q_3 + C_6^{\not H} Q_6 \right)\ \ ,
\end{eqnarray}
where $C_1^{\not H}=(3 N_c-1)/(2 N_c)$, $C_2^{\not H}
=r_{LR} (2 N_c-1)/(2 N_c)$, 
$C_3^{\not H}=-r_{LR}$, $C_6^{\not H}= r_{LR}^2 C_1^{\not H}$, and
$r_{LR}=(c^c_R s^c_R)/(c^c_L s^c_L)$. Performing the RG running, 
we obtain the effective hamiltonian at the scale $m_c$ with 
the Wilson coefficients 
\begin{eqnarray}
& & {\rm C}_1(m_c) = r_1(m_c,M) {\rm C}_1(M)\ , 
\nonumber \\
& & {\rm C}_2(m_c)= r_2(m_c,M) {\rm C}_2(M)  \ , 
\nonumber \\
& & {\rm C}_3(m_c)= \frac{2}{3} \left[
r_2(m_c,M) - r_3(m_c,M)\right] {\rm C}_2(M)+ r_3(m_c,M) C_3(M)\ , 
\nonumber \\
& & {\rm C}_6(m_c)= r_6(m_c,M) {\rm C}_6(M) \ \ .
\label{Higgsless3}
\end{eqnarray}
This, in turn, implies for the mixing amplitude,
%
%\begin{eqnarray}\label{Higgsless4}
%x^{(\not H)}_{\rm D}=\frac{1}{3} \frac{f_D^2 M_D}{\Gamma_D} 
%\left(c^c_L s^c_L\right)^2
%\frac{g^2}{M^2} \left[2 C_1 (m_c) B_1 - \frac{1}{2} C_2 (m_c) B_2 -
%\frac{5}{4} C_3 (m_c) B_3 + 2 C_6 (m_c) B_6 \right] \ .
%\end{eqnarray}
%
%Since most of the B-parameters in Eq.~(\ref{Higgsless4}) are not known,
%we once again employ the modified vacuum saturation approximation. 
%In this case,
%
\begin{eqnarray}
& & x^{(\not H)}_{\rm D} = {f_D^2 M_D B_D \over \Gamma_D} 
\left(c^c_L s^c_L\right)^2 \frac{g^2}{M^2} \left[ 
{2 \over 3} \left( C_1 (m_c) + C_6 (m_c) \right) + 
- C_2 (m_c) \left( {1\over 2} + {\eta \over 3} \right) \right. \nonumber \\
& & \left. \hspace{3.0cm} + 
\frac{1}{12} C_3 (m_c) \left( 1 + 6 \eta \right) \right] \ \ ,
\label{Higgsless5}
\end{eqnarray}
where $\eta$ is as in Eq.~(\ref{eta}).  
The available experimental data can be used to constrain 
the mass of the gauge boson $M$
for different values of the coupling constant $g$, as shown 
in Fig.~\ref{HiggslessFig}. It is 
clear that unless $g$ is very small, it is unlikely 
that any of the gauge bosons
of the new flavor interactions of Higgsless models will be 
directly seen at the LHC. 
%
%%%%%%%%%%%%
%%%%%%%%%%%%%%%%%%%%%%%%%%%%%%%%%%%%%%%%%%%%%%%%%%%%%%%%%%%%%%%%%%%%%
\section{Extra Space Dimensions}

Recent speculation that the geometry of spacetime could resolve
the hierarchy problem have led to theories with extra spatial 
dimensions that have verifiable consequences at the TeV scale.  
There are several such models~\cite{jlhmaria} and the size and 
geometry of the additional spatial dimensions, 
as well as the field content that is allowed to propagate within 
them, varies between the different scenarios.  When the extra
dimensions are compactified, the fields that
reside in the higher dimensional space (known as the bulk) expand 
into Kaluza-Klein (KK) towers of states.
The masses of these KK states correspond to the extra dimensional
components of the bulk field momentum and are related to the bulk
geometry.  The observation, either directly or by indirect effects,
of these KK states signals the existence of extra dimensions.  The
properties of the KK states reveal the underlying geometry of the
higher dimensional spacetime.

The extra dimensional theories which yield contributions to 
$\Delta F = 2$ processes are those in which the SM fermion fields 
reside in the bulk.  Here, we consider three such scenarios:  
(i) universal extra dimensions, (ii) localized fermions in a flat 
extra dimension, and (iii) warped extra dimensions.

%%%%%%%%%%%%%%%
\subsection{Universal Extra Dimensions}

The possibility of TeV$^{-1}$-sized extra dimensions naturally
arises in braneworld theories~\cite{ant}.  By themselves, they
do not allow for a reformulation of the hierarchy problem, but
they may be incorporated into a larger structure in which this
problem is solved.  The scenario which places all Standard Model
fields in the bulk is known as Universal Extra Dimensions~\cite{ued}.
The simplest model of this type contains a single extra dimension 
compactified on an $S_1/Z_2$ orbifold.  Since branes 
are not present in this case, translational invariance in the 
higher dimensional space would be  preserved without the presence of
the orbifolding.  This leads to the tree-level conservation of the
extra dimensional momentum of the bulk fields, which in turn implies 
that KK number is conserved at tree-level while KK parity, $(-1)^n$
where $n$ denotes the KK level, 
is conserved to all orders in interactions involving the KK 
states.  Two immediate consequences of KK number 
and parity conservation are that the KK states must be produced
in pairs, and the lightest
KK particle is stable and is a dark matter candidate~\cite{timt}.
The former results
in a substantial reduction of the sensitivity to such states in 
precision electroweak and collider data.  The present bound 
from Run I at the Tevatron on the mass of the first KK excitation 
is of order 250 GeV~\cite{cdfued}.

Since all SM fields reside in the bulk in this model, every SM field
expands into a KK tower of states.  The KK reduction
of the 5-dimensional fermion fields leaves a chiral zero-mode and a
vector-like tower of KK states for each flavor.  There is one KK
tower for each SM gauge boson, as well as the Higgs, and 
additional towers of KK scalars, 
$a^0_{(n)}$ and $a^\pm_{(n)}$, which correspond to the physical
eigenstates of the mixing between the KK towers associated with the
SM Goldstone fields and 
the $W_5\,, Z_5$ remnants from the electroweak gauge KK reduction.  
This mixing also generates scalar KK towers which behave as Goldstone
fields, which are eaten by 
the gauge boson KK towers and provide masses for the gauge KK states.
The additional physical scalar KK towers $a^{0,\pm}_{(n)}$ do not 
have zero-modes.  The masses of the 
KK states are roughly degenerate and are given at tree-level by
\begin{equation}
m_{n} = (m_0^2+n^2/R_c^2)^{1/2}\ \ ,
\label{kkmass}
\end{equation}
where $R_c$ represents the compactification radius of the extra
dimension and $m_0$ is the
zero-mode mass.  The KK states clearly become more degenerate with increasing
KK-level (increasing $n$).  These masses are modified~\cite{Cheng:2002iz}
by loop-induced localized kinetic terms and non-local radiative
corrections.  Given that the effect of these mass corrections 
occurs at two-loop order in $D$ meson mixing, one would expect them to have 
a small effect.  We thus neglect them in our initial analysis, 
but will return to this issue at the end of this Section.

The contributions to D mixing in this model are box diagrams with
the $W^\pm$ boson KK tower, its associated KK Goldstone modes 
$G^\pm_{(n)}$, 
and the $a^\pm_{(n)}$, all in exchange with the KK towers associated with
the $d$-, $s$-, and $b$-quarks; the zero-mode analogues of these diagrams 
are shown in Fig.~\ref{chhiggsdiag}. Note that the conservation of KK parity 
restricts the KK levels of the 
KK quark and boson being exchanged.  In addition, only the quark
KK towers that are even under the $Z_2$ symmetry (and thus have
a zero-mode) couple to the external zero-mode quarks in the box
diagram.   
The relevant hamiltonian at the compactification scale is then
\begin{equation}
{\cal H}_{UED} = {G_F^2M_W^2\over 4\pi^2}\sum_{\vec n=1}^\infty
\sum_{i,j} ~ \lambda_i\lambda_j S(x_i^{(n)},x_j^{(n)}) Q_1\ \ ,  
\end{equation}
which has the same structure as that occurring in Eq.~(\ref{smham}).  
Here, $i,j$ run over $d,s,b$.  Note that at all KK levels, the 
CKM structure is the same as that in the SM.  Using unitarity of the
CKM matrix, the function $S(x_i^{(n)},x_j^{(n)})$ becomes
\begin{equation}
S(x_i^{(n)},x_j^{(n)})=\sum_{XY}\left(F_{XY}(x_i^{(n)},x_j^{(n)})
+F_{XY}(x_d^{(n)},x_d^{(n)}) - F_{XY}(x_i^{(n)},x_d^{(n)})
-F_{XY}(x_j^{(n)},x_d^{(n)})\right)\ , 
\end{equation}
with $x_i^{(n)}=(m_n^i)^2/(m_n^W)^2$ where now $i,j=s,b$, and the sum 
extends over the bosons 
$X,Y=W^\pm_{(n)},G^\pm_{(n)},a^\pm_{(n)}$.  The functions
$F_{XY}(x_i^{(n)},x_j^{(n)})$ are given in the Appendix of
Ref.~\cite{Buras:2002ej}, with the appropriate substitutions of
quark flavors relevant for $D$ mixing.  After the RG evolution of the
hamiltonian to the charm-quark scale, this leads to
\begin{equation}
x_{\rm D}^{(UED)}={G_F^2M_W^2\over 6\pi^2\Gamma_D}f_D^2M_DB_D
r_1(m_c,m_1)
\sum_{\vec n=1}^\infty
\sum_{i,j} ~ \lambda_i\lambda_j S(x_i^{(n)},x_j^{(n)})\ \ .
\end{equation}

Looking at the expression in Eq.~(\ref{kkmass}) 
for the KK masses, we see that the $d$- and $s$-quark KK towers are
degenerate and the mass splittings between the $b-$ and $d-,s$-quark
towers are non-zero, yet small, for the first couple of 
KK levels and then effectively
vanish for higher KK excitations.  The GIM cancellation is thus exact 
in the case of the $s$-quark KK tower contributions, level by level
in the KK tower, and leaves a
tiny contribution from the first few $b$-quark KK states.  However, factoring
in that $\lambda_b\sim {\cal O}(10^{-4})$, we see that
the contributions to $D$ mixing from the $b$-quark KK states 
are numerically negligible.  Hence, this
model is not probed by $D^0$-$\bar D^0$ mixing.

We now return to the case where mass splittings are generated for
the KK states via localized boundary terms or loop-induced gauge
interactions.  Since the above one-loop contributions to $D$ mixing
essentially vanish due to the degeneracy of the KK towers, perhaps
a non-negligible effect is obtained once the KK degeneracy is lifted.
Examining the latter effect first, we see from Ref.~\cite{Cheng:2002iz}
that the non-local radiative corrections yield two classes of mass
splittings for the fermion fields: ($i$) a term which is dependent on
the gauge couplings and is flavor independent, and ($ii$) a term
which depends on the fermion's Yukawa couplings.  The latter term
takes the form
\begin{equation}
\delta m^f_n = m^f_n~\left( {-3 h_f^2\over 16\pi^2X}\ln{\Lambda^2
\over\mu^2}\right)\ \ ,
\end{equation}
where $X=(2,4)$ for fermion (singlets, doublets), respectively, $h_f$ is
the fermion Yukawa coupling, $\Lambda$ represents a cut-off scale
which absorbs the logarithmic divergences and $\mu$ is the
renormalization scale.  If $1/R_c$ is of order a few hundred GeV, the third 
generation quark doublet and top quark singlet 
thus receives a correction from the Yukawa term 
of order $10$-$20$~GeV 
for the first KK state, while the $b$-quark singlet KK excitation
remains essentially unaffected.  Given the small CKM factor for the
$b$-quark KK contributions to $D$ meson mixing, and the effectiveness
of the GIM mechanism, we find that this mass splitting
is not enough to generate a sizable contribution to $x_{\rm D}$.  The
second possibility of including the localized boundary terms holds
the promise of inducing large mass splittings between the KK
states associated with the various quark flavors.  However, these
boundary terms may take on essentially any value with no predictivity, 
leaving a virtual continuum of possible contributions to
$D^0$-$\bar D^0$ mixing.

%%%%%%%%%%%%%%%
\subsection{Split Fermion Models}

\begin{figure}[tbp]
\centerline{
\includegraphics[width=3.4cm,angle=0]{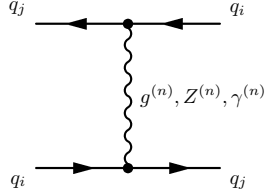}}
\vspace*{0.1cm}
\caption{Tree-level gauge KK exchange that
mediates neutral meson oscillations.}
\label{treegraph}
\end{figure}

In this scenario, the Standard Model fermions are localized at specific 
points, $y_i$, where $0\le y_i\le R_c$,
in extra TeV$^{-1}$-sized flat dimensions.  The fermions 
have narrow Gaussian-like wavefunctions in the extra dimensions with the
width of their wavefunction being much smaller than the compactification 
radius $R_c$ of the additional dimensions.
The placement of the different fermions at distinct locations in the
additional dimensions, along with the narrowness of their wavefunctions,
can then naturally suppress~\cite{nimamartin} operators mediating dangerous
processes such as proton decay and also provide a mechanism for generating
the fermion mass hierarchy~\cite{splitf}.

This split fermion scenario is capable of generating large flavor 
changing neutral currents~\cite{splitfcnc,ben}.  
In contrast to the fermion sector, the
gauge bosons are free to propagate throughout the extra dimensions.
The gauge KK states have cosine profiles which have different heights
at the various distinct fermion locations, generating non-universal
couplings to different fermion species.  This leads to tree-level
FCNC as depicted in Fig.~\ref{treegraph}, with the gluon KK states
clearly giving the largest contributions.

With one extra dimension, the coupling of the $n^{th}$ KK gluon to
a quark localized at the scaled position $y_q$ is determined by 
the overlap of wavefunctions in the additional dimension
\begin{equation}
\int_0^1 dy\, \bar\psi(y)\psi(y)G^{(n)}(y) \sim
\int_0^1 dy\, \cos(n\pi y)e^{-(y-y_q)^2R_c^2/\sigma^2} \sim
\cos\left({n\pi y_q\over R_c}\right)e^{-n^2\sigma^2/R_c^2} \ ,
\label{splitoverlap}
\end{equation}
where $\sigma$ represents the width of the quark's localized
wavefunction with $\sigma/R_c\ll 1$, and $y_q$ has been normalized
to $R_c$ so that it lies in the range $0\le y_q \le 1$.  
The interaction lagrangian in the quark mass eigenstate
basis is then
\begin{equation}
{\cal L} = \sum_{n=1}^\infty\left[ \sqrt 2 g_s G_\mu^{A(n)}
\left( {\bf \bar u}_L\gamma^\mu T^A V_L^u C_L^{(n)}
V_L^{u\dagger}{\bf u}_L
+{\bf \bar u}_R\gamma^\mu T^A V_R^u C_R^{(n)}
V_R^{u\dagger}{\bf u}_R + 
({\bf u}\rightarrow {\bf d} \right)\right]\ ,
\label{splitL}
\end{equation}
where the product $V_L^{u\dagger}V_L^d$ is the usual CKM matrix,
the diagonal matrices $C_{L,R}^{(n)}$ are the wavefunction 
overlaps given above in Eq.~(\ref{splitoverlap}), 
and the factor of $\sqrt 2$ arises from the
renormalization of the KK gauge kinetic terms to the canonical
value.  ${\bf u_i}$ refers to the set $(u_i,c_i.t_i)$.

The effective hamiltonian mediating D meson mixing is given by
(taking the contributions from the first two generations to be
dominant)
\begin{equation}
{\cal H}_{split} = {2\over 3} g_s^2\sum_{\vec n=1}^\infty
{1\over M_n^2} \left( U_{L(cu)}^{u(n)\dagger}U_{L(uc)}^{u(n)}Q_1
+ 2U_{L(cu)}^{u(n)\dagger}U_{R(uc)}^{u(n)}Q_2
+ U_{R(cu)}^{u(n)\dagger}U_{R(uc)}^{u(n)}Q_6 \right) \ ,
\end{equation}
where $U_i^{u(n)}\equiv V_i^{u\dagger}C_i^{(n)}V_i^u$ with
$i=L,R$ and $M_n$ is the mass of the $n^{th}$ gluon KK state
with $M_n=n/R_c$.  For the case of one additional dimension, 
the sum over the gluon KK tower converges, and for the scenario
with numbers of extra dimensions $>1$, the sum is naturally
cut-off from the finite width of the fermion wavefunction.  
Performing this sum~\cite{ben} and making use of the unitarity
properties of the $V^q_{L,R}$, we can write the effective
hamiltonian at the compactification scale as
\begin{eqnarray}
{\cal H}_{split} & = & {2\over 3}g_s^2R_c^2\left(
|V^u_{L\, 11}V^{u*}_{L\, 12}|^2F(y_{u_L},y_{c_L})Q_1
+2|V^u_{L\, 11}V^{u*}_{L\, 12}V^u_{R\, 11}V^{u*}_{R\, 12}|
G(y_{u_L},y_{c_L},y_{u_R},y_{c_R})Q_2\right.
\nonumber\\
& & \hspace{1.7cm} \left. + |V^u_{R\, 11}V^{u*}_{R\, 12}|^2
F(y_{u_R},y_{c_R})Q_6\right)\ \ ,
\end{eqnarray}
with $y_{u_i,c_i}$ being the positions of the up- and charm-quark
fields, and
\begin{eqnarray}
F(x,y) & = & {\pi^2\over 2}|x-y| \ ,
\nonumber\\
G(x_1,y_1,x_2,y_2) & = & {-\pi^2\over 4}\left(|x_1-x_2|+
|y_1-y_2|-|x_1-y_2|-|x_2-y_1|\right)\ .
\end{eqnarray}
Although some cancellations could occur between the $Q_{1,6}$
and $Q_2$ terms by finely-tuning the quark positions,
and thus decreasing the KK gluon contribution to
D meson mixing, operator mixing from QCD renormalization
would spoil this possibility. 
The RG running of the above effective hamiltonian is the same as
that performed for the case of flavor changing $Z'$ bosons in
Section~\ref{ZprimeSect}, with the appropriate replacement of the Wilson
coefficients, since the same operator basis of $Q_{1,2,6}$ is
present.

In order to explore the magnitude of the KK gluon FCNC effects 
we examine a single term in the above hamiltonian.  This will reduce
the number of parameters in the computation without significantly
changing the results.  Choosing the term proportional to $Q_1$ yields
an effective hamiltonian at the charm scale of
\begin{equation}
{\cal H}_{split} = {g_s^2R_c^2\pi^2\Delta y\over 3}~ r_1(m_c,M) 
|V^u_{L\, 11}V^{u*}_{L\, 12}|^2 Q_1\ \ .
\end{equation}
Here, $\Delta y\equiv |y_{u_L}-y_{c_L}|$ 
is the separation between the localized $u_L$
and $c_L$ quarks, scaled to the compactification radius. This leads 
to a contribution to $x_{\rm D}$ of
\begin{equation}
x^{(split)}_{\rm D} = {2\over 9 \Gamma_D}g_s^2R_c^2\pi^2\Delta y 
~r_1(m_c,M) |V^u_{L\, 11}V^{u*}_{L\, 12}|^2 f_D^2 M_D B_1\ \ .
\end{equation}

\begin{figure}[tbp]
\centerline{
\includegraphics[width=6cm,angle=90]{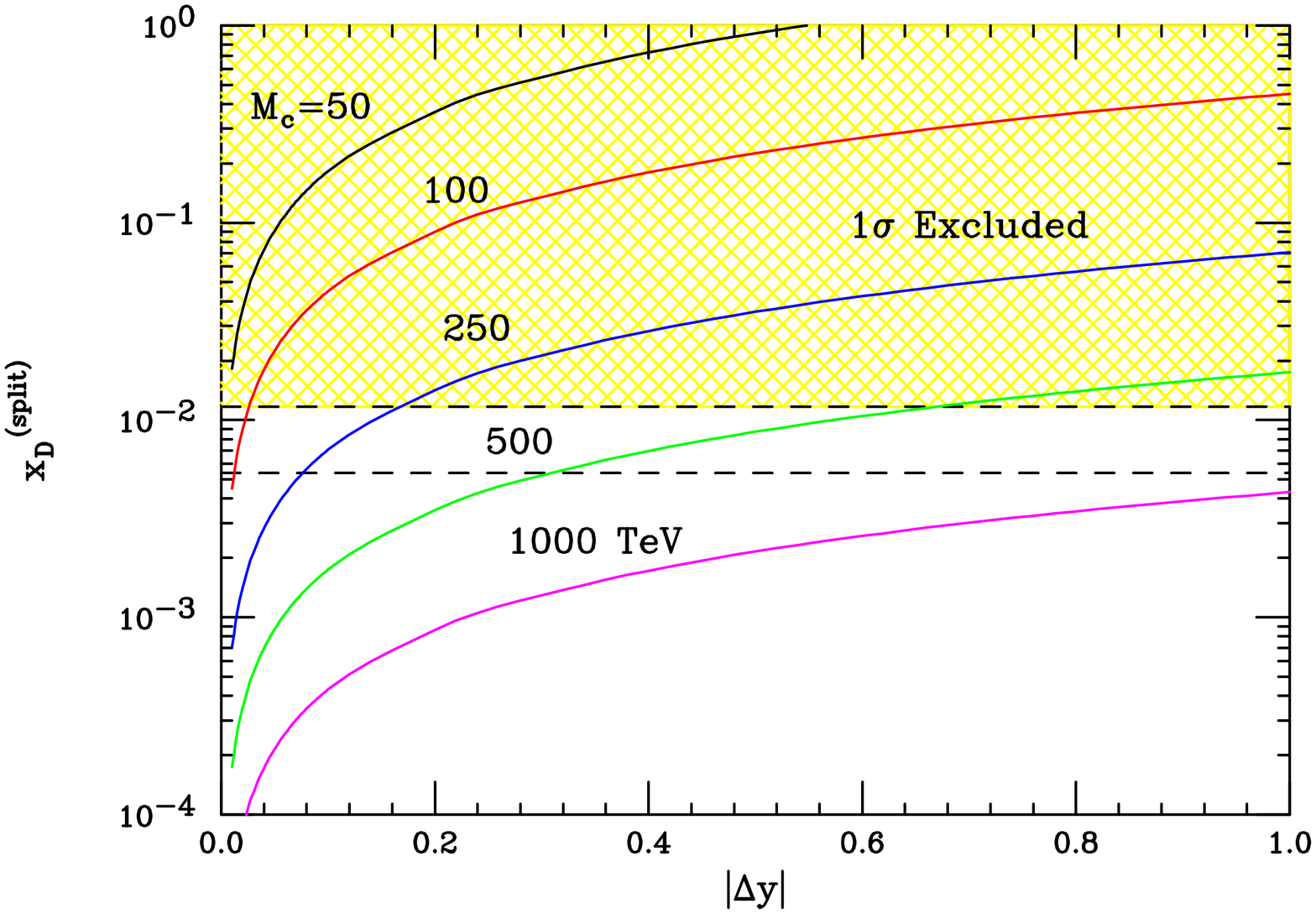}
%\hspace*{5mm}
\includegraphics[width=6cm,angle=90]{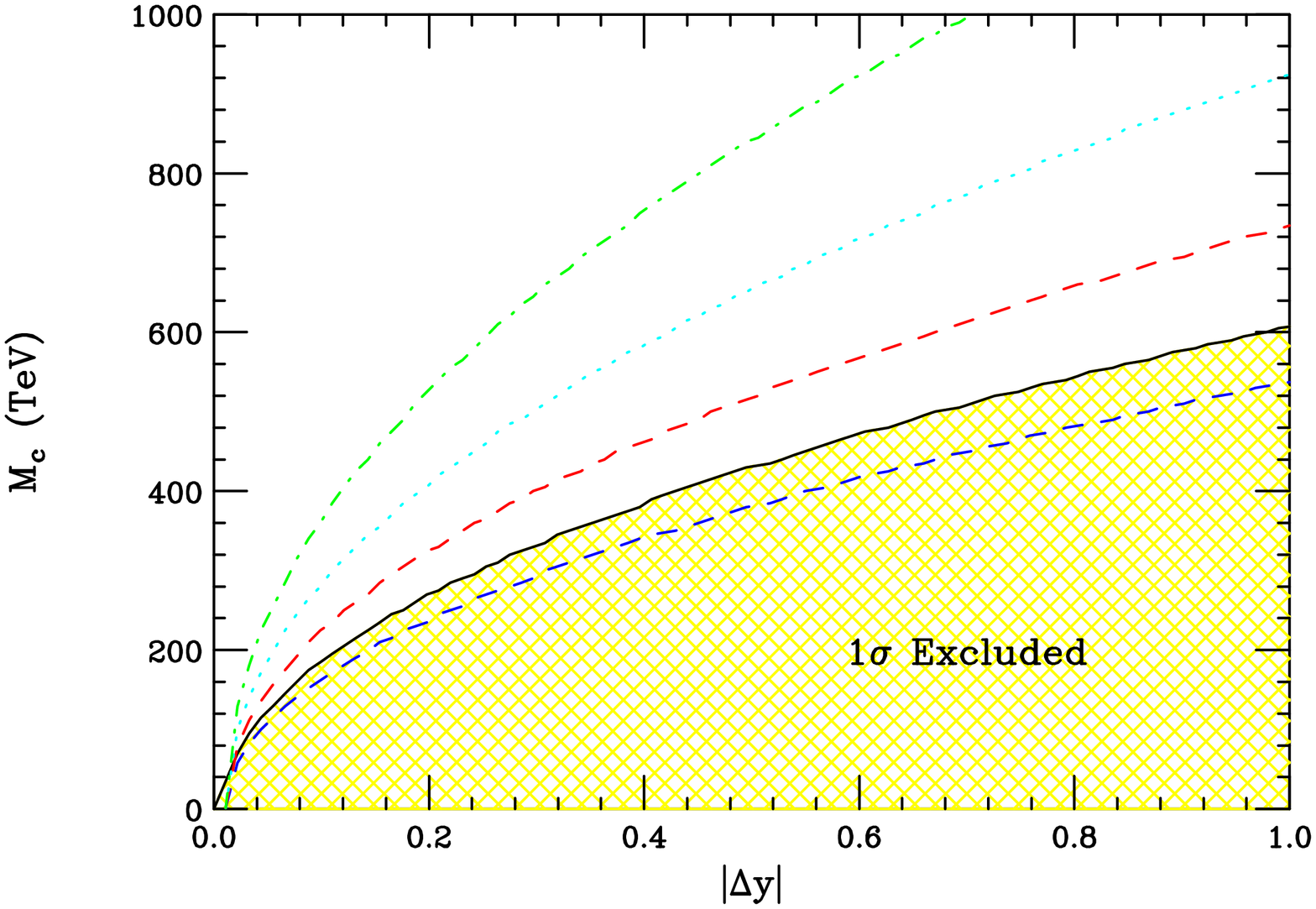}}
\vspace*{0.1cm}
\caption{Left: $x_{\rm D}$ in the Split Fermion model as a function of the
separation between the left-handed $u$- and $c$-quark states in the
extra dimension for various values of the compactification scale.
The $1\sigma$ experimental
bounds are as indicated, with the yellow shaded region depicting the
region that is excluded. \\  
Right: $1\sigma$ excluded region in the $u_L-c_L$ separation and
compactification 
scale parameter plane, as well as possible future contours taking
$x_{\rm D}< (15.0\,, 8.0\,, 5.0\,, 3.0)\times 10^{-3}$,
corresponding to the blue dashed, red dashed, cyan dotted, and 
green dot-dashed curves, respectively.}
\label{splitfig}
\end{figure}

Figure~\ref{splitfig} shows the range of values for $x^{(split)}_{\rm D}$
as a function of the separation between the $u_L$ and $c_L$ states
for various values of the compactification scale, where
$M_c=1/R_c$.  In our numerical work we have used the natural
assumption that $(V_L)_{ij}=
(V_{CKM})_{ij}$.  We see that
$x_{\rm D}^{(split)}$ vanishes as the separation of the 2 fermions 
tends to zero as
expected.  However, for most of the range of $\Delta y$, we find that
compactification scales of order $100-500$ TeV are excluded by the 
observation of $D^0$-$\overline D^0$ mixing and hence $D$ mixing provides
severe constraints on the localization of the up-type fermions
within this model.  Note that these constraints are dependent on the
choice of values for the elements of the quark diagonalization
matrices $V_{L,R}$, which are {\it a priori} unknown, and could be
reduced if quark mixing is tiny in the up-quark sector.  The worst case
scenario would be if the $V^u_{L,R}$ are diagonal and all quark mixing
occurs in the down-quark sector.\footnote{This is frequently the case 
in models of quark mass matrices where the up-quark mass matrix is 
taken to be diagonal and all mixing is assigned to the down-quark 
sector.  A rationale for this is given, for example, in Ref.~\cite{q7}.}
In this case, strong bounds on the compactification scale, similar
to those presented here, would be obtained from $K$ meson 
mixing~\cite{splitfcnc,ben}.

%%%%%%%%%%%%%%%
\subsection{Warped Geometries}\label{warped}

In the simplest scenario with warped extra dimensions~\cite{rs}, 
known as the Randall-Sundrum (RS) model,
the hierarchy between the electroweak and Planck scales is generated
geometrically via a large curvature of a single extra dimension.
The geometry is that of a 5-dimensional Anti-de-Sitter 
space (AdS$_5$), where the extent of the
$5^{th}$ dimension is $y=\pi r_c$ ($r_c$ is the 
compactification radius), and every slice of the additional
dimension corresponds to a 4-d Minkowski metric.  Two 3-branes
reside at the boundaries of the AdS$_5$ slice, with the 3-brane
located at the fixed point $y=\pi r_c$ being known as the TeV brane, 
while the opposite brane at the other boundary $y=0$ is 
referred to as the Planck brane.  Within this framework, 
gravity is localized about the Planck brane, and 
electroweak symmetry breaking can take place either with the Higgs 
field being localized on or near the 
TeV brane, or via boundary conditions imposed 
at the fixed points as in the Higgsless models 
discussed above.  The FCNC effects 
considered here are independent of this choice.

FCNC effects are induced~\cite{warpfc,soni2} when the SM fermions
and gauge bosons are localized in the warped $5^{th}$ 
dimension~\cite{rsgauge,rsfermion,rswall}. As in the case of 
flat TeV$^{-1}$-sized extra dimensions with split fermions discussed 
above, the observed fermion masses and mixings are 
automatically explained by the geometry, with the 5-dimensional
Yukawa couplings all being of order unity.  Localizing the light fermions near 
the Planck brane results in small 4-dimensional Yukawa couplings for 
these fields, whereas if the top-quark field is localized near the 
TeV brane a large 4-d top Yukawa coupling is induced.  This
localization scheme also naturally suppresses higher dimensional
flavor changing operators that are problematic when the SM is
confined to the TeV brane.  This flavor breaking fermion
localization leads to FCNC interactions via non-universal
couplings of the zero-mode fermions to the gauge boson KK states.  
Since the SM gauge bosons are localized near the TeV brane (in 
order to acquire their masses) and have exponentially decaying 
wavefunctions towards the Planck brane, we expect FCNC in
the light quark sector to be suppressed.

The action for fermion fields in the RS bulk is given by~\cite{rsfermion}
\begin{equation}
S = \int d^4x~dy \sqrt{G} \left( {i\over 2}\bar\Psi
\gamma^MD_M\Psi  + sgn(y)M_f\bar\Psi\Psi + h.c.
\right)\ \ ,
\end{equation}
where $G$ represents the determinant of the 5-dimensional metric, 
$D_M$ is the covariant derivative in
curved space, and $\gamma^M\equiv V_\mu^M\gamma^\mu$ with $V_\mu^M$
being the inverse vierbein.  The parameter of importance to us here
is $M_f$ which is the 5-dimensional bulk mass for the fermion $f$.
It given by $M_f=kc_f$, where $k$ is
the parameter describing the curvature of 
the AdS$_5$ space and is of order of the 5-d Planck scale.  The
constants $c_f$ indicate the position of the fermion's localized
wavefunction in the bulk, with $c_f>1/2\quad (c_f<1/2)$ 
corresponding to the
fermion being localized near the Planck (TeV) brane.  These 
constants determine the flavor structure of the theory.

The KK decomposition of the bulk fermion fields yields the 
normalized zero-mode wavefunction (a discrete symmetry ensures that 
the zero-mode fields are chiral),
\begin{eqnarray}
f^{(0)} =  \sqrt{ {kr_c(1-2c_f)\over e^{\pi kr_c (1-2c_f)}
-1}}\, e^{-c_fky}\ \equiv \  \sqrt{kr_c}\, Y_f\, e^{-c_fky}\ \ .
\end{eqnarray}

\begin{table}
\centering
\begin{tabular}{|c|c|} \hline\hline
 $Y_f^{2}$ & Range of $c_f$ \\ \hline
${1\over 2} - c_f$ & $c_f < {1\over 2} - \epsilon$\\
$\displaystyle{1\over 2\pi kr_c} $ & $c_f\to {1\over 2}$\\
$\left( c_f-{1\over 2}\right) e^{\pi kr_c (1-2c_f)}$ & $c_f>
{1\over 2}+\epsilon$\\ \hline\hline
\end{tabular}
\caption{\label{tab:rstab}
The asymptotic behavior of the square of the parameter
$Y_f$ for various localization points $\{ c_f \}$ of the fermion's 
wavefunction.}
\end{table}
The asymptotic behavior of the $Y_f$ on the localization 
parameters $c_f$ are listed in Table~\ref{tab:rstab}.  We see that
these factors become exponentially small when the fermions are
localized near the Planck brane.
In the basis where the 5-d bulk masses, $M_f$, are diagonal the 
fermion Higgs interactions yield the 4-dimensional Yukawa
couplings, 
\begin{equation}
\lambda^f_{4(ij)}= \lambda^f_{5(ij)} Y_{f_{Li}}Y_{f_{Rj}}
e^{\pi kr_c(1-c_{f_{Li}}-c_{f_{Rj}})}\ \ ,
\end{equation}
for the zero-mode fermions
in terms of the 5-d Yukawa couplings $\lambda_5^f$.  We take the elements of
$\lambda_5^f$ to be complex
and of order unity.  Note that $kr_c\approx 11.3$ in order to resolve the
gauge hierarchy problem.  The elements of the matrices that diagonalize 
the up and down quark fields to their 4-d mass eigenstates have 
magnitude
\begin{equation}
|V_L^{u,d}|_{ij} \simeq {Y_{f_{Li}}\over Y_{f_{Lj}}}
\simeq |V_{CKM}|_{ij} \qquad\qquad\quad  (i<j)\ \ ,
\end{equation} 
with $L\to R$ for the matrices that diagonalize the right-handed
fields.  For the elements with $j\le i$ one should interchange
$i\leftrightarrow j$.

The wavefunctions for the gauge KK states are given by the first
order Bessel functions $J_1\,, Y_1$, and the mass of the $n^{th}$ 
gauge KK mode is $M_n=x_nke^{-\pi kr_c}$ where $x_n$ is related to
the roots of Bessel functions~\cite{rsgauge}.  The first few 
values of $x_n$ are 2.45, 5.57, 8.70, and 11.84. Precision 
electroweak data places severe bounds on the masses
of the gauge KK states~\cite{rsgauge,rsew}. However these bounds
can be reduced to $M_1\gsim 3$ TeV if the gauge symmetry in the
bulk is expanded to $SU(2)_L\times SU(2)_R\times U(1)_{B-L}$,
which restores custodial symmetry~\cite{rscust}. The couplings
of these states to the zero-mode fermions, $\{ C_f^{(n)}\} $, 
are determined by the
overlap of their wavefunctions in the additional dimension.
They are given (as a ratio to the SM coupling) by
\begin{equation}
C_f^{(n)}={g^{(n)}\over g_{SM}} = \sqrt{2\pi kr_c}
Y_f^2I^{(n)}_f\ \ ,
\end{equation}
where $I^{(n)}_f$ is an integral over $J_1$ Bessel functions,
and are given explicitly in Ref.~\cite{rswall} and in the Appendix.
As displayed in Fig.~\ref{rscoupl},  
this coupling weakens substantially as the gauge KK level, $n$, increases
for $c_f<1/2$, while for $c_f>1/2$ the couplings tend to a small fixed
value for all KK levels.  The interaction
lagrangian for the gluon KK states 
in the quark mass eigenstate basis is as given in
Eq.~(\ref{splitL}) with the substitution of the prefactor
$\sqrt 2\to \sqrt{2\pi kr_c}$, which arises from the renormalization
of the KK gauge kinetic terms to the canonical value.

\begin{figure}[htbp]
\centerline{
\includegraphics[width=6cm,angle=90]{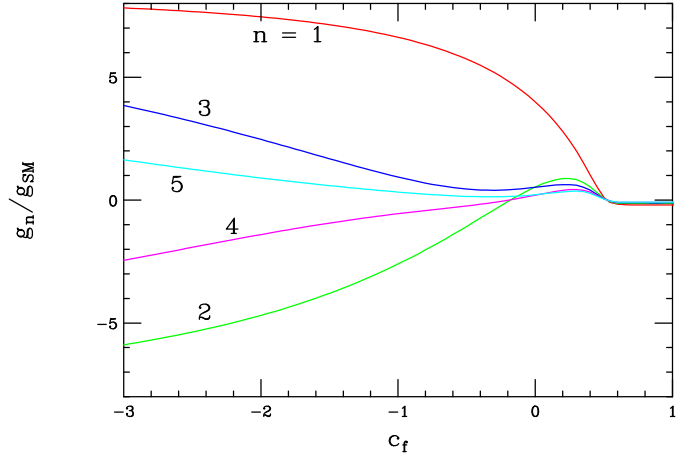}}
\vspace*{0.1cm}
\caption{The coupling strength, scaled to the SM strong coupling constant, 
of the zero-mode fermions to the first five gluon KK excitations (as labeled) 
as a function of the fermion bulk mass parameter $c_f$.}
%\label{gncoupl}
\label{rscoupl}
\end{figure}

$D^0$-$\bar D^0$ mixing is then mediated via tree-level flavor
changing interactions of the KK gauge boson states as 
depicted in Fig.~\ref{treegraph}.  In analogy to the previous
section, the effective hamiltonian for this process
is given by (for the exchange of gluon KK states which yield the largest
contribution)
\begin{equation}
{\cal H}_{RS}  =  {2\pi kr_c \over 3}g_s^2\sum_{\vec n=1}^\infty 
{1\over M_n^2} \left( U_{L(cu)}^{u(n)\dagger}U_{L(uc)}^{u(n)}Q_1
+ 2U_{L(cu)}^{u(n)\dagger}U_{R(uc)}^{u(n)}Q_2
+ U_{R(cu)}^{u(n)\dagger}U_{R(uc)}^{u(n)}Q_6 \right) \ \ ,
\end{equation}
where $U^{u(n)}_{L,R}\equiv 
V^{u\dagger}_{L,R}C^{(n)}_fV^u_{L,R}$.  Writing this explicitly
for $U^{u(n)}_L$ yields, 
\begin{eqnarray}
U^{u(n)}_{L(uc)}  =  V^{u\dagger}_{L(uj)}C^{(n)}_{jk}
V^u_{L(kc)}\delta_{jk} = 
V^{u\dagger}_{L(uj)} Y^2_{f_j} V^u_{L(jc)}I^{(n)_{f_j}}\ \ ,
\end{eqnarray}
since the $C_f^{(n)}$ are diagonal, and
where the index $j$ sums over the generations.  Looking at
the asymptotic values of $Y_f$ in Table~\ref{tab:rstab}, we see that
$U^{u(n)}_{L(uc)}$ is only sizable when the fermion is localized
towards the TeV brane.  Unitarity of the $V^u_{L,R}$ results in
\begin{equation}
{\cal H}_{RS} = {2\pi kr_c \over 3M_1^2}g_s^2
\left( C_1(M_n)Q_1 + C_2(M_n)Q_2 + C_6(M_n)Q_6
\right)\ \ ,
\end{equation}
where $M_1$ is the mass of the first gluon KK excitation, and
with the Wilson coefficients being given by
\begin{eqnarray}
C_1(M_1) & = & 2\pi kr_c\sum_{\vec n=1}^\infty 
{M_1^2\over M_n^2} \left( V^{u\dagger}_{L(13)}V^u_{L(32)}
[Y_{t_L}^2 (I^{(n)}_{t_L})^2-Y_{u_L}^2(I^{(n)}_{u_L})^2]\right.\nonumber\\
& & \left. \hspace{0.3cm} + V^{u\dagger}_{L(12)}V^u_{L(22)}
[Y_{c_L}^2 (I^{(n)}_{c_L})^2-Y_{u_L}^2(I^{(n)}_{u_L})^2]\right)^2
\,, \nonumber \\
C_2(M_1) & = & 4\pi kr_c\sum_{\vec n=1}^\infty 
{M_1^2\over M_n^2} \left(V^{u\dagger}_{L(13)}V^u_{L(32)}
[Y_{t_L}^2 (I^{(n)}_{t_L})^2-Y_{u_L}^2(I^{(n)}_{u_L})^2]\right. \nonumber\\
& & \hspace{0.3cm} + \left. V^{u\dagger}_{L(12)}V^u_{L(22)}
[Y_{c_L}^2 (I^{(n)}_{c_L})^2-Y_{u_L}^2(I^{(n)}_{u_L})^2]\right)
\left( V^{u\dagger}_{R(13)}V^u_{R(32)}
[Y_{t_R}^2 (I^{(n)}_{t_R})^2-Y_{u_R}^2(I^{(n)}_{u_R})^2]\right. \nonumber\\
& & \hspace{0.3cm} +  \left. V^{u\dagger}_{R(12)}V^u_{R(22)}
[Y_{c_R}^2 (I^{(n)}_{c_R})^2-Y_{u_R}^2(I^{(n)}_{u_R})^2]\right)
\,, \nonumber \\
C_6(M_1) & = & 2\pi kr_c\sum_{\vec n=1}^\infty 
{M_1^2\over M_n^2} \left(
V^{u\dagger}_{R(13)}V^u_{R(32)}
[Y_{t_R}^2 (I^{(n)}_{t_R})^2-Y_{u_R}^2(I^{(n)}_{u_R})^2]\right. \nonumber\\
& & \hspace{0.3cm} +  \left. V^{u\dagger}_{R(12)}V^u_{R(22)}
[Y_{c_R}^2 (I^{(n)}_{c_R})^2-Y_{u_R}^2(I^{(n)}_{u_R})^2]\right)^2\ ,
\label{rswilsons}
\end{eqnarray}
with $M_1^2/M_n^2=x_1^2/x_n^2$ where $x_n$ are the Bessel function
roots described above.

The RG evolution proceeds as in Section~\ref{RGSect} and results in
the effective hamiltonian at the charm quark scale
\begin{equation}
{\cal H}_{RS}={g_s^2 \over 3M_1^2}\left(
C_1(m_c)Q_1+C_2(m_c)Q_2+C_3(m_c)Q_3+C_6(m_c)Q_6\right)\ \ ,
\end{equation}
where additional operators have been generated due to mixing
in the RG evolution.  The evolved Wilson coefficients at the charm
scale are as given in Eq.~(\ref{zwilsons}) with the appropriate
substitution of $M_{Z'}\to M_1$.  
Upon evaluating the matrix elements
we obtain the contribution to $x_{\rm D}$ from warped extra dimensions
\begin{equation}
x^{\rm (RS)}_{\rm D}={g_s^2 \over 3M_1^2} {f_D^2B_DM_D\over\Gamma_D}
\left({2\over 3}[C_1(m_c)+C_6(m_c)]-{5\over 6}C_2(m_c)+{7\over 12}C_3(m_c)
\right)
\end{equation}
in the modified vacuum saturation approximation.  Here, we have
taken the factor $\eta$ of Eq.~(\ref{eta}) to be unity.

To obtain numerical results, we need to specify the fermion locations
in the warped dimension.  We examine three popular scenarios in the
literature that correctly generate the 4-d Yukawa hierarchy for the
SM fermions.  As mentioned above, localizing the fields near the UV
(ultraviolet or Planck) brane
generates an exponentially small 4-d Yukawa coupling.  In all three
models, all of the light quarks are localized such that
their bulk mass parameters take on values with $c_f>1/2$.  Special attention
must be paid to the localization of the third generation quarks; in
order to generate a large top-quark mass, the corresponding SU(2) singlet 
field is usually taken to reside close to the TeV brane.  The third
generation SU(2) doublet fields and $b$-quark SU(2) singlet field are
located as close to the IR (infrared, or TeV) 
brane while maintaining consistency with the
experimentally determined $Zb\bar b$ coupling.  The three scenarios that we
follow are fairly uniform in their treatment of the third generation,
differing only slightly in the location of the SU(2) top-quark singlet.
The scenarios are: (I)  A study of flavor physics in the Randall-Sundrum
model~\cite{frankp}, (II)  A scenario that has been constructed in order to 
generate fermion masses within the 5-d picture of Higgsless 
models~\cite{csakiferm},  (III) The up-quark singlet field is taken 
to lie even closer to the UV brane~\cite{hooman} in order to solve the
strong CP problem with warped geometries.  The numerical values of the 
bulk mass parameters are summarized in Table~\ref{tab:cfvals} for the 
three cases.

\begin{table}
\centering
\begin{tabular}{|c|c|c|c|} \hline\hline
  & Model I & Model II & Model III \\ \hline
$c_{u_L}$ & $> 1/2$ & 0.6 & 0.5 \\
$c_{u_R}$ & $> 1/2$ & 0.6 & 1.4 \\
$c_{c_L}$ & $> 1/2$ & 0.52 & 0.5 \\
$c_{c_R}$ & $> 1/2$ & 0.52 & 0.53 \\
$c_{t_L}$ & 0.45 & 0.4 & 0.46 \\
$c_{t_R}$ & 0 & 0.3 & on the IR brane \\ \hline\hline
\end{tabular}
\caption{\label{tab:cfvals}
The values of the bulk mass parameters for the three models described
in the text.}
\end{table}

\begin{figure}[htbp]
\centerline{
\includegraphics[width=6cm,angle=90]{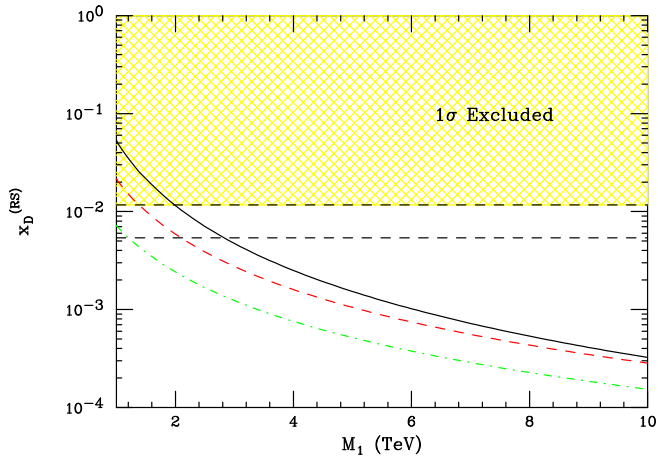}}
%\hspace*{5mm}
%\includegraphics[width=6cm,angle=90]{warp2.ps}}
\vspace*{0.1cm}
\caption{The contribution to $x_{\rm D}$ from a warped extra dimension
with the SM fields in the bulk as a function of the mass for the 
first gluon KK excitation.
The $1\sigma$ experimental
bounds are as indicated, with the yellow shaded region depicting the
region that is excluded. The curves correspond to Model I
(dashed-dot, green), Model II (dashed, red) and Model III
(solid) as described in the text.}
\label{warpedfig}
\end{figure}

Our results for $x_{\rm D}^{RS}$ for these three models 
are presented in Fig.~\ref{warpedfig},
where as above, we assume the quark diagonalization matrices take on
CKM-like values.  In the figure, the dot-dashed green, dashed
red, and solid black curves correspond to the bulk mass
parameters of Model I, II, and III, respectively.  We see from
Fig.~\ref{rscoupl} that fermions localized towards the Planck brane have
very small couplings to the KK gluon states and thus do not substantially
contribute to $x_{\rm D}^{RS}$.  This simplifies the expressions in
Eq.~(\ref{rswilsons}) in this case, as only the $t_{L,R}$ terms have
sizable contributions.  Looking at the figure we see the mass
of the first gluon KK excitation is constrained to lie $\gsim 1-2$ TeV,
which is essentially
the same value as the bound obtained from the precision electroweak
data in warped models with bulk custodial symmetry~\cite{rscust}.  Lastly,
we recall from discussion in the previous section, that
these constraints can be evaded if the
matrix which diagonalizes the up-quark sector is essentially diagonal.

%%%%%%%%%%%%
%%%%%%%%%%%%%%%%%%%%%%%%%%%%%%%%%%%%%%%%%%%%%%%%%%%%%%%%%%%%%%%%%%%%%
\section{Extra Symmetries}

In this section, we focus on supersymmetry.  Weak
scale supersymmetry is a possible solution to the gauge hierarchy
problem, leads to the unification of the gauge couplings at high energies,
and provides a natural Dark Matter candidate.  It is thus a very well
motivated theory of physics beyond the SM.   Supersymmetry is an
extension of the Poincare symmetry, relating fermions and bosons at
a fundamental level.  All SM particles have supersymmetric partners 
(`{\it sparticles}') with the same mass and gauge
interactions, but with spin differing by one-half unit.  
Since the supersymmetric particles have 
yet to be discovered, we know that supersymmetry is broken; in this section,
we will be agnostic as to which supersymmetry breaking mechanism Nature may
have chosen.  We note
that in non-broken supersymmetry the rates for all loop-induced processes 
would vanish due to an exact cancellation between the SM and supersymmetric
contributions.  It is thus due to the breaking of supersymmetry that
contributions to FCNC are generated in these theories.  

Here, we examine the contributions to $D^0$-$\overline D^0$ mixing in
four supersymmetric scenarios: the Minimal Supersymmetric
Standard Model (MSSM), models with alignment in the quark-squark 
mass matrices,
models with R-parity violating couplings, and Split Supersymmetry.
Other scenarios with extended non-supersymmetric 
symmetries have been considered elsewhere in this paper.

%%%%%%%%%%%%%%%
\subsection{Minimal Supersymmetric Standard Model}

As the name implies, the MSSM is the simplest version of supersymmetry
as it contains the minimal number of new particles.  The SM fermions
are placed in chiral supermultiplets, the SM gauge bosons lie in vector
supermultiplets, and the Higgs sector
takes the form of the flavor conserving two-Higgs-doublet Model II
discussed above.  A discrete symmetry, R-parity, is imposed to forbid
unwanted terms in the superpotential that would mediate proton decay at
a dangerous level.  Conservation of R-parity implies that only pairs
of sparticles can be produced or exchanged in loops.  Collider searches
for direct squark and gluino pair production place the bound
$m_{\tilde q,g}\gsim 330$~GeV~\cite{PDG} in the MSSM with gravity
mediated Supersymmetry breaking.

As mentioned above, we will not assume any particular supersymmetry breaking 
mechanism in our discussion, and so we employ a model independent 
parameterization of
all possible soft supersymmetry breaking terms.   This soft supersymmetry
breaking sector generally includes three gaugino masses, trilinear
scalar interactions, as well as Higgs and sfermion masses, and thus
contains many potential sources of flavor violation.  Here, we are interested
in the flavor violating sources that arise in the up-squark sector.  In
what is known as the super-CKM basis, the squark fields are rotated
by the same matrices that diagonalize the quark masses, giving
rise to non-diagonal squark mass matrices.  The squark propagators
are then expanded such that the non-diagonal mass terms result in mass
insertions that change the squark 
flavor~\cite{Ellis:1981ts,Nilles:1983ge,Georgi:1986ku,Hall:1985dx}.  
This source of flavor violation
differs from that of the SM and many NP models discussed 
earlier.  Here, the quark-squark-gaugino neutral couplings are flavor
conserving, while flavor violation arises from the non-diagonality
of the squark mass propagators.   The $6\times 6$ mass matrix for
the $Q = +2/3$ squarks can be divided into $3\times 3$ sub-matrices,
\beq
\widetilde{M}^2 = 
\left(
\begin{array}{c c}
\widetilde{M}^2_{LL} & \widetilde{M}^2_{LR} \cr
\widetilde{M}_{LR}^{2~T} & \widetilde{M}^2_{RR}
\end{array}
\right) \ \ ,
\eeq
and the mass insertions can be parameterized in a model independent
fashion as
\beq
\left(\delta_{ij}\right)_{MN} = 
{\left(V_M \widetilde{M}^2 V_N^\dagger\right)_{ij} \over 
m_{\tilde q}^2} \ \ .
\label{squarkmat}
\eeq
%
%We could add a sentence after eq 115 saying that the standard notation is
%\delta^u_{ij})_{MN} and that hereafter we will follow this.
Here, $i,j$ are flavor indices, $M,N$ refers to the helicity choices 
$LL$, $LR$, $RR$, and $m_{\tilde q}$ represents the average
squark mass.  Although this source of flavor violation is present
in general, and in particular in 
models with gravity mediated supersymmetry breaking, 
it can be avoided if supersymmetry is broken by
gauge or anomaly mediation.  These mass insertions are thought to be
small in the MSSM, but can be large in non-minimal supersymmetric models.

\begin{figure} [tb]
\centerline{
\includegraphics[width=9cm,angle=0]{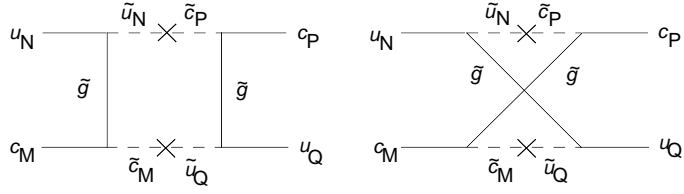}}
\caption{Contributions to $D$ mixing from mass insertions in the squark 
propagator in MSSM.  $N\,, M\,, P,$ and $Q$ label the helicity
$(L\,, R)$.
\label{SUSYbox}}
\end{figure}

In this scenario, the virtual exchange of squarks and gluinos in the
box diagrams depicted in Fig.~\ref{SUSYbox} can have a
strong contribution to $D^0$-${\bar D}^0$ mixing.  
Note that the second diagram in 
the figure is due to the Majorana nature of the gluino.
The effective hamiltonian at the
supersymmetric scale is given by
\begin{equation}
{\cal H}_{MSSM} = {\alpha_s^2\over 2 m^2_{\tilde q}}\sum_{i=1}^8
C_i(m_{\tilde q})Q_i\ \ ,
\end{equation}
where all eight operators in the independent basis contribute.  The matching
conditions at the supersymmetric mass scale are~\cite{bigsusyfcnc}
\begin{eqnarray}
C_1(m^2_{\tilde q}) & = & {1\over 18}
\left(\delta^u_{12}\right)^2_{LL} [4xf_1(x)+11f_2(x)]\ ,
\nonumber\\
C_2(m^2_{\tilde q})& = & {1\over 18}\left\{
\left(\delta^u_{12}\right)_{LR}\left(\delta^u_{12}\right)_{RL}\, 15f_2(x)
-\left(\delta^u_{12}\right)_{LL}\left(\delta^u_{12}\right)_{RR}
[2xf_1(x)+10f_2(x)]\right\}\ ,\nonumber\\
C_3(m^2_{\tilde q})& = & {1\over 9}\left\{
\left(\delta^u_{12}\right)_{LL}\left(\delta^u_{12}\right)_{RR}
[42xf_1(x)-6f_2(x)]-
\left(\delta^u_{12}\right)_{LR}\left(\delta^u_{12}\right)_{RL}\, 11f_2(x)
\right\}\ ,\nonumber\\
C_4(m^2_{\tilde q})& = & {1\over 18}
\left(\delta^u_{12}\right)^2_{RL} 37xf_1(x)\ ,\nonumber\\
C_5(m^2_{\tilde q})& = & {1\over 24}
\left(\delta^u_{12}\right)^2_{RL}\, xf_1(x)\ ,\\
C_6(m^2_{\tilde q})& = & {1\over 18}
\left(\delta^u_{12}\right)^2_{RR}[4xf_1(x)+11f_2(x)]\ ,\nonumber\\
C_7(m^2_{\tilde q}) & = & {1\over 18}
\left(\delta^u_{12}\right)^2_{LR}37xf_1(x)\ ,\nonumber\\
C_8(m^2_{\tilde q}) & = & {1\over 24}
\left(\delta^u_{12}\right)^2_{LR}\, xf_1(x)\ ,\nonumber
\end{eqnarray}
where $x \equiv m^2_{\tilde g}/m^2_{\tilde q}$, 
with $m_{\tilde g}$ being the
mass of the gluino.  The functions $f_1(x)$ and $f_2(x)$ are
given in the Appendix.  Note that these conditions are symmetric
under the interchange $L\leftrightarrow R$.  We also note
that the NLO expressions for these
matching conditions have been computed in Ref.~\cite{Ciuchini:2006dw}.

The RG evolution to the charm-quark scale results in
\begin{eqnarray}
C_1(m_c) & = & r_1(m_c, m_{\tilde q})C_1(m_{\tilde q})\ ,\nonumber\\
C_2(m_c) & = & r_2(m_c, m_{\tilde q})C_2(m_{\tilde q})\ ,\nonumber\\
C_3(m_c) & = & {2\over 3}\left[ r_2(m_c, m_{\tilde q})-
r_3(m_c, m_{\tilde q})\right] C_2(m_{\tilde q}) +
r_3(m_c, m_{\tilde q})C_3(m_{\tilde q}) \ ,\nonumber\\
C_4(m_c) & = & {8\over\sqrt{241}}\left[ r_5(m_c, m_{\tilde q})- 
r_4(m_c, m_{\tilde q})\right]\left[C_4(m_{\tilde q}) +{15\over 4}
C_5(m_{\tilde q})\right]\nonumber\\
& & \hspace{0.5cm} + {1\over 2}\left[ r_4(m_c, m_{\tilde q})+
r_5(m_c, m_{\tilde q})\right]C_4(m_{\tilde q}) \ ,\nonumber\\
C_5(m_c) & = & {1\over 8\sqrt{241}}\left[ r_4(m_c, m_{\tilde q})- 
r_5(m_c, m_{\tilde q})\right]\left[C_4(m_{\tilde q})+ 64C_5(m_{\tilde q})
\right]\nonumber\\
& & \hspace{0.5cm} + {1\over 2}\left[ r_4(m_c, m_{\tilde q})+
r_5(m_c, m_{\tilde q})\right]C_5(m_{\tilde q})\ ,\nonumber\\
C_6(m_c) & = & r_6(m_c, m_{\tilde q})C_6(m_{\tilde q})\ ,\\ 
C_7(m_c) & = &  {8\over\sqrt{241}}\left[ r_8(m_c, m_{\tilde q})- 
r_7(m_c, m_{\tilde q})\right]\left[C_7(m_{\tilde q}) +{15\over 4}
C_8(m_{\tilde q})\right]\nonumber\\
& & \hspace{0.5cm} + {1\over 2}\left[ r_7(m_c, m_{\tilde q})+
r_8(m_c, m_{\tilde q})\right]C_7(m_{\tilde q}) \ ,\nonumber\\
C_8(m_c) & = & {1\over 8\sqrt{241}}\left[ r_7(m_c, m_{\tilde q})- 
r_8(m_c, m_{\tilde q})\right]\left[C_7(m_{\tilde q})+ 64C_8(m_{\tilde q})
\right]\nonumber\\
& &  \hspace{0.5cm}  + {1\over 2}\left[ r_7(m_c, m_{\tilde q})+
r_8(m_c, m_{\tilde q})\right]C_8(m_{\tilde q})\ ,\nonumber 
\end{eqnarray}
which agrees in form with that in Ref.~\cite{Bagger:1997gg}.
Here, we have assumed that the squarks and gluinos are integrated
out at roughly the same scale.
Upon evaluating the matrix elements in the modified vacuum
saturation approximation we obtain the MSSM contribution
to $x_{\rm D}$, 
\begin{eqnarray}
x_{\rm D}^{(MSSM)} = {\alpha_s\over 2m^2_{\tilde q}}{f_D^2B_Dm_D\over
\Gamma_D}& & \left[ {2\over 3}[C_1(m_c)+C_6(m_c)] - {5\over 12}
[C_4(m_c)+C_7(m_c])+ {7\over 12}C_3(m_c) \right. \nonumber\\
& & \left. -{5C_2(m_c)\over
6}+[C_5(m_c)+C_8(m_c)] \right]\ \ .
\end{eqnarray}
Here, we have taken the factor $\eta$ of Eq.~(\ref{eta}) to be unity.

\begin{figure}[htbp]
\centerline{
\includegraphics[width=6cm,angle=90]{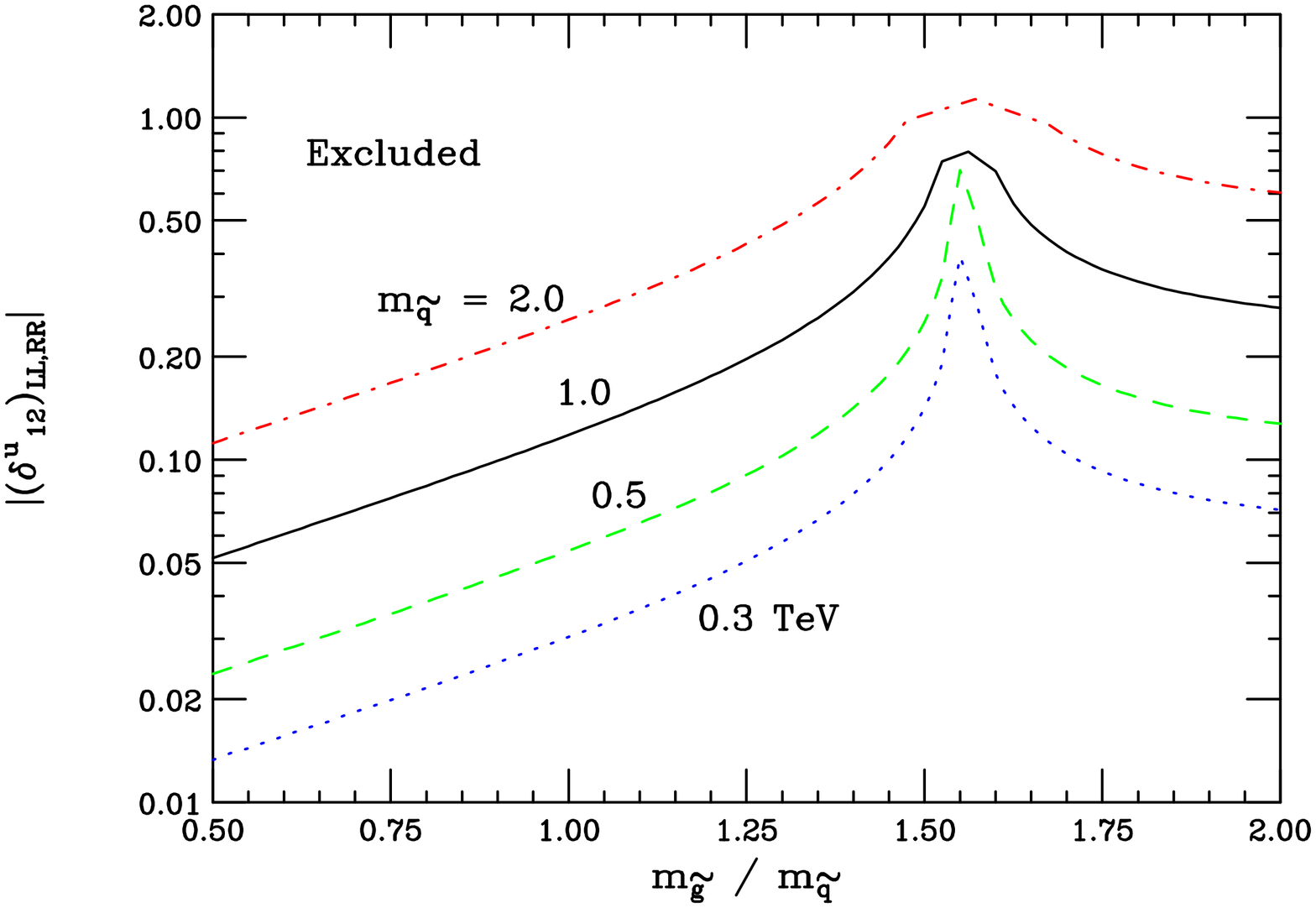}
%\hspace*{5mm}
\includegraphics[width=6cm,angle=90]{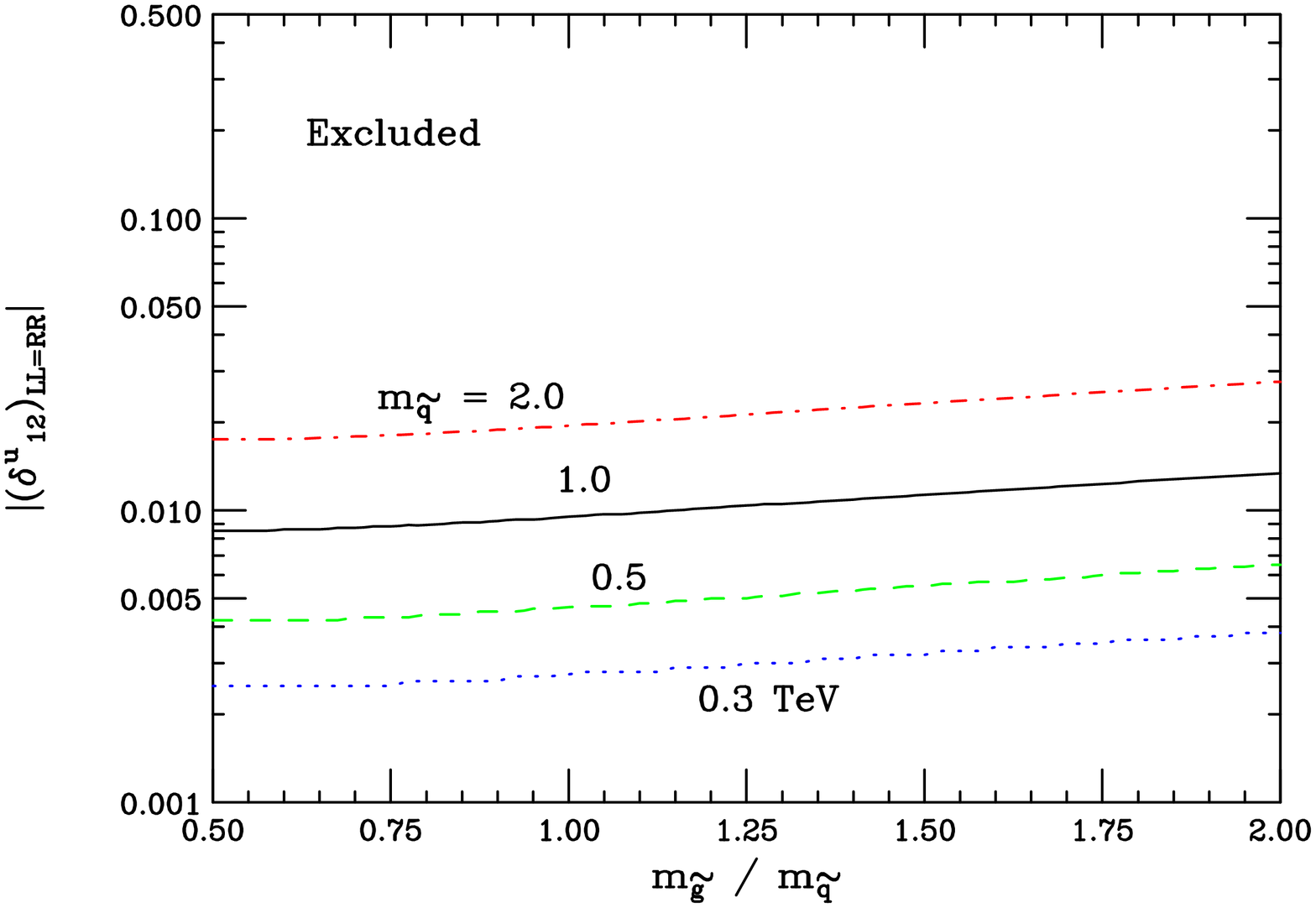}}
\centerline{
\includegraphics[width=6cm,angle=90]{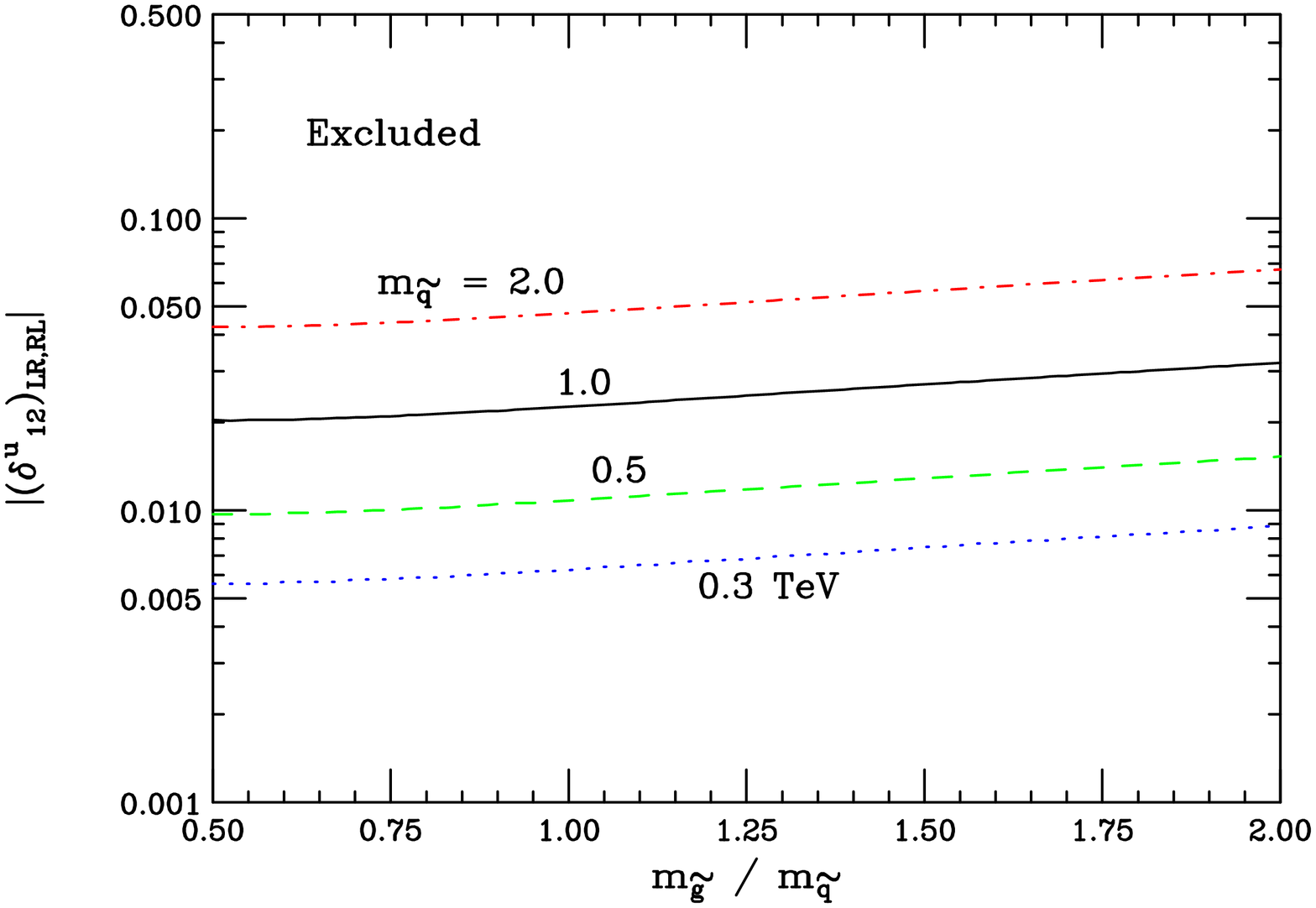}
%\hspace*{5mm}
\includegraphics[width=6cm,angle=90]{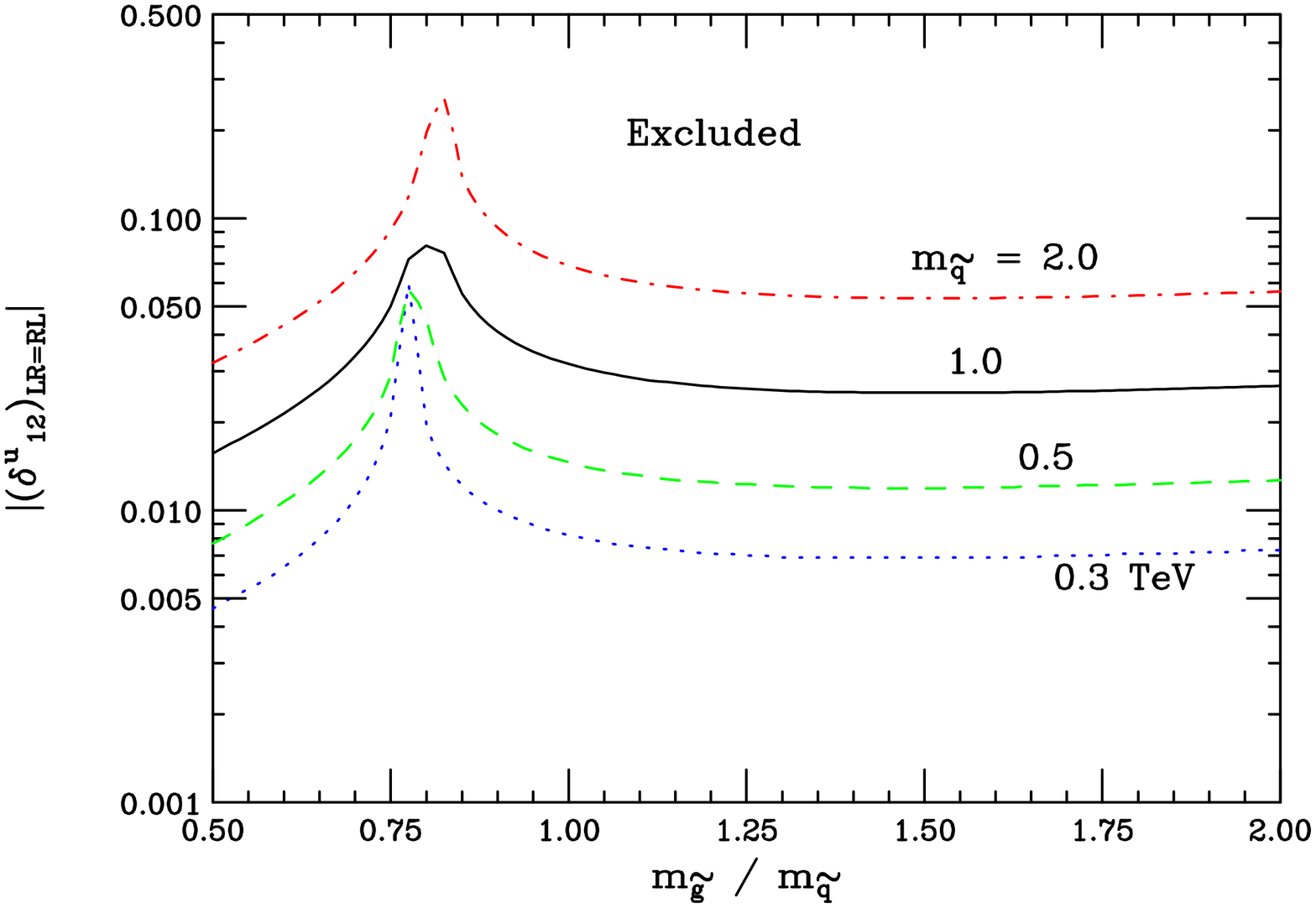}}
\vspace*{0.1cm}
\caption{The constraints 
on the absolute value of the mass insertions with different helicities
as a function of the mass ratio $m_{\tilde g}/m_{\tilde q}$ for
various values of the average squark mass. The $1\sigma$ excluded
region, corresponding to 
$x_{\rm D}<11.7\times 10^{-3}$, lies above the curves.  
\label{mssmfig1}}
\end{figure}

\begin{figure}[htbp]
\centerline{
\includegraphics[width=6cm,angle=90]{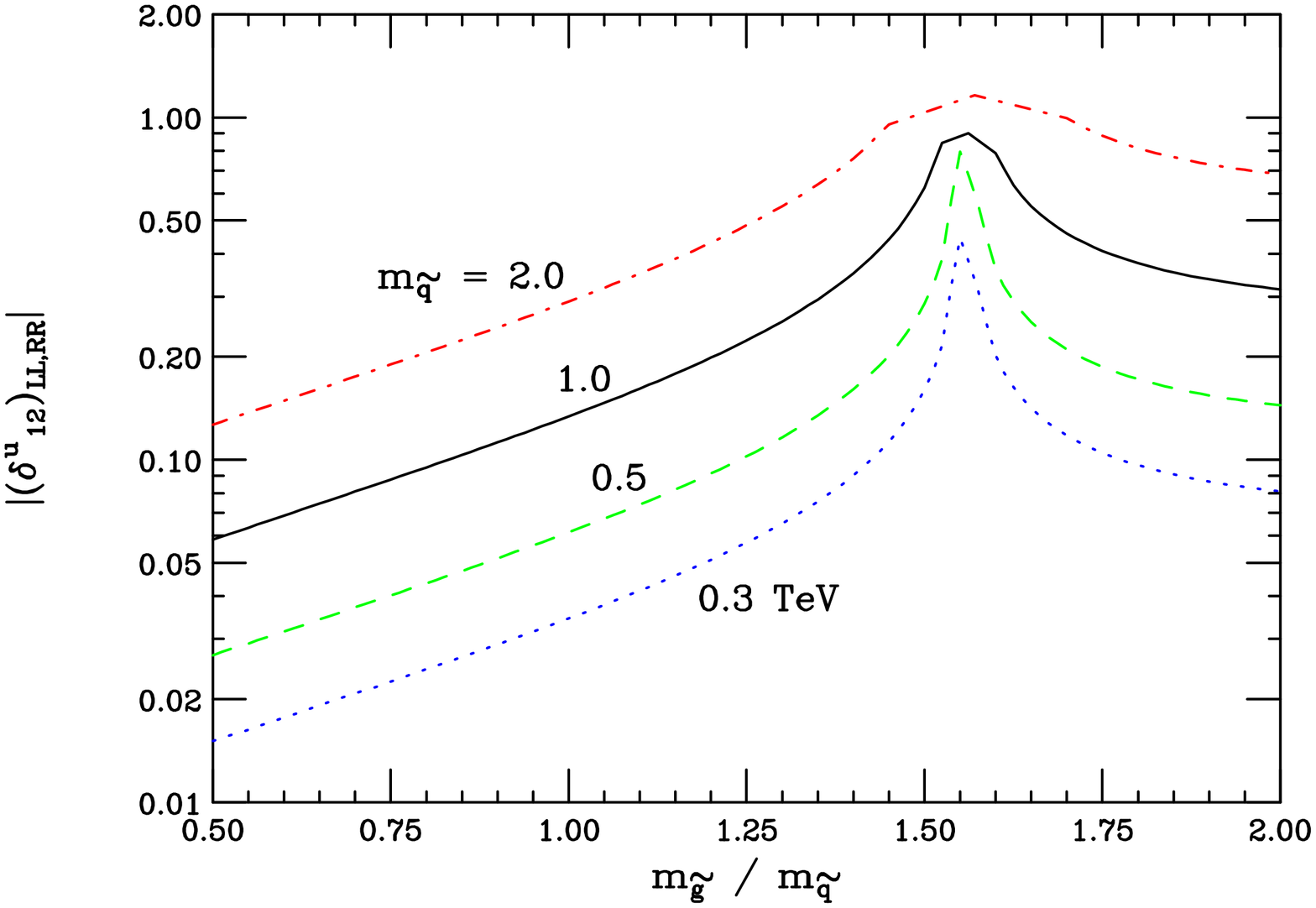}
%\hspace*{5mm}
\includegraphics[width=6cm,angle=90]{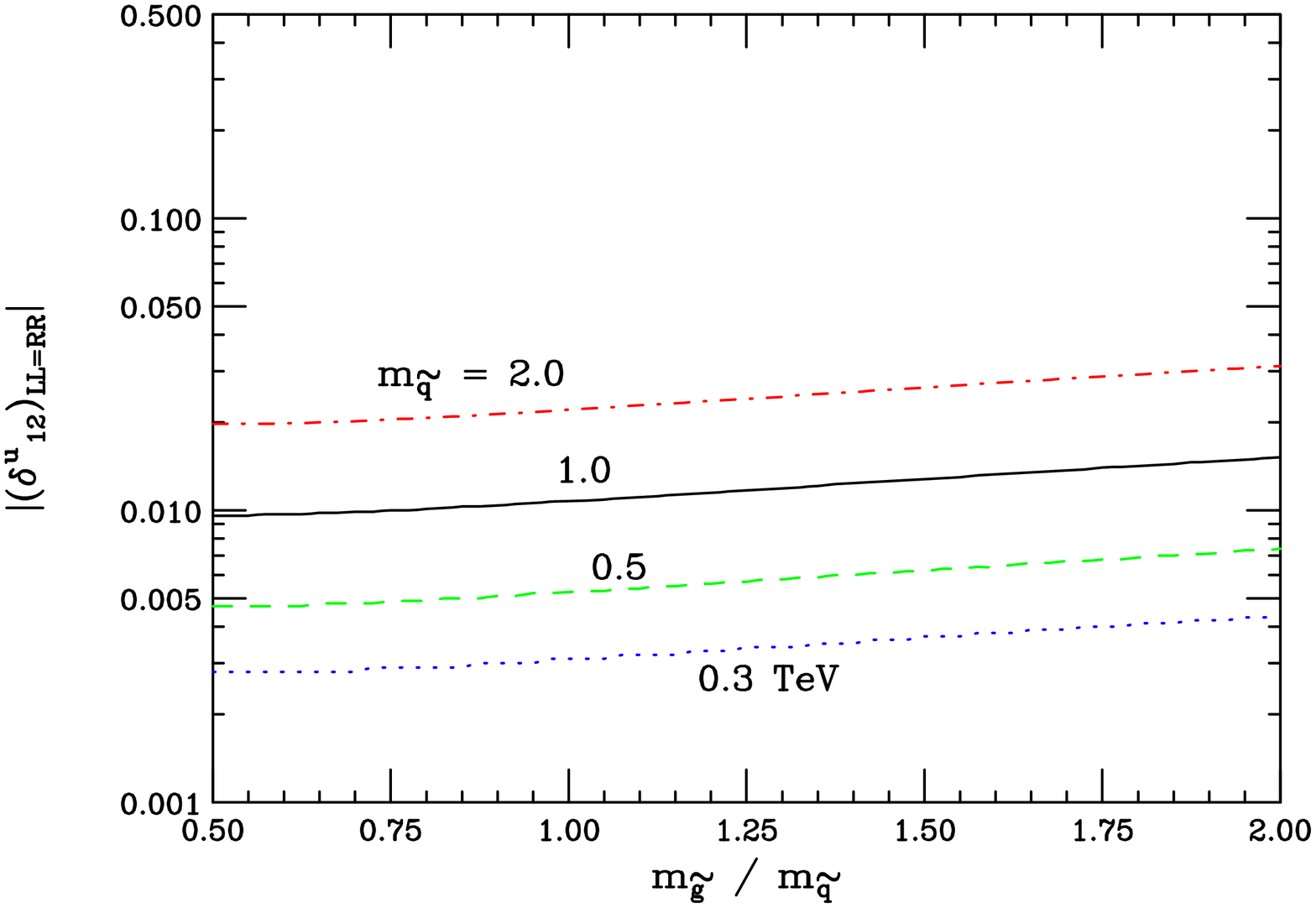}}
\centerline{
\includegraphics[width=6cm,angle=90]{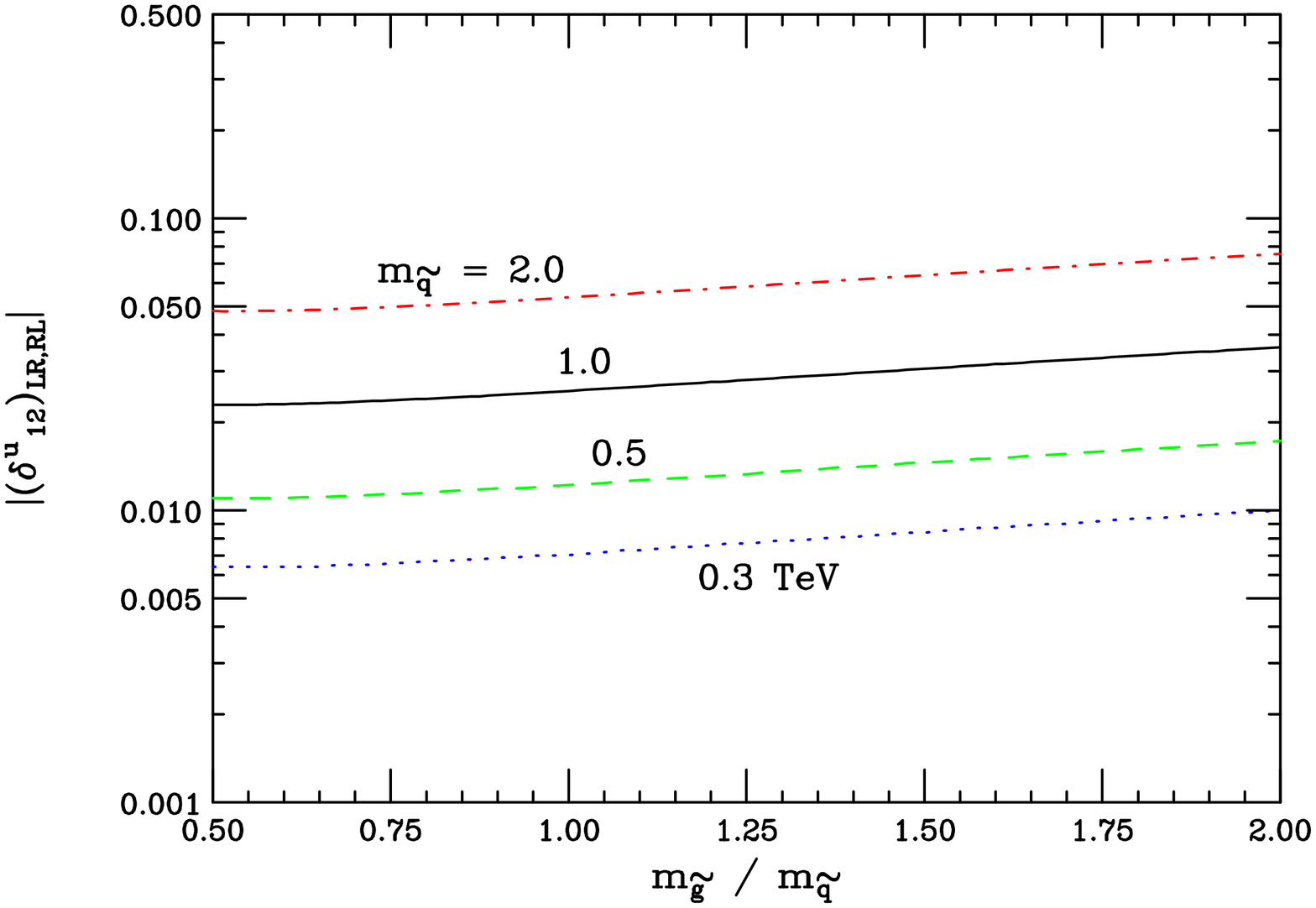}
%\hspace*{5mm}
\includegraphics[width=6cm,angle=90]{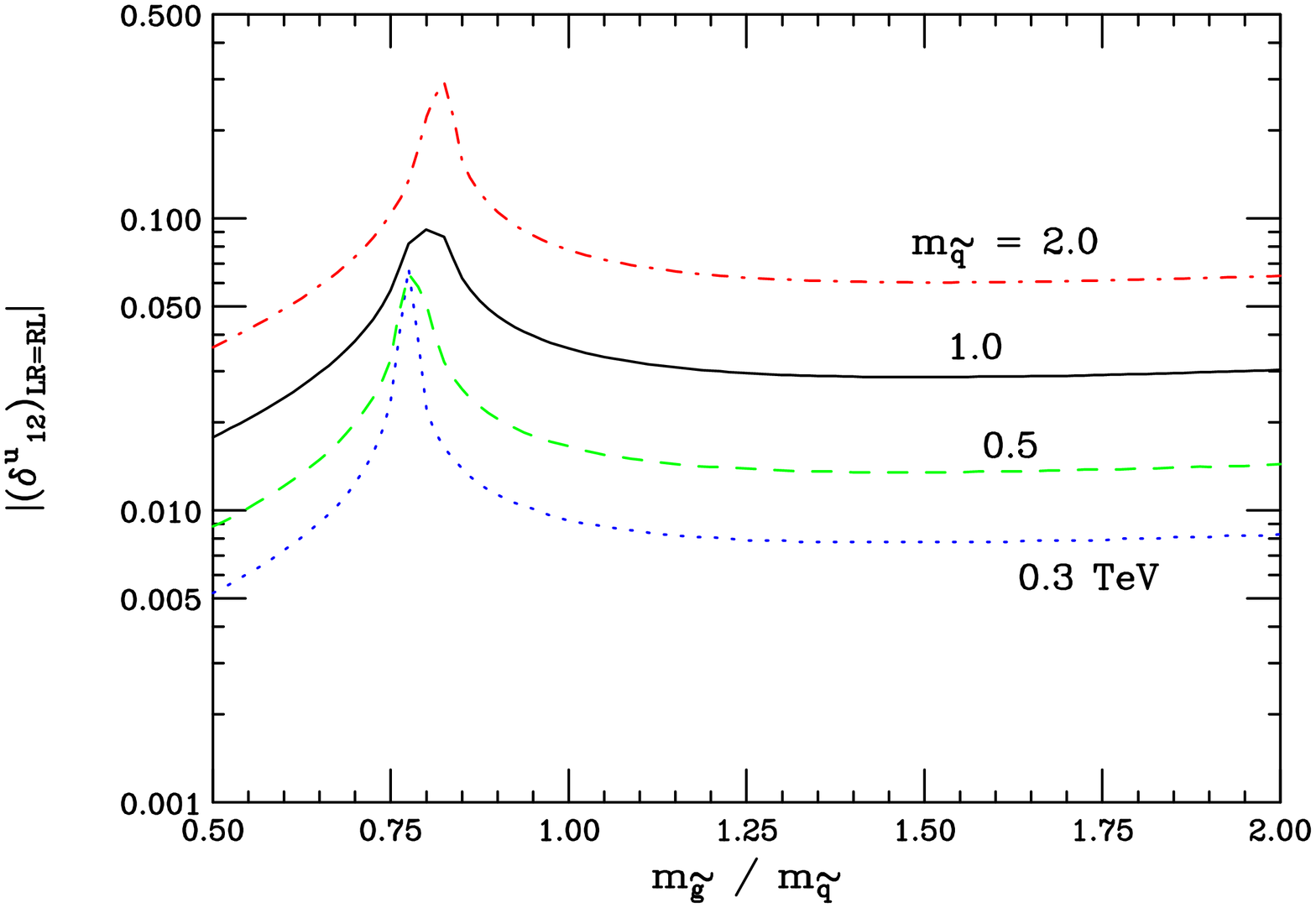}}
\vspace*{0.1cm}
\caption{Contours, corresponding to $x_{\rm D}=15.0\times 10^{-3}$, for
the absolute value of the mass insertions with different helicities
as a function of the mass ratio $m_{\tilde g}/m_{\tilde q}$ for
various values of the average squark mass.  The
region above the curves corresponds to larger values of $x_{\rm D}$.
\label{mssmfig2}}
\end{figure}

\begin{figure}[htbp]
\centerline{
\includegraphics[width=6cm,angle=90]{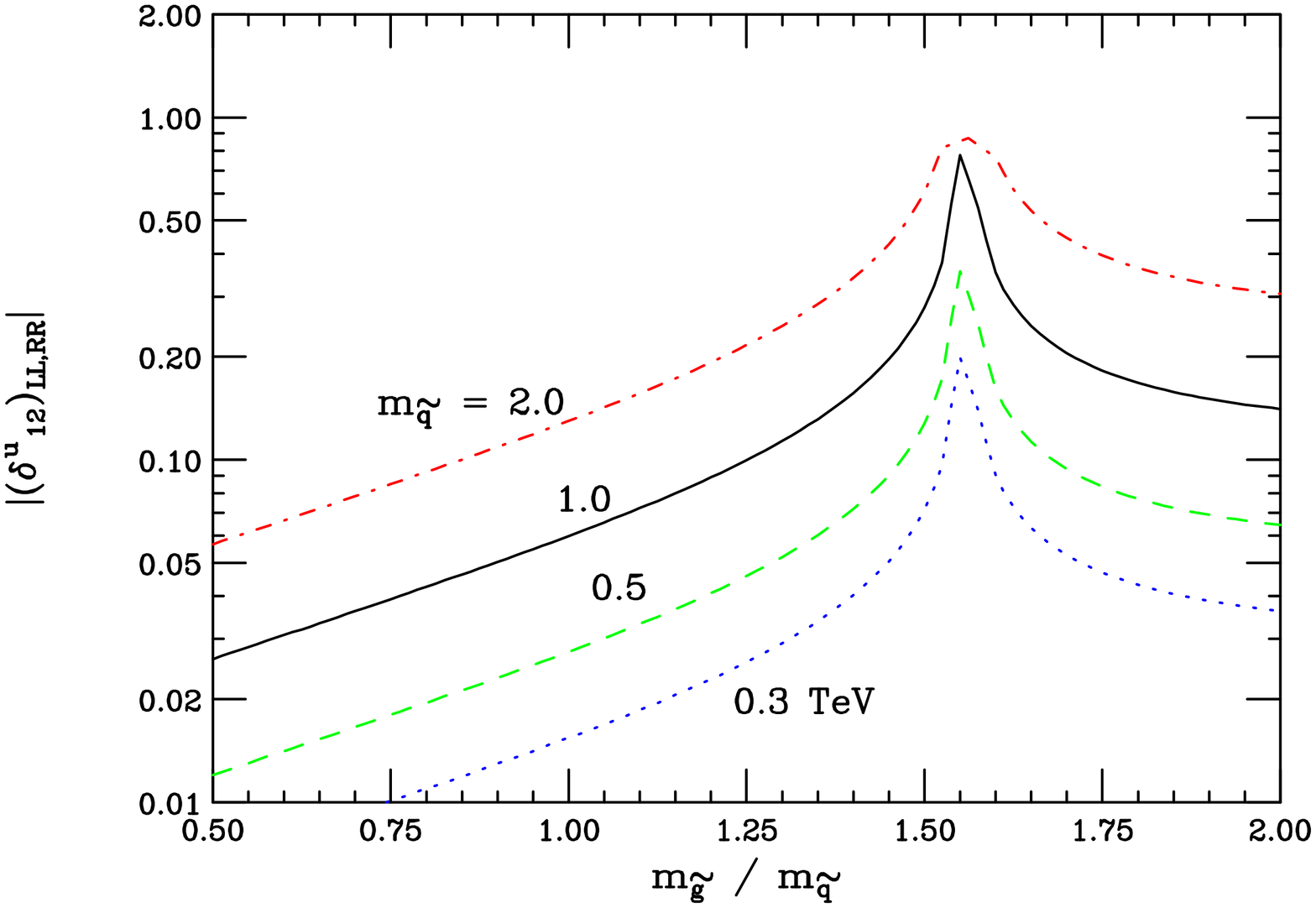}
%\hspace*{5mm}
\includegraphics[width=6cm,angle=90]{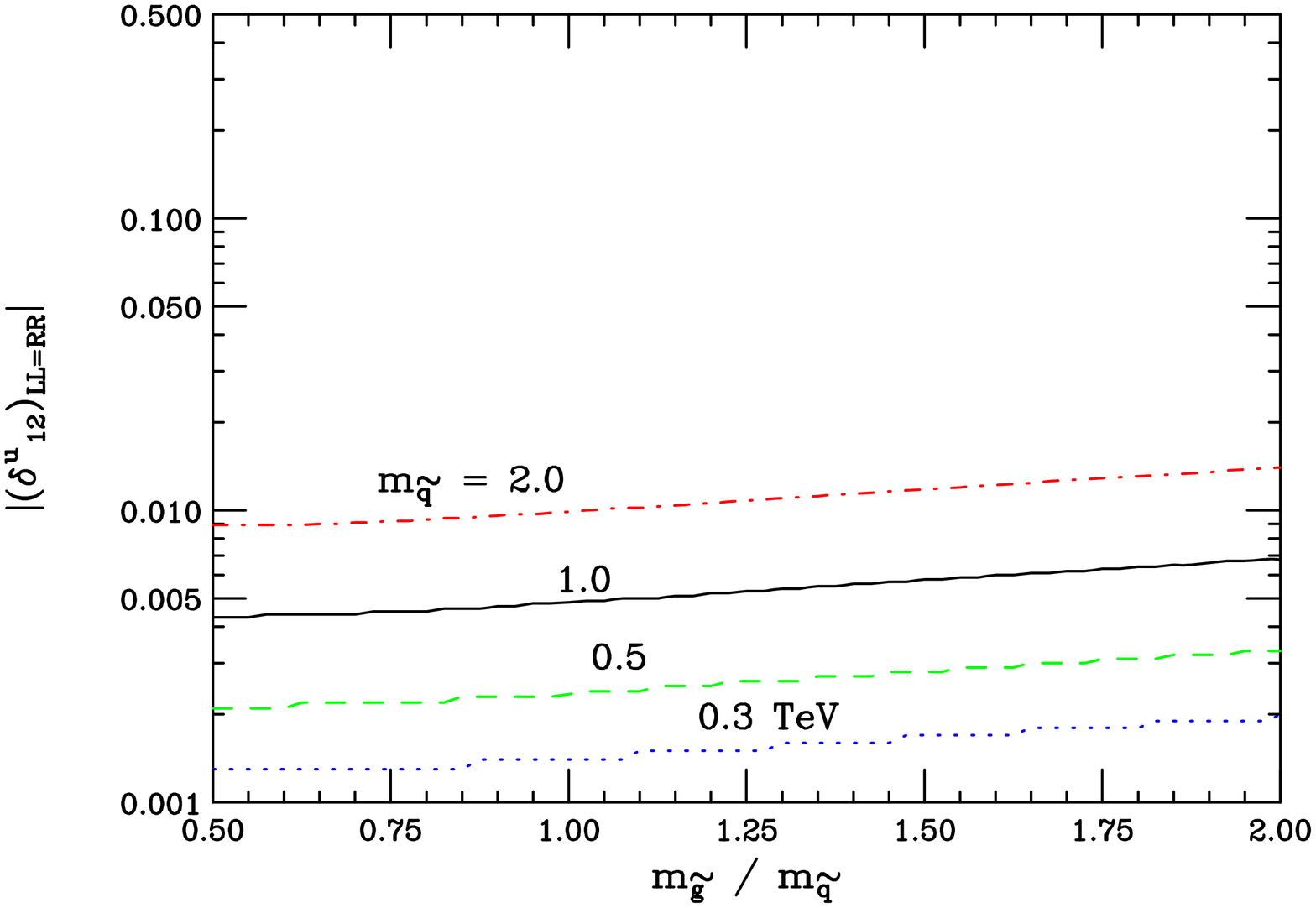}}
\centerline{
\includegraphics[width=6cm,angle=90]{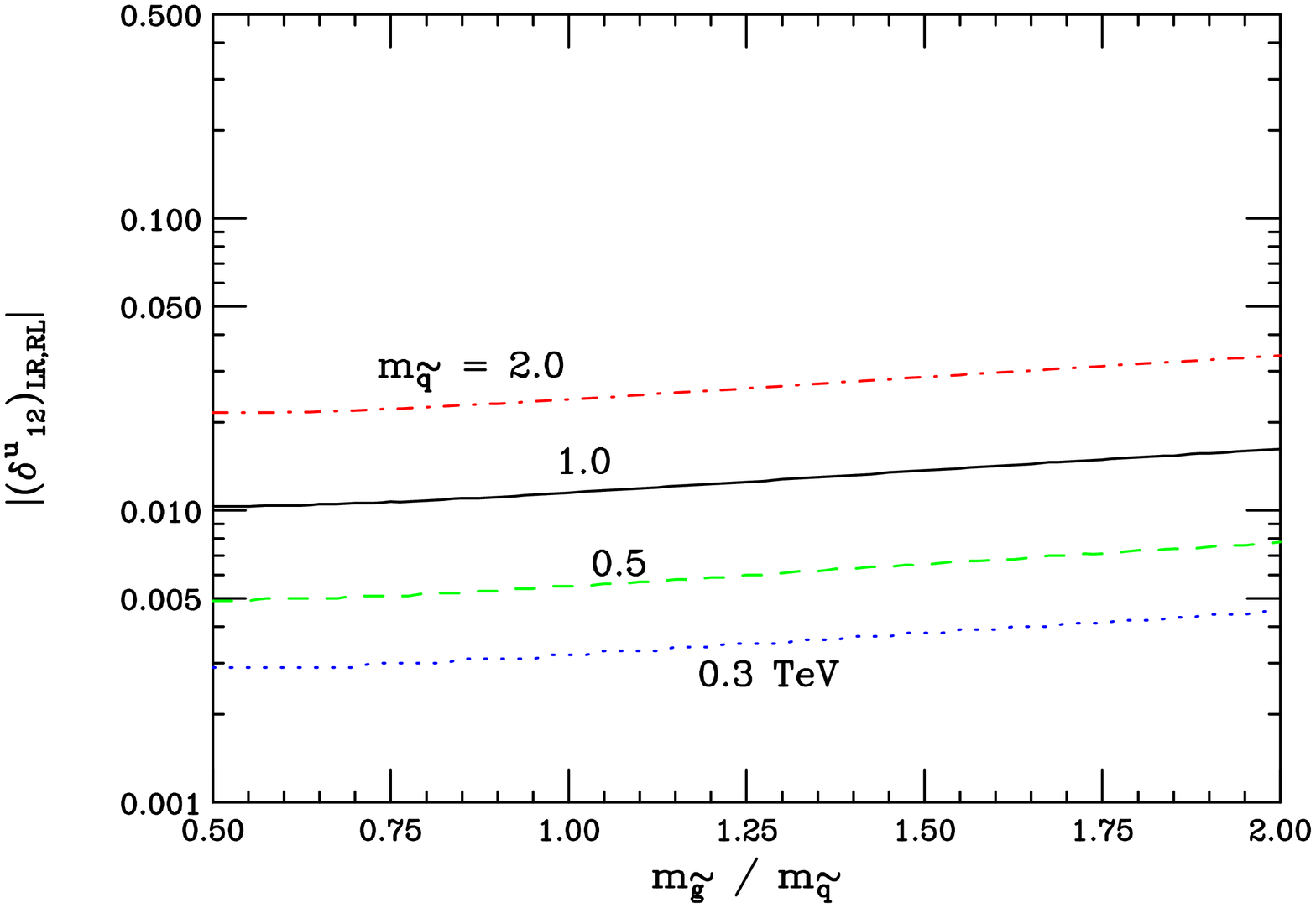}
%\hspace*{5mm}
\includegraphics[width=6cm,angle=90]{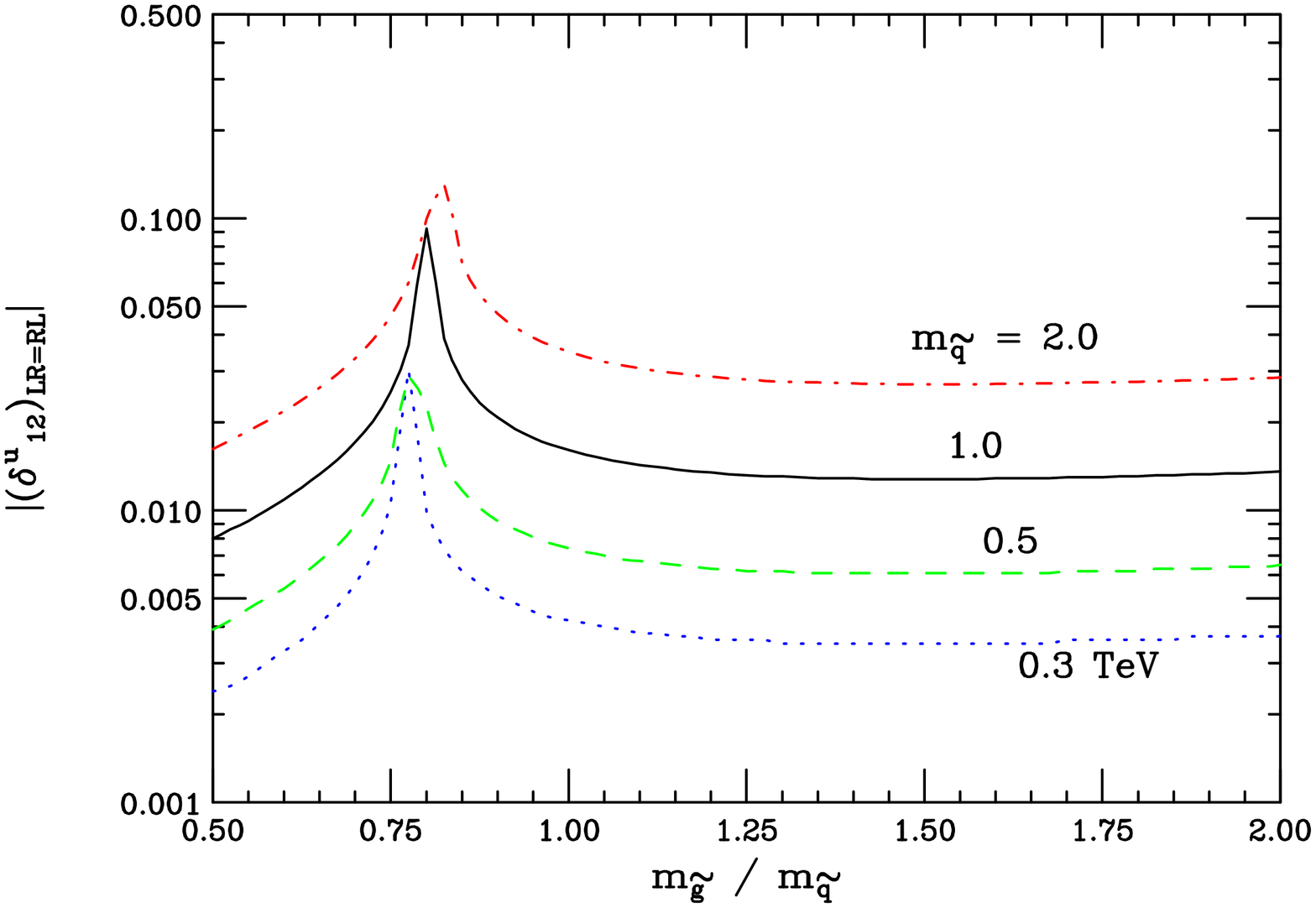}}
\vspace*{0.1cm}
\caption{Contours, corresponding to $x_{\rm D}=3.0\times 10^{-3}$, for
the absolute value of the mass insertions with different helicities
as a function of the mass ratio $m_{\tilde g}/m_{\tilde q}$ for
various values of the average squark mass.  The
region above the curves corresponds to larger values of $x_{\rm D}$.
\label{mssmfig3}}
\end{figure}

Our results for $x_{\rm D}^{(MSSM)}$ are presented in 
Figs.~\ref{mssmfig1}, \ref{mssmfig2}, \ref{mssmfig3}.
In these figures, we show contours for the absolute value
of the up-charm squark mass insertions for various helicities as a 
function of the
ratio $m_{\tilde g}/m_{\tilde q}$ for different average squark
masses.  These contours correspond to $x_{\rm D}=(11.7\,, 15.0\,,
3.0)\times 10^{-3}$ in the three figures.  In figure \ref{mssmfig1},
the region above the contours represents the current $1\sigma$
excluded region.  In these figures we take one or two of the mass
insertions to be non-vanishing, as indicated.
Due to the $L\leftrightarrow R$ symmetry of the
matching conditions, the constraints on $|\delta^u_{12}|_{LL}$
and $|\delta^u_{12}|_{RR}$, as well as $|\delta^u_{12}|_{LR}$ and
$|\delta^u_{12}|_{RL}$ are identical.
We see that $D$ meson mixing restricts the 
up and charm squark masses to be degenerate at the $1-10\%$ level
for most of the parameter space.  We note that our results 
numerically agree
with those recently computed by Ciuchini \etal~\cite{Ciuchini:2007cw}.

There are several other contributions to $D$ meson mixing within
the MSSM.  These are all mediated via box diagrams with internal
sparticle exchange and we now discuss each one in turn.  (i) The 
exchange of any of the 4 neutralinos $\chi^0_i$ ($i=1,4$) with the
up and charm squarks.  This contribution proceeds via mass insertions
in the squark propagators with flavor diagonal 
quark-squark-neutralino couplings as in the case of internal
squark-gluino exchange discussed above.  Since the couplings are of weak
interaction strength in this case, the magnitude of this contribution
is suppressed by the ratio $g^4/g_s^4$ compared to the squark-gluino
results and is thus numerically insignificant.  
(ii) The exchange of one of the neutralinos and
one gluino with the up and charm squarks.  This again proceeds via
the non-diagonal squark mass insertions at a rate of $g^2/g_s^2$
compared to the pure gluino-squark contribution.  Although larger
than the pure neutralino-squark contribution, it is still a 
sub-leading
effect.  (iii) The exchange of charginos $\chi^\pm_i$ ($i=1,2$) 
and all three down-type squarks inside the box diagram.  Here, the 
squark propagators are diagonal (mass insertions do not contribute 
in this case since the internal squarks are $Q=-1/3$) and the flavor
violation is given by the CKM structure of the 
quark-squark-chargino vertices.  However, the $Q=-1/3$ squarks
are constrained to be highly degenerate from their contributions 
(with gluino exchange) to $K\,, B_d\,, B_s$ meson mixing.  Thus a
supersymmetric-GIM mechanism is in effect, yielding nearly exact
cancellations, and rendering this contribution negligible.  This is
in contrast to the chargino-squark contributions to $K\,, B_d\,, 
B_s$ meson mixing, where the potentially non-degenerate stop
squark participates and can induce large contributions.  
(iv) The charged Higgs 
contribution of the two-Higgs-Doublet model of type II discussed
in Section~\ref{2hdm}.  As shown in that section, these
contributions are numerically small, even
in the case of large $\tan\beta$.  In summary, we see that 
all other supersymmetric
contributions to $D^0$-$\overline D^0$ mixing are numerically
insignificant compared to the squark-gluino exchange.  It is
interesting to note that stop-squarks do not contribute to
$D$ meson mixing.

%%%%%%%%%%%%%%%%%%%%%%%%%%%%%%%%%%%%%%%%%%%%%%%%%%
\subsection{Quark-Squark Alignment Models}

As we saw in the previous Section, in the MSSM 
there is a new ``flavor problem,'' namely, how to keep 
the contributions from the supersymmetric particles
to FCNC as small as the observations.  The conventional
solution is to impose constraints, such
as those derived above, of 
(i) degeneracy in the squark sector (except
for the special case of stop squarks), {\it i.e.} 
the diagonal sub-matrices $M_{LL}$ and $M_{RR}$ in 
Eq. (\ref{squarkmat}) should be 
proportional to the unit matrix, 
and (ii) the non-diagonal sub-matrices $M_{LR}$ should be proportional 
to the corresponding quark matrix. 

Nir and Seiberg~\cite{Nir:1993mx} have proposed an alternative to this 
picture where the quark and squark mass 
matrices are approximately aligned with each other. 
Their proposal is as follows:  if for some symmetry reason the
matrices corresponding to the
squark mass insertions, $\delta_{MN}$,  are themselves diagonal, 
then the squark contributions to FCNC vanish, regardless of 
the mass spectrum of the squarks.  
Corrections to this approximation are expected to remain
tolerably small and it should be possible to simultaneously
diagonalize the quark mass matrices and the squark mass-squared
matrices while essentially preserving flavor diagonal gluino 
interactions.

Within this framework, it is somewhat problematic to satisfy the 
constraints from $K^0$-$\overline K^0$ mixing.  Specific 
implementations of this proposal, based on Abelian horizontal 
symmetries, restrict the
supersymmetric contributions to Kaon mixing via a unique structure
for the down quark mass matrix using holomorphic 
zeros~\cite{Nir:2002ah}.  This implies that
Cabibbo mixing between the first and second generation quarks
must be induced by mixing in the up-quark sector, which in turn
leads to sizable supersymmetric contributions to 
$D^0$-$\overline D^0$ mixing.
In this case, mixing in the up-charm squark sector gives 
\beq
\left(\delta_{LL}\right)_{uc} = 
{\left(V_L^u \widetilde{M}^2 V_L^{u\dagger}\right)_{uc} \over \widetilde{m}^2}
\approx \theta_c \frac{\Delta \widetilde{m}^2_{uc}}{\widetilde{m}^2_q}\ \ ,
\label{sqalign}
\eeq
where $\theta_c$ is the Cabibbo angle, while
the $(\delta_{LR})_{uc}$ mass insertions can naturally remain small.
Mirroring the above discussion for MSSM, 
this leads to the effective Hamiltonian that mediates $D$ mixing
\begin{equation}
{\cal H}_A= {\alpha_s^2\over 2 m^2_{\tilde q}}C_1(m_{\tilde q})Q_1\ \ ,
\end{equation}
with
\begin{equation}
C_1(m_{\tilde q})={1\over 18} (\delta_{12}^u)^2_{LL}
[4xf_1(x)+11f_2(x)]\ \ ,
\end{equation}
where $f_{1,2}(x)$ with $x\equiv m^2_{\tilde g}/m^2_{\tilde q}$
are again given in the Appendix.  The RG evolution is
simple and yields
\beq
x_{\rm D}^{(A)} = {\alpha_s\over 3m^2_{\tilde q}}{f_D^2B_Dm_D\over
\Gamma_D} r_1(m_c,m_{\tilde q}) C_1(m_{\tilde q})\ \ .
\eeq
The bounds on $(\delta^u_{12})_{LL}$ from the current measurement
of $D$ meson mixing are given in the upper
left-hand panel of Fig.~\ref{mssmfig1}.  Using Eq.~(\ref{sqalign})
above, this results in the constraint on squark and gluino masses
of (assuming $m_{\tilde{q}} \approx m_{\tilde{g}}$ for simplicity)
$m_{\tilde g,\tilde q}\gsim 2$~TeV, which agrees with the results
in Refs.~\cite{Nir:2007ac,Ciuchini:2007cw}.  This would exclude
early discovery of Supersymmetry at the LHC, but leaves a discovery
window with higher luminosities as the LHC detectors are expected
to have a search reach of $m_{\tilde g,\tilde q}$ up to $2.5-3.0$~TeV
with 300~fb$^{-1}$ of integrated luminosity.

%%%%%%%%%%%%%%%
\subsection{Supersymmetry with R-Parity Violation}\label{SUSYRsection}

The conventional gauge symmetries of supersymmetry allow for the 
existence of additional terms in the superpotential that violate
baryon and lepton number.  The assumption of R-parity conservation
in the MSSM prohibits these terms, ensuring that baryon and lepton
number are conserved, and forbids related dangerous operators, {\it e.g.,} 
those that mediate proton decay.  However, it is possible to construct
alternative discrete symmetries~\cite{Ibanez:1991pr}, 
such as baryon-parity
or lepton-parity, that allow terms which violate either baryon or 
lepton number, but not both.  These symmetries also
forbid unwanted operators, and there is no strong theoretical
motivation to prefer R-parity over these alternative scenarios.  The R-parity
violating terms in the superpotential can be written as
\begin{equation}
W_{R_p}= {1\over 2}\lambda_{ijk}L_iL_j\bar E_k
+\lambda'_{ijk}L_iQ_j\bar D_k+{1\over 2}\lambda''_{ijk}
\bar U_i\bar D_j\bar D_k  \ \ .
\label{super}
\end{equation}
$i,j,k$ are generation indices and symmetry demands $i\neq j$
($j\neq k$) in the
terms proportional to $\lambda$ ($\lambda''$).  The quantities 
$L,E,Q,D,U$ in Eq.~(\ref{super}) 
are the chiral superfields in the MSSM, and the $SU(2)_L$, 
$SU(3)_C$ indices have been suppressed.  A bilinear term may also be
present, but it
can be rotated away and will not be considered here.  The
lepton number violating terms, $\lambda$
and $\lambda'$, cannot exist simultaneously with the
$\Delta B\neq 0$ term containing $\lambda''$.  The $\lambda'$ terms
have the same structure as the couplings for scalar leptoquarks, as
discussed in Section~\ref{ScalarLQSect}.  This model still contains
the minimal superfield content, but leads to a markedly different
supersymmetric phenomenology as sparticles can now be produced singly
and can mediate FCNC at tree-level.

The superfields are in the weak basis and should be
rotated to their mass eigenstates.  
The $\Delta L\neq 0$ $\lambda'$ term becomes~\cite{Agashe:1995qm}
\begin{equation}
W_{R_p} = \tilde\lambda'_{ijk}[N_iV_{jl}D_l-E_iU_j]\bar D_k\ \ ,
\end{equation}
with the definition
\begin{equation}
\tilde\lambda'_{ijk}\equiv \lambda'_{irs}{\cal U}^L_{rj}
{\cal D}^{*R}_{sk}\ \ .
\end{equation}
Here, ${\cal U}^L$ and ${\cal D}^R$ are the matrices which
rotate the left-handed up- and right-handed down-quark fields to
their mass basis.  Written in terms of component fields, the second
term in this superpotential contains the interactions
\begin{eqnarray}
W_{\lambda'}& = & \tilde\lambda'_{ijk}\left\{V_{jl}\left[ \tilde\nu^i_L
\bar d^k_Rd^l_L+\tilde d^l_L\bar d^k_R\nu^i_L
+(\tilde d^k_R)^*(\bar\nu^i_L)^cd^l_L\right]\right.\\
& & \left. \hspace{1.5cm} -\tilde e^i_L\bar d^k_Ru^j_L
-\tilde u^j_L\bar d^k_Re_L^i-(\tilde d_R^k)^*(\bar e^i_L)^cu_L^j
\right\}\ \ , \nonumber
\label{lrpar}
\end{eqnarray}
where the second line involving the up-quark sector is relevant for
$D^0$-$\overline D^0$ mixing.

Constraints on the size of these R-parity violating couplings have 
been obtained in the literature.  These limits are derived from
considerations of various processes~\cite{Allanach:1999ic} such as
charged current universality, semi-leptonic meson decays, rare
meson decays, atomic parity violation, double nucleon decay, neutron
oscillations, and $Z$ boson decays.  A compilation
of the $2\sigma$ bounds on the couplings relevant for $D^0$-$\bar D^0$
mixing are given in Table~\ref{rparcoupl}.  In addition, the recently
improved upper bound on the branching fraction for the process 
$D^+\to\pi^+e^+e^-$ of ${\cal B}<7.4\cdot 10^{-6}$ from 
CLEO-c~\cite{He:2005iz} yields the stringent 
restriction~\cite{Burdman:2001tf} on the product of
couplings $\tilde\lambda'_{12k}\tilde\lambda'_{11k}< 0.003 
\left(m_{\tilde d_{R,k}}/(100~{\rm GeV})\right)^2$. 
 
It is possible for the quark flavor 
rotations to generate flavor violation in the down- or up-quark
sectors, but not both.  In the case where the flavor rotations
occur in the up-quark sector only, large flavor changing effects
are expected in the $D$ meson system and the limits on the R-parity
violating couplings shown in Table~\ref{rparcoupl}  
become modified~\cite{Allanach:1999ic}.  However,  
this scenario is rather model dependent, 
we will adopt a more conservative, model-independent formalism 
in the following.

\begin{table}
\centering
\begin{tabular}{|c|c|c|c|c|c|} \hline\hline
$\tilde\lambda'_{11k}$ & $\tilde\lambda'_{12k}$ & 
$\tilde\lambda'_{21k}$ & $\tilde\lambda'_{22k}$ &
$\tilde\lambda'_{31k}$ & $\tilde\lambda'_{32k}$ \\ \colrule
$5\times 10^{-4} - 0.021$ & 0.043 & $0.021-0.059$ & $0.18-0.21$ 
& 0.11 & 0.52 \\ \hline\hline
$\tilde\lambda''_{11k}$ & $\tilde\lambda''_{21k}$ & 
$\tilde\lambda''_{12k}$ & $\tilde\lambda''_{22k}$ &
$\tilde\lambda''_{13k}$ & $\tilde\lambda''_{23k}$ \\ \colrule
$10^{-15}-10^{-4}$ & 1.23 & $10^{-15}-1.23$ & 1.23 & 
$10^{-4}-1.23$ & 1.23 \\ \hline\hline
\end{tabular}
\vskip .05in\noindent
\caption{\label{rparcoupl}
$2\sigma$ constraints on the R-parity violating couplings
which participate in $D$ mixing.  Here, $k=1,2,3$ with the exception that
$k\neq j$, where $j$ represents the middle index, for the $\lambda''$
couplings.  All numbers are scaled by the factor
$(m_{\tilde d_{R,k}}/100$ GeV).
Details of the derivation of these restrictions are given in
Refs.~\cite{Burdman:2001tf,Allanach:1999ic}.}
\end{table}

\begin{figure} [tb]
\centerline{
\includegraphics[width=9cm,angle=0]{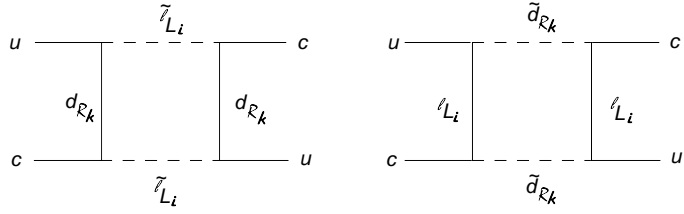}}
\caption{Contributions to $D^0$-$\overline D^0$ mixing from the
$\lambda'$ superpotential terms in supersymmetric
models with R-parity violation.
\label{rpardiag}}
\end{figure}

For the lepton number violating coupling $\tilde\lambda'$,
the first and third terms in the second line of Eq.~(128) mediate
$D^0$-$\bar D^0$ mixing via box diagrams where either the pair
$(\tilde \ell_{L,i} -d_{R,k})$ or $(\ell_{L,i} -\tilde d_{R,k})$ 
are exchanged internally with the assignment 
of the generational index $j=1,2$~\cite{Agashe:1995qm}.  The corresponding
Feyman diagrams are depicted in Fig.~\ref{rpardiag}.  Note that
there are no tree-level contributions as in the case of meson mixing
in the down-quark sector.
This is described at the high mass scale by the 
effective hamiltonian 
\begin{equation}
{\cal H}_{R_p}={1\over 128\pi^2}(\tilde\lambda'_{i2k}
\tilde\lambda'_{i1k})^2\left[ {1\over m_{\tilde\ell_{L,i}}^2}
+{1\over m_{\tilde d_{R,k}}^2}\right] Q_1\ \ ,
\end{equation}
where the dependence on the operator $Q_1$ is induced due to the fermion
propagator.
This interaction will yield constraints on the product of couplings 
$\tilde\lambda'_{i2k}\tilde\lambda'_{i1k}$.  Here, we have
assumed that only one set of the R-parity violating
couplings $\tilde\lambda'_{i2k}\tilde\lambda'_{i1k}$ (\ie, only one
value of $i$ and $k$)  is large
and dominant.  This is equivalent to saying that, {\it e.g.}, both 
sleptons and both down-type quarks being exchanged 
in the first box diagram shown in Fig.~\ref{rpardiag} 
are from the same generation.  In general, this
need not be the case and, for example, the coupling factor would then be the
product $\tilde\lambda'_{i2k}\tilde\lambda'_{m1k}
\tilde\lambda'_{m2n}\tilde\lambda'_{i1n}$,
with, {\it e.g.}, the set of $\tilde\ell_{L,i}, d_{R,k}, 
\tilde\ell_{L,m}, d_{R,n}$ 
being exchanged.

Matching at the
SUSY scale yields the Wilson coefficient
\begin{equation}
C_1(m_{\tilde q})={1\over 64\pi^2}(\tilde\lambda'_{i2k}
\tilde\lambda'_{i1k})^2\left(1+
{m^2_{\tilde d_{R,k}}\over m^2_{\tilde\ell_{L,i}}}\right)\ \ .
\end{equation}
Computing the evolution to the charm-quark scale yields 
\begin{equation}
{\cal H}_{R_p}= {1\over 2 m^2_{\tilde d_{R,k}}} C_1(m_c)Q_1\ ,
\end{equation}
with
\begin{equation}
C_1(m_c)=r_1(m_c,m_{\tilde q})C_1(m_{\tilde q})\ \ .
\end{equation} 
Evaluating the appropriate matrix element gives the 
$D$ mixing contribution for the R-parity violating $\lambda'$ 
terms, 
\begin{equation}
x^{(R_p)}_{\rm D} = {f_D^2B_D M_D\over 
3\Gamma_Dm^2_{\tilde d_R}}C_1(m_c)\ \ .
\end{equation}
Taking $m_{\tilde\ell_{L,i}}\simeq m_{\tilde d_{R,k}}$ for simplicity,
we obtain the constraint
\begin{equation}
{(\tilde\lambda'_{i2k}\tilde\lambda'_{i1k})^2\over m^2_{\tilde d_{R,k}} }
  \leq  x_{\rm D}^{\rm (expt)} {96\pi^2\Gamma_D\over 
f_D^2B_D M_Dr_1(m_c,m_{\tilde q})}\ \ ,
\end{equation}
which yields numerically
\begin{equation}
\tilde\lambda'_{i2k}\tilde\lambda'_{i1k}\leq 0.085 
\sqrt{x_{\rm D}^{\rm (expt)}} 
\left({m_{\tilde d_{R,k}}}\over 500~{\rm GeV}\right)
\ \ . 
\label{lamlim}
\end{equation}
We find that relaxing our assumption on the slepton mass and taking 
$m_{\tilde\ell_{L,i}}\leq m_{\tilde d_{R,k}}$ strengthens this
bound at most by a factor of 3.7 when $m_{\tilde\ell_{L,i}}=100$ GeV.
In computing the RGE evolution we used the value $m_{\tilde q}=500$ GeV 
and find little sensitivity in the evolution on the
squark mass once it is above the current experimental limit from
HERA of $\sim 300$ GeV~\cite{PDG}.  
It is trivial to scale our result in Eq.~(\ref{lamlim}) to
compare to the limits in Table~\ref{rparcoupl} which are based 
on setting $m_{\tilde d_{R,k}} =100$~GeV.  Taking, 
$m_{\tilde d_{R,k}} =500$~GeV, we see that the bounds on
$\tilde\lambda'_{i2k}\tilde\lambda'_{i1k}$ from 
$D^0$-$\overline D^0$ mixing are a factor of 50 (250) times stronger than
those in the Table for $i=2$ ($i=3$).

\begin{figure}[htbp]
\centerline{
\includegraphics[width=6cm,angle=90]{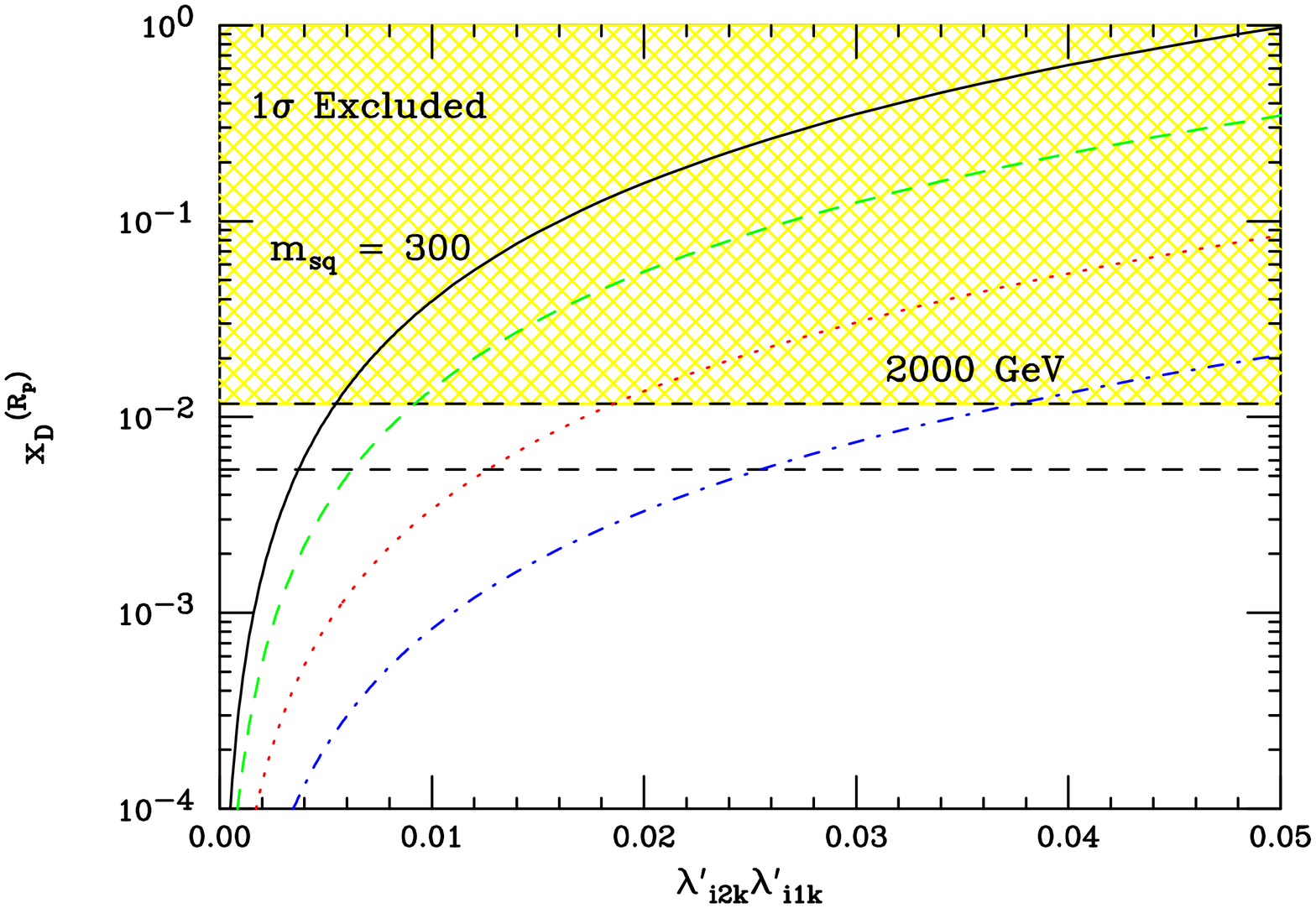}
%\hspace*{5mm}
\includegraphics[width=6cm,angle=90]{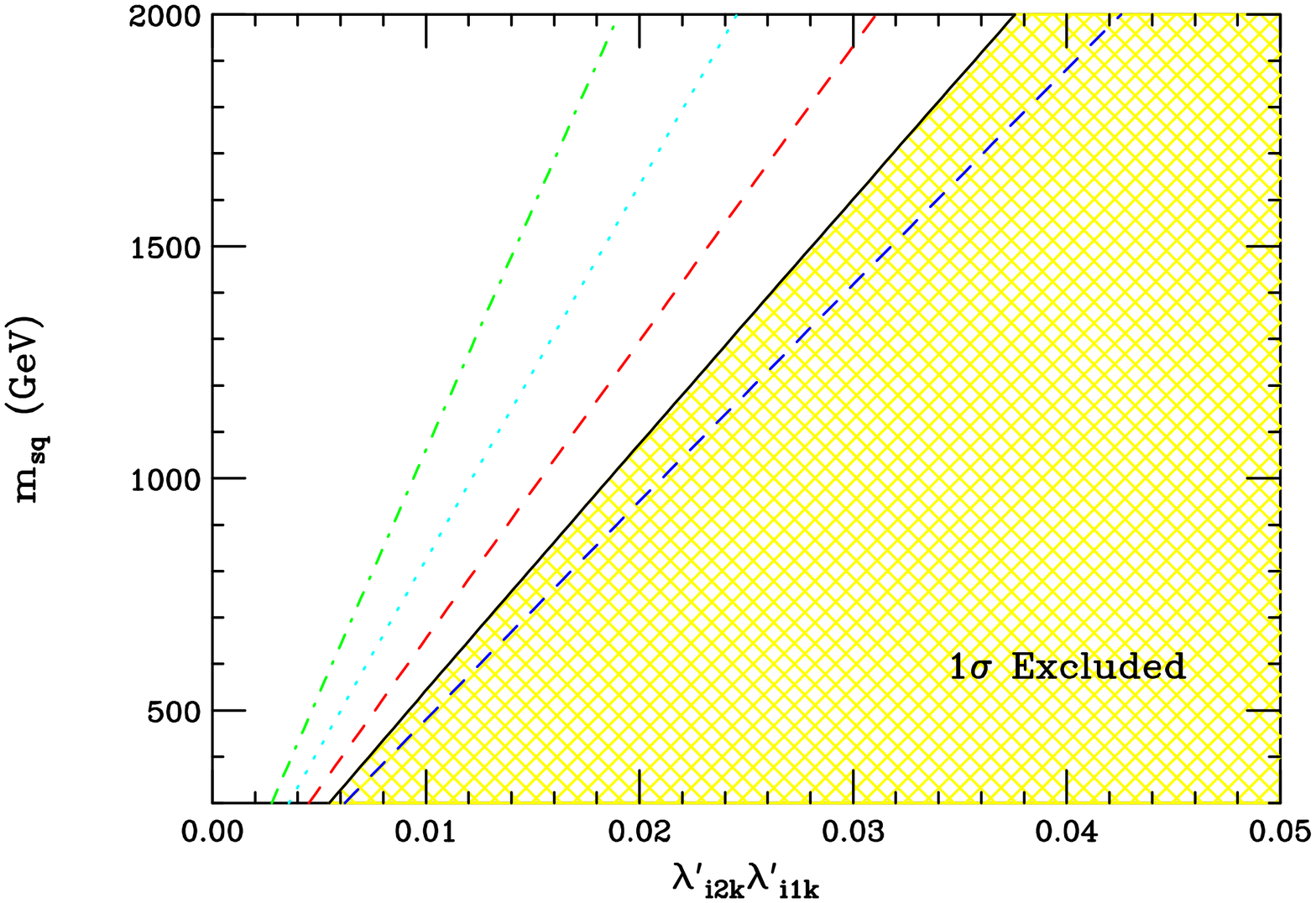}}
\vspace*{0.1cm}
\caption{Left: $x_{\rm D}$ in supersymmetry with R-parity violation 
as a function of the product of R-parity violating couplings 
$\tilde\lambda'_{i2k}\tilde\lambda'_{i1k}$ taking $m_{\tilde d_{R,k}}=\
m_{\tilde\ell_{L,i}}$, with $m_{\tilde d_{R,k}}=300\,, 500\,, 1000$
and 2000 GeV corresponding to the solid, green dashed, red dotted, and
blue dashed-dot curves, respectively.
The $1\sigma$ experimental
bounds are as indicated, with the yellow shaded region depicting the
region that is excluded.   
Right: $1\sigma$ excluded region in the R-parity violating coupling -
squark mass plane, as well as possible future contours taking
$x_{\rm D}< (15.0\,, 8.0\,, 5.0\,, 3.0)\times 10^{-3}$,
corresponding to the blue dashed, red dashed, cyan dotted, and 
green dot-dashed curves, respectively.}
\label{rparfig}
\end{figure}

The full numerical results for $x_{\rm D}$ and the constraints obtained in 
the R-parity violating coupling, squark mass parameter plane  are 
presented in Fig.~\ref{rparfig} in the limit 
$m_{\tilde\ell_{L,i}}\simeq m_{\tilde d_{R,k}}$.  
We see that $D$ meson mixing provides stringent constraints
on R-parity violating couplings.  These bounds can be directly translated
to constraints on the couplings of scalar leptoquarks as discussed
in a previous Section.

The baryon number violating $\lambda''$ couplings can also contribute
to $D^0$-$\bar D^0$ mixing.  We remind the reader that they cannot
exist simultaneously with the lepton number violating terms in the
superpotential. They participate in $D$ mixing via $d_R$-quark and
$\tilde d_R$ exchange in the box diagram.  The formalism is
analogous to the $\Delta L\neq 0$ case above.  The effective hamiltonian
at the SUSY mass scale is
\begin{equation}
{\cal H}_{R_p}= {1\over 128\pi^2}(\tilde\lambda''_{1jk}
\tilde\lambda''_{2jk})^2
\left[{1\over m^2_{\tilde d_{R,j}}} +{1\over m^2_{\tilde d_{R,k}}}
\right] Q_1\ \ .
\end{equation}
Recall that symmetry dictates $j\neq k$.
After completing the RGE evolution as described above, and 
assuming the charged $-1/3$ squarks are degenerate, we have
\begin{equation}
x^{(R_p)}_{\rm D} = {f_D^2B_D M_D\over 96\pi^2m^2_{\tilde d_R}\Gamma_D}
r_1(m_c,m_{\tilde d_R})(\tilde\lambda''_{1jk}
\tilde\lambda''_{2jk})^2\ \ .
\end{equation}
This yields the constraint
\begin{equation}
\tilde\lambda''_{1jk}\tilde\lambda''_{2jk}\leq 0.085 
\sqrt{x_{\rm D}^{\rm (expt)}}
 \left({m_{\tilde d_{R}}}\over 500{\rm GeV}\right)\ \ ,
\end{equation}
which mirrors that for the lepton number violating scenario.

%%%%%%%%%%%%%%%

%%%%%%%%%%%%%%%
\subsection{Split Supersymmetry}

Lastly, for completeness, we briefly 
discuss the case of Split Supersymmetry~\cite{nimagian}.  This
scenario postulates that supersymmetry breaking occurs at a very high
scale, $m_S\gg 1000$ TeV.  The scalar particles all acquire masses at
this high scale, except for a single neutral Higgs boson, whose mass is
either finely-tuned or is preserved by some other mechanism.  Split
Supersymmetry proponents argue that this tuning may, indeed, be present
in Nature, perhaps being related to the cosmological constant problem 
(which suffers an even greater degree of fine-tuning).  The fermions
in this theory, including the gauginos, are assumed to be protected by
chiral symmetries and thus can have weak-scale masses.  This feature
preserves the gauge coupling unification found in supersymmetric models,
and provides a natural Dark Matter candidate in the lightest neutralino.
One important consequence of this scenario is that since all the scalar
fields are present at only a very high scale, they decouple from physics
at the TeV scale and their contributions to
FCNC in the flavor sector are negligible.  
Since all contributions to $D^0$-$\overline D^0$
mixing in supersymmetry involve the internal exchange of scalar quarks,
we expect these effects to essentially vanish in this scenario.

%%%%%%%%%%%%%%%
%%%%%%%%%%%%%%%%%%%%%%%%%%%%%%%%%%%%%%%%%%%%%%%%%%%%%%%%%%%
\section{Conclusions}
The recent BaBar and Belle findings on $D^0$-${\bar D}^0$ 
mixing have brought the long standing search for 
this phenomenon to a successful conclusion, 
although much remains to be done.  Compared to mixing in the 
other flavor sectors, the observed value for charm 
({\it cf} Eq.~(\ref{hfag})) is by far the smallest,
\begin{equation}\label{flmix}
x_{\rm K} \simeq 0.47 \ , \qquad x_{{\rm B}_d} \simeq 0.776 \ ,
\qquad x_{{\rm B}_s} \simeq 26. \ , \qquad x_{\rm D} \simeq 0.009 \ \ .
\end{equation} 
In our opinion, the measured value for $x_{\rm D}$ is in accord 
with expectations of the Standard Model, with the proviso that hadronic 
(rather than quark-level) effects are the dominating influence 
({\it cf} Sect.~III).  We have argued that the relatively small magnitude 
of charm mixing could afford New Physics an enhanced chance to compete 
successfully with the Standard Model.  It should be kept 
in mind, however, that the current experimental value of 
$x_{\rm D}$ is relatively imprecise and that 
the SM theoretical determination contains hadronic 
uncertainties.  These facts tend to frustrate the attempt to disentangle 
any potentially large NP contribution from that of the SM.  

By design, our study of NP contributions has addressed a rather broad 
spectrum of possibilities.  We have avoided playing 
favorites among the NP models contained in this paper, letting 
the results speak for themselves.  Since the average reader is unlikely 
to be conversant with the details of such 
a large array of NP models, our presentation has been 
pedagogical in nature.  We have tried to precede any formula 
for $x_{\rm D}^{\rm (NP)}$ with a summary of the relevant background. 

A work such as this is meant to constrain the parameter spaces of 
NP models.  The case of $D^0$ mixing is especially interesting 
because the intermediate states which generate the 
$D^0$-to-${\bar D}^0$ transitions are distinct from those occurring in 
$K$, $B_d$ and $B_s$ mixing.  
Typically, NP parameters will involve the masses of 
yet-to-be-discovered particles and their coupling strengths to 
ordinary matter.   In some cases (Left-Right Symmetric Model,
Split Supersymmetry, Universal 
Extra Dimensions, Flavor Conserving Two-Higgs Doublets) 
we have found that the NP model will not generate a 
$D^0$-${\bar D}^0$ signal at the observed level for any values 
of its parameters.   More often, however, this is not the case and 
for some models (Split Fermions, Flavor Changing Neutral Higgs) 
the constraints can be strong.    

The main quantitative conclusions for this work appear 
in the set of figures which appear throughout the paper.
For convenience, we have compiled a summary of our results 
in Table~\ref{tab:bigtableofresults}, using the 
$1\sigma$ value $x_{\rm D} < 11.7 \cdot 10^{-3}$ to mark the  boundary
between allowed and excluded regions.  
Such a list is by nature approximate, and we refer the reader to 
the body of the paper for a more precise presentation of our 
results.  
\begin{table}[t]
\begin{tabular}{|c||c|}
\colrule\hline 
Model & Approximate Constraint 
\\ \hline\hline
Fourth Generation (Fig.~\ref{4gen}) &\ \ $|V_{ub'} V_{cb'}|\cdot m_{b'}  
<   0.5 $~(GeV)  
\ \ \\
$Q=-1/3$ Singlet Quark (Fig.~\ref{SingletQuark13Fig}) 
&  $s_2\cdot m_S  < 0.27$~(GeV) 
\\
$Q=+2/3$ Singlet Quark (Fig.~\ref{SingletQuark23Fig}) 
&  $|\lambda_{uc}| < 2.4 \cdot 10^{-4}$ \\
Little Higgs  &  Tree: See entry for $Q=-1/3$ Singlet Quark \\
& Box: Region of parameter space can reach 
observed $x_{\rm D}$
\\
Generic $Z'$ (Fig.~\ref{ZprimeFig}) 
&  $M_{Z'}/C > 2.2\cdot 10^3$~TeV  \\
Family Symmetries (Fig.~\ref{Family}) & $m_1/f>1.2\cdot 10^{3}$~TeV 
 (with $m_1/ m_2 = 0.5$)   \\
Left-Right Symmetric  (Fig.~\ref{lrmfig}) & No constraint   \\
Alternate Left-Right Symmetric (Fig.~\ref{alrmfig}) &
$M_R>1.2$~TeV ($m_{D_1}=0.5$~TeV)   \\
 & ($\Delta m/m_{D_1})/M_R>0.4$~TeV$^{-1}$\\
Vector Leptoquark Bosons (Fig.~\ref{VectorLeptoquarksFig}) 
& $M_{VLQ} > 55 (\lambda_{PP}/0.1) $~TeV  \\
Flavor Conserving Two-Higgs-Doublet (Fig.~\ref{2hdmfig}) 
&   No constraint \\
Flavor Changing Neutral Higgs  (Fig.~\ref{NeutralHiggsFig}) 
&  $m_H/C>2.4\cdot 10^3$~TeV \\
FC Neutral Higgs (Cheng-Sher ansatz) (Fig.~\ref{chengsher}) & 
$m_H/|\Delta_{uc}|>600$~GeV\\
Scalar Leptoquark Bosons  & See entry for RPV SUSY \\
Higgsless (Fig.~\ref{HiggslessFig})  & $M > 100$~TeV  \\
Universal Extra Dimensions & No constraint \\
Split Fermion (Fig.~\ref{splitfig})  
& $M / |\Delta y| > (6 \cdot 10^2~{\rm GeV})$
 \\
Warped Geometries (Fig.~\ref{warpedfig}) &  $M_1 > 3.5$~TeV \\
Minimal Supersymmetric Standard (Fig.~\ref{mssmfig1})  & 
$|(\delta^u_{12})_{\rm LR,RL}| 
< 3.5 \cdot 10^{-2}$ for ${\tilde m}\sim 1$~TeV  \\
  & 
$|(\delta^u_{12})_{\rm LL,RR}| < .25 $ for ${\tilde m}\sim 1$~TeV  
\\
Supersymmetric Alignment & ${\tilde m} > 2$~TeV  \\
Supersymmetry with RPV (Fig.~\ref{rparfig}) & 
$\lambda'_{12k} \lambda'_{11k}/m_{\tilde d_{R,k}} < 
1.8 \cdot 10^{-3}/100$~GeV
\\
Split Supersymmetry & No constraint  \\
\hline\hline
\end{tabular}
\vskip .05in\noindent
\caption{Approximate constraints on NP models from $D^0$ mixing.}
\label{tab:bigtableofresults}
\end{table}

We recommend further experimental study of this subject on two 
fronts.  First, of course, is the need to reduce error bars in the 
measured values of $y_{\rm D}$ and especially $x_{\rm D}$. 
Equally important is continuing the search for evidence 
of CP violation in mixing for the $D^0$ system.  CP violation 
provides an interesting contrast with $D^0$ mixing because 
it provides an independent arena for competition between 
the SM and NP signals.  There is especially room for improvement 
in the SM analysis of charm CP violation, 
and work on this is underway.

%%%%%%%%%%%%%%%%%%%%%%%%%%%%%%%%%%%%%%%%%%%%%%%%%%%%%%%%%%%
\acknowledgments

The work of E.G. was supported in part by the U.S.\ National Science
Foundation under Grant PHY--0555304, J.H. was supported by the U.S.
Department of Energy under Contract DE-AC02-76SF00515, 
S.P. was supported by the U.S.\ Department of 
Energy under Contract DE-FG02-04ER41291 and 
A.P.~was supported in part by the U.S.\ National Science Foundation under
CAREER Award PHY--0547794, and by the U.S.\ Department of Energy 
under Contract DE-FG02-96ER41005.  J.H. would like to thank the
theoretical physics group at Fermilab for their hospitality 
and E.G., J.H., and A.P. thank
the High Energy Physics Group at the University of Hawaii for
their hospitality while part of this work was performed.  We would like
to thank C.F. Berger, S. Chivukula, B. Dobrescu, J. Donoghue, 
K.C. Kong, B. Lillie, E. Lunghi, E. Simmons and 
T. Tait for discussions related to this work 
and especially T. Browder, T. Rizzo, and X. Tata for a careful reading of the 
manuscript.
%%%%%%%%%%%%%%%%%%%%%%%%%%%%%%%%%%%%%%%%%%%%%%%%%%%%%%%%%%%%
\appendix\section{Collected Formulae}

Here, we collect the formulae used throughout the manuscript to compute the
contributions to $D^0$-$\overline D^0$ mixing in the various New Physics
models.

{\it Inami-Lim}: (from Ref.~\cite{Inami:1980fz})

The loop functions, first calculated by Inami Lim, apply for several NP 
scenarios discussed in the text. For a contribution from two internal
quarks of the same flavor in the box diagram, the loop function is 
\begin{equation} 
 S(x) = x\left[{1\over 4}+{9\over 4(1-x)}-{3\over 2(1-x)^2}\right]
-{3x^3\over 2(1-x)^3}\ln x\ \ , 
\end{equation}
and for two quarks of different flavors,
\begin{eqnarray}
& & S(x_i,x_j) = x_i x_j \left( {\ln x_i \over x_i - x_j} 
\left[ {1\over 4}+ {3\over 2(1-x_i)}-{3\over 4(1-x_i)^2}\right] 
+ (x_i \leftrightarrow x_j) \right. \nonumber 
\\
& & \left. \hspace{5.0cm} - {3\over 4(1-x_i)(1-x_j)} \right) \ \ .
\end{eqnarray}

{\it Little Higgs}: (from Ref.~\cite{Hubisz:2005bd})

The loop function for the case where the mirror fermions and heavy
gauge bosons are exchanged in the box diagram is given by
\begin{eqnarray}
 F_{LH}(z_i,z_j) & = & {1\over (1-z_i)(1-z_j)}\left( 1-{7\over 4}z_iz_j\right)
+{z_i^2\log z_i)\over (z_i-z_j)(1-z_i)^2}\left(1-2z_j+{z_iz_j\over 4}\right)
\nonumber\\
& & -{z_j^2\log z_j\over (z_i-z_j)(1-z_j)^2}\left(1-2z_i+{z_i z_j\over 4}\right)
\nonumber\\
& & -{3\over 4}\left( {1\over (1-z_i)(1-z_j)}+
{z_i^2\log z_i\over (z_i-z_j)(1-z_i)^2}
-{z_j^2\log z_j\over (z_i-z_j)(1-z_j)^2} \right)\\
& & - { 3\over 100a } \left( {1\over (1-z_i')(1-z_j') }
+ { z_i'z_i\log z_i'\over (z-I-z_j)(1-z_i')^2)}
- {z_j'z_j\log z_j'\over (z_i-z_j)(1-z_j')^2} \right) \nonumber \\
& & -{3\over 10}\left( {\log a\over (a-1)(1-z_i')(1-z_j')}
+{z_i^2\log z_i\over (z_i-z_j)(1-z_i)(1-z_i')}\right. \nonumber\\
& & \left. -{z_j^2\log z_j\over (z_i-z_j)(1-z_j)(1-z_j')}\right)\ \ , \nonumber
\end{eqnarray}
with
\begin{equation}
z_i={m_{M_i}^2\over M^2_{W_H}}\quad\quad
z_i'={m_{M_i}^2\over M^2_{A_H}}\quad\quad
a={5\over\tan^2\theta_w} \ \ .
\end{equation}

{\it Left-Right Symmetric Model}:

The loop function with one $W_L$ and one $W_R$ boson being exchanged in
the box diagram is
\begin{equation}
J(x,\beta) = {x\beta\ln\beta\over (1-\beta)(1-\beta x)^2}-
{x+x\ln x\over (1-x)(1-\beta x)} \ \ .
\end{equation}

{\it Charged Higgs}: (from Ref.~\cite{bhp})

The loop functions with one $H^\pm$ and one SM $W$ boson and with two
$H^\pm$ being exchanged in the box diagram are
\begin{eqnarray}
A_{\rm HH}(x,y) & = & {x^2\over 4}\left[ {x+y\over (x-y)^2}-{2xy\over
(x-y)^3}\ln {x\over y}\right]\ , \nonumber \\
A_{\rm WH}(x,y) & = & 2x^2\left[{1\over (x-y)(1-x)}+{y\ln y\over (x-y)^2(1-y)}
+{(x^2-y)\ln x\over (x-y)^2(1-x)^2}\right. \nonumber\\
& & \left. -{1\over 4}\left( {x\over (x-y)(1-x)} + {y^2\ln y\over (1-y)(x-y)^2}
+{x(x+xy-2y)\ln x\over (x-y)^2(1-x)^2}\right)\right]\ \ .
\end{eqnarray}

{\it Cheng-Sher box}: 

The loop function in the Cheng-Sher ansatz with flavor changing neutral
Higgs bosons for a top quark and neutral Higgs being exchanged in a
box diagram is
\begin{equation}
F_{tH}(x)={-1\over 1-x} -{\ln x\over (1-x)^2} + {x^2-4x+3+2\ln x\over
2(1-x)^3}\ \ .
\end{equation}

{\it Universal Extra Dimensions}: 

The expressions for the case of Universal Extra Dimenions are
given in full in Ref.~\cite{Buras:2002ej}.

{\it Warped Extra Dimensions}: (from Ref.~\cite{rswall})

The coupling of two zero-mode fermions to the $n^{th}$ gauge
boson KK state, $A^{(n)}$, relative to the SM coupling strength is 
\begin{equation}
C^{f\bar fA}_{oon}={g^{(n)}\over g_{SM}}= \sqrt{2\pi kr_c}
\left[ {1-2c_f\over 1-e^{-\pi kr_c(1-2c_f)}}\right]
\int_{e^{-\pi kr_c}}^1 dz\, z^{(1-2c_f)}{J_1(x_nz)+\alpha_nY_1(x_nz)
\over |J_1(x_n)+\alpha_nY_1(x_n)|}\ \ ,
\end{equation}
where the roots $x_n$ for the gauge boson KK spectrum are given by
\begin{equation}
J_1(x_n)+x_nJ'_1(x_n)+\alpha_n\left[ Y_1(x_n)+
x_nY'_1(x_n)\right] =0\ \ ,
\end{equation}
and $\alpha_n$ is defined by
\begin{equation}
\alpha_n={J_1(m_n/k)+(m_n/k)J'_1(m_n/k)\over
Y_1(m_n/k)+(m_n/k)Y'_1(m_n/k)}\ \ .
\end{equation}

{\it Supersymmetry}: (from Ref.~\cite{bigsusyfcnc}) 

The loop functions from squark and gluino exchange in a box diagram
with squark mass insertions are given by
\bea
f_1(x) &=& \frac{6(1+3x)\ln x + x^3 - 9x^2 - 9x+17}{6(1-x)^5}\ \ , 
\nonumber \\
f_2(x) &=& \frac{6x(1+x)\ln x - x^3 - 9x^2 + 9x+1}{3(1-x)^5}\ \ .
\eea

%%%%%%%%%%%%%%%%%%%%%%%%%%%%%%%%%%%%%%%%%%%%%%%%%%%%%%%%%%%

\end{document}